\input harvmac
\input epsf

\def\CF{{\cal F}}
%
\let\includefigures=\iftrue
%
%
%
%
%
\input rotate
\noblackbox
%
%
\includefigures
\message{If you do not have epsf.tex (to include figures),}
\message{change the option at the top of the tex file.}
\def\figin{\epsfcheck\figin}\def\figins{\epsfcheck\figins}
\def\epsfcheck{\ifx\epsfbox\UnDeFiNeD
\message{(NO epsf.tex, FIGURES WILL BE IGNORED)}
\gdef\figin##1{\vskip2in}\gdef\figins##1{\hskip.5in}
\else\message{(FIGURES WILL BE INCLUDED)}%
\gdef\figin##1{##1}\gdef\figins##1{##1}\fi}
\def\DefWarn#1{}
\def\N{{\cal N}}
\def\figinsert{\goodbreak\midinsert}
\def\ifig#1#2#3{\DefWarn#1\xdef#1{fig.~\the\figno}
\writedef{#1\leftbracket fig.\noexpand~\the\figno}%
\figinsert\figin{\centerline{#3}}\medskip\centerline{\vbox{\baselineskip12pt
\advance\hsize by -1truein\noindent\footnotefont{\bf
Fig.~\the\figno:} #2}}
\bigskip\endinsert\global\advance\figno by1}
\else
\def\ifig#1#2#3{\xdef#1{fig.~\the\figno}
\writedef{#1\leftbracket fig.\noexpand~\the\figno}%
\global\advance\figno by1} \fi

\def\tilde{\widetilde}

\def\subsubsec#1{\bigskip\noindent{\it #1}}
\def\yboxit#1#2{\vbox{\hrule height #1 \hbox{\vrule width #1
\vbox{#2}\vrule width #1 }\hrule height #1 }}
\def\fillbox#1{\hbox to #1{\vbox to #1{\vfil}\hfil}}
\def\ybox{{\lower 1.3pt \yboxit{0.4pt}{\fillbox{8pt}}\hskip-0.2pt}}

\def\rightarrowbox#1#2{
  \setbox1=\hbox{\kern#1{${ #2}$}\kern#1}
  \,\vbox{\offinterlineskip\hbox to\wd1{\hfil\copy1\hfil}
    \kern 3pt\hbox to\wd1{\rightarrowfill}}}

\def\half{{1\over 2}}
\def\Tr{{{\rm Tr~ }}}
\def\tr{{\rm tr\ }}

\def\Re{{\rm Re\hskip0.1em}}

\def\CF{{\cal F}}

\def\CN{{\cal N}}

\def\tilde{\widetilde}

\def\II{\relax{I\kern-.10em I}}

\def\bar{\overline}

\def\IZ{\relax\ifmmode\mathchoice
{\hbox{\cmss Z\kern-.4em Z}}{\hbox{\cmss Z\kern-.4em Z}}
{\lower.9pt\hbox{\cmsss Z\kern-.4em Z}} {\lower1.2pt\hbox{\cmsss
Z\kern-.4em Z}}\else{\cmss Z\kern-.4em Z}\fi}
\def\IB{\relax{\rm I\kern-.18em B}}
\def\IC{{\relax\hbox{$\inbar\kern-.3em{\rm C}$}}}
\def\ID{\relax{\rm I\kern-.18em D}}
\def\IE{\relax{\rm I\kern-.18em E}}
\def\IF{\relax{\rm I\kern-.18em F}}
\def\IG{\relax\hbox{$\inbar\kern-.3em{\rm G}$}}
\def\IGa{\relax\hbox{${\rm I}\kern-.18em\Gamma$}}
\def\IH{\relax{\rm I\kern-.18em H}}
\def\II{\relax{\rm I\kern-.18em I}}
\def\IK{\relax{\rm I\kern-.18em K}}
\def\IN{\relax{\rm I\kern-.18em N}}
\def\IP{\relax{\rm I\kern-.18em P}}

%
\def\inbar{\,\vrule height1.5ex width.4pt depth0pt}

\font\cmss=cmss10 \font\cmsss=cmss10 at 7pt
\def\IR{\relax{\rm I\kern-.18em R}}

\def\lp10{l_P^{10}}
\def\lp11{l_P^{11}}
\def\R11{R_{11}}

\def\sig{\sigma}


\def\a{\alpha}
\def\b{\beta}

\def\GG{{\rm Gal}(\bar\Bbb{Q}/\Bbb{Q})}

\def\doublefig#1#2#3#4{\DefWarn#1\xdef#1{fig.~\the\figno}
\writedef{#1\leftbracket fig.\noexpand~\the\figno}%
\figinsert\figin{\centerline{#3\hskip1.0cm#4}}\medskip\centerline{\vbox{
\baselineskip12pt\advance\hsize by -1truein
\noindent\footnotefont{\bf Fig.~\the\figno:} #2}}
\bigskip\endinsert\global\advance\figno by1}

\newbox\tmpbox\setbox\tmpbox\hbox{\abstractfont
}
 \Title{\vbox{\baselineskip12pt\hbox
{hep-th/0611072}\hbox{RUNHETC-2006-27}}}
 {\vbox{\centerline{Children's Drawings From Seiberg-Witten Curves} }}
\smallskip
\centerline{Sujay K. Ashok$^a$, Freddy Cachazo$^a$, and Eleonora Dell'Aquila$^b$}
\smallskip
\bigskip
\centerline{\it $^a$Perimeter Institute for Theoretical Physics}
\centerline{\it Waterloo, Ontario, ON N2L 2Y5, Canada}
\bigskip
\centerline{\it $^b$NHETC, Department of Physics, Rutgers University}
\centerline{\it 136, Frelinghuysen Road, Piscataway NJ 08854, USA}
\bigskip
\vskip 1cm \noindent

\input amssym.tex

We consider $\CN=2$ supersymmetric gauge theories perturbed by tree
level superpotential terms near isolated singular points in the Coulomb
moduli space. We identify the Seiberg-Witten curve at these points
with polynomial equations used to construct what Grothendieck called
``dessins d'enfants" or ``children's drawings" on the Riemann sphere.
From a mathematical point of view, the dessins are important because
the absolute Galois group ${\rm Gal}(\bar \Bbb{Q}/\Bbb{Q})$ acts
faithfully on them. We argue that the relation between the dessins
and Seiberg-Witten theory is useful because gauge theory criteria
used to distinguish branches of $\CN=1$ vacua can lead to
mathematical invariants that help to distinguish dessins belonging
to different Galois orbits. For instance, we show that the
confinement index defined in hep-th/0301006 is a Galois invariant.
We further make some conjectures on the relation between
Grothendieck's programme of classifying dessins into Galois orbits
and the physics problem of classifying phases of $\CN=1$ gauge theories.

\Date{November 2006}

\lref\leila{Leila Schneps, "The Grothendieck Theory of Dessins d'Enfants," London Mathematical Society Lecture Note Series, vol 200, 1994.}

\lref\melanie{M.~Wood, ``Belyi-extending maps and the Galois action on dessins d'enfants," math.NT/0304489.
}

\lref\AshokCD{
 S.~K.~Ashok, F.~Cachazo and E.~Dell'Aquila,
   ``Strebel Differentials With Integral Lengths And Argyres-Douglas Singularities,'' hep-th/0610080.
}

\lref\MSone{
  G.~W.~Moore and N.~Seiberg,
  ``Polynomial Equations for Rational Conformal Field Theories,''
  Phys.\ Lett.\ B {\bf 212}, 451 (1988).
}

\lref\MStwo{
  G.~W.~Moore and N.~Seiberg,
  ``Naturality in Conformal Field Theory,"
  Nucl.\ Phys.\ B {\bf 313}, 16 (1989).
}

\lref\MSthree{
  G.~W.~Moore and N.~Seiberg,
  ``Classical and quantum Conformal Field Theory,''
  Commun.\ Math.\ Phys.\  {\bf 123}, 177 (1989).
}

\lref\SWone{
  N.~Seiberg and E.~Witten,
   ``Electric - magnetic duality, monopole condensation, and confinement in N=2
  supersymmetric Yang-Mills theory,''
  Nucl.\ Phys.\ B {\bf 426}, 19 (1994)
  [Erratum-ibid.\ B {\bf 430}, 485 (1994)], hep-th/9407087.
}

\lref\SWtwo{
  N.~Seiberg and E.~Witten,
   ``Monopoles, duality and chiral symmetry breaking in N=2 supersymmetric QCD,''
  Nucl.\ Phys.\ B {\bf 431}, 484 (1994), hep-th/9408099.
}

\lref\CachazoDSW{
  F.~Cachazo, M.~R.~Douglas, N.~Seiberg and E.~Witten,
  ``Chiral rings and anomalies in supersymmetric gauge theory,''
  JHEP {\bf 0212}, 071 (2002), hep-th/0211170.
}

\lref\CachazoSWtwo{
  F.~Cachazo, N.~Seiberg and E.~Witten,
  ``Chiral rings and phases of supersymmetric gauge theories,''
  JHEP {\bf 0304}, 018 (2003), hep-th/0303207.
}

\lref\AD{
  P.~C.~Argyres and M.~R.~Douglas,
  ``New phenomena in SU(3) supersymmetric gauge theory,''
  Nucl.\ Phys.\ B {\bf 448}, 93 (1995), hep-th/9505062.
}

\lref\Douglas{
  M.~R.~Douglas and S.~H.~Shenker,
  ``Dynamics of SU(N) supersymmetric gauge theory,''
  Nucl.\ Phys.\ B {\bf 447}, 271 (1995), hep-th/9503163.
}

\lref\CachazoIV{
  F.~Cachazo, K.~A.~Intriligator and C.~Vafa,
  ``A large N duality via a geometric transition,''
  Nucl.\ Phys.\ B {\bf 603}, 3 (2001), hep-th/0103067.
}

\lref\CachazoV{
 F.~Cachazo and C.~Vafa,
  ``N = 1 and N = 2 geometry from fluxes,''
  arXiv:hep-th/0206017.
}

\lref\CachazoYC{
  F.~Cachazo, N.~Seiberg and E.~Witten,
  ``Phases of N = 1 supersymmetric gauge theories and matrices,''
  JHEP {\bf 0302}, 042 (2003), hep-th/0301006.
}

\lref\Belyi{
G~.V~.Belyi, ``On Galois extensions of a maximal cyclotonic field," Math. U.S.S.R. Izvestija {\bf 14} (1980), 247-256 }

\lref\G{
A. Grothendieck, ``Esquisse d'un Programme," Preprint 1985.
}

\lref\Strebel{
K. ~Strebel, ``Quadratic Differentials," Springer-Verlag, 1984
}

\lref\Bala{
  V.~Balasubramanian, B.~Feng, M.~x.~Huang and A.~Naqvi,
  ``Phases of N = 1 supersymmetric gauge theories with flavors,''
  Annals Phys.\  {\bf 310}, 375 (2004), hep-th/0303065.
}

\lref\WittenM{
  E.~Witten,
  ``Monopoles and four manifolds,''
  Math.\ Res.\ Lett.\  {\bf 1}, 769 (1994), hep-th/9411102.
}

\lref\HananyO{
  A.~Hanany and Y.~Oz,
   ``On the quantum moduli space of vacua of N=2 supersymmetric SU(N(c)) gauge theories,''
  Nucl.\ Phys.\ B {\bf 452}, 283 (1995), hep-th/9505075.
}

\lref\ArgyresPS{
  P.~C.~Argyres, M.~R.~Plesser and A.~D.~Shapere,
  ``The Coulomb phase of N=2 supersymmetric QCD,''
  Phys.\ Rev.\ Lett.\  {\bf 75}, 1699 (1995), hep-th/9505100.
}

\lref\JonesS{
G.~ Jones and M.~Streit, ``Galois Groups, Monodromy Groups and Cartographic Groups Geometric Galois Actions: 2. The Inverse Galois Problem, Moduli Spaces and Mapping Class
Groups," London Math. Soc. Lecture Notes Series {\bf 243}, Cambridge Univ. Press 1997, 25-65.
}

\lref\Shih{
  D.~Shih,
  ``Singularities of N = 1 supersymmetric gauge theory and matrix models,''
  JHEP {\bf 0311}, 025 (2003), hep-th/0308001.
}

\lref\DVone{
  R.~Dijkgraaf and C.~Vafa,
  ``Matrix models, topological strings, and supersymmetric gauge theories,''
  Nucl.\ Phys.\ B {\bf 644}, 3 (2002), hep-th/0206255.
}

\lref\DVtwo{
 R.~Dijkgraaf and C.~Vafa,
  ``On geometry and matrix models,''
  Nucl.\ Phys.\ B {\bf 644}, 21 (2002), hep-th/0207106.
}

\lref\DVthree{
 R.~Dijkgraaf and C.~Vafa,
  ``A perturbative window into non-perturbative physics,'' hep-th/0208048.
}

\lref\Shabat{
G.~Shabat and A.~Zvonkin, ``Plane Trees and Algebraic Numbers," in Jerusalem Combinatorics '93, 178 AMS Publications.
}

\lref\fraleigh{John B. Fraleigh, ``First Course in Abstract Algebra"
Addison-Welsley, Third Edition.}

\lref\EguchiS{T.~Eguchi and Y.~Sugawara, ``Branches of $\CN=1$ Vacua and Argyres-Douglas Points," JHEP {\bf 0305}:063, 2003, hep-th/0305050.}

\lref\SeibergF{
 N.~Seiberg,
   ``Adding fundamental matter to 'Chiral rings and anomalies in supersymmetric gauge theory',''
  JHEP {\bf 0301}, 061 (2003), hep-th/0212225.
}

\lref\Kapustin{
A.~Kapustin,
  ``The Coulomb branch of N = 1 supersymmetric gauge theory with adjoint  and fundamental matter,''
  Phys.\ Lett.\ B {\bf 398}, 104 (1997), hep-th/9611049.
}


\newsec{Introduction}

The interest in $\CN=2$ supersymmetric gauge theories is especially
due to the seminal work of Seiberg and Witten in the mid nineties
\refs{\SWone,\SWtwo}. They showed that the low energy action and
infrared dynamics of the gauge theory on the Coulomb branch can be
completely solved and that all the relevant  information about the
low energy theory is encoded in a hyperelliptic curve and in an
associated meromorphic differential. This work led to a tremendous
amount of progress in the understanding of the physical aspects of
$\CN=2$ gauge theories, including the vacuum structure of related
$\CN=1$ theories. At the same time, there have also been fascinating
connections between Seiberg-Witten theory and mathematics,
especially to the Donaldson theory of four manifolds \WittenM.

In this article, we unearth a new connection between a particular
class of Seiberg-Witten curves and what Grothendieck called
``dessins d'enfants" or ``children's drawings". We will refer to
these simply as ``dessins" in what follows. We will formally define
a dessin later on, but for the moment by a dessin we simply mean a
connected graph on a two dimensional surface, with vertices of two
kinds - say black and white - that alternate along the graph.

The original reason for studying such drawings in mathematics was
that there is a natural action of the absolute Galois group ${\rm
Gal}(\bar \Bbb{Q}/\Bbb{Q})$ on them. Moreover, the action is
faithful, i.e. there is no group element, other than the identity,
that leaves invariant all dessins. The absolute Galois group is one
of the central objects of interest in mathematics, especially in
number theory. Surprisingly, this object has already made its
appearance in the physics literature in the context of rational
conformal field theory due to work by Moore and Seiberg
\refs{\MSone,\MStwo,\MSthree}. It has been known that the solutions
of the Moore-Seiberg equations lead to a projective representation
of the so called Teichm\"uller tower. As noted by Grothendieck in
\G, the absolute Galois group also acts on this tower. Amusingly
enough, both the Teichm\"uller tower and the dessins were introduced
by Grothendieck in the same letter \G.

In the rest of the introduction we will summarize recent progress
made in \refs{\CachazoYC, \CachazoSWtwo} in understanding the vacuum
structure of $\CN=1$ gauge theories obtained by deforming an $\N=2$
theory by a tree level superpotential. We will see that these gauge
theory techniques and the existing results could have implications
for the study of the action of the Galois group on the dessins and,
conversely, we would like to argue that there is a wealth of
information on the mathematical side that could potentially lead to
an improved understanding of the phases of $\CN=1$ gauge theories.

More generally, we conjecture that there is an intimate relation
between Grothendieck's programme of classifying dessins into Galois
orbits and the physics problem of classifying certain special phases
of $\CN=1$ vacua. The precise form of the conjecture is given in
Section 4.4. The meaning of the different physical and mathematical
elements that go into this conjecture will be explained in the remaining part of
Section 1 and in Section 2, respectively.

This paper is organized as follows: In Section 2, we discuss the
dessins and related mathematical material. This includes Belyi maps,
the action of ${\rm Gal}(\bar \Bbb{Q}/\Bbb{Q})$ on dessins via Belyi
maps and the correspondence with Seiberg-Witten curves. In Section
3, we review the concept of invariants (order parameters) which are
used to distinguish orbits of dessins under the action of ${\rm
Gal}(\bar \Bbb{Q}/\Bbb{Q})$ ($\N=1$ gauge theory phases). In Section
4, we give examples of cross fertilization that arise from our
identification of the physics and mathematics programmes. In
particular, we explain how the confinement index introduced in
\CachazoYC\ leads to a Galois invariant. In Sections 5 and 6, we
give some illustrative examples in full detail. In Section 7, we
conclude with many open questions and directions for future work. In
Appendix A, we give a very basic introduction to algebraic concepts
in Galois theory with the goal of understanding the definition of
${\rm Gal}(\bar \Bbb{Q}/\Bbb{Q})$ and its action. In the same
appendix, a small glossary of terms used in the text is given.
Finally, in Appendices B and C we expand on some interesting issues
discussed in the rest of the paper.

\subsec{Physics Preliminaries}

In the Seiberg-Witten solution of $\CN=2$ gauge theories it is
natural to study the loci in the moduli space where the SW curve
develops singularities. For a pure $U(N)$ gauge theory - the most
well studied case -  the curve has the form \eqn\SW{ y^2=P_N^2(z)
-4\Lambda^{2N} \,, } where $P_N(z)=\langle\det (z\II - \Phi) \rangle\,$ and $\Phi$ is the adjoint scalar in the vector multiplet. (We adopt the usual
notation, with the subscripts denoting the degree of the
polynomials.) A well studied \CachazoYC\ factorization is
\eqn\nonrigid{
P_N^2(z) -4\Lambda^{2N} = F_{2n}(z)\, H_{N-n}^2(z) \,,
}
which corresponds to a family of degenerate curves.
Let us briefly discuss the physical information that is encoded in this kind of polynomial equation.

The $\CN=2$ Coulomb moduli space of the gauge theory is parametrized by $N$ parameters $u_k = {1\over k} \langle\Tr \Phi^k\rangle\,$, constructed from the adjoint scalar $\Phi$. The problem studied in the physics literature is to find and classify the $\CN=1$ supersymmetric vacua obtained by perturbing the $\CN=2$ theory by a tree level superpotential
\eqn\superpot{
W_{\rm tree} = \sum_{k=1}^{n+1} {g_k\over k}\, \Tr \Phi^{k}\,.
}
Once the tree level potential is introduced, all points in the Coulomb moduli space are lifted except those for which $N-n$ mutually local magnetic monopoles become massless. The superpotential triggers the condensation of monopoles and the magnetic Higgs mechanism leads to confinement of the electric charges. The points at which this occurs are precisely those that solve the factorization problem \nonrigid. At low energies at these points, out of the original $U(1)^N$ only a $U(1)^n$ subgroup remains unbroken and its coupling constants are given by the reduced Seiberg-Witten curve
\eqn\reduced{
y^2 = F_{2n}(z) \,.
}

From a simple counting of parameters, we see that \nonrigid\
defines an $n$-dimensional subspace of the moduli space. Plugging
the $ u_k$'s as functions of these $n$ parameters
into the superpotential $W_{\rm tree}\,$, one gets an effective
superpotential
\eqn\effsuperpot{
W_{\rm eff} = \sum_{k=1}^{n+1} g_k\,  u_k
}
on the moduli space. Thus, given the parameters $g_k\,$, one can
vary with respect to the coordinates on the moduli space and obtain
{\it all} the $ u_k$'s  as functions of the $g_k$'s
and $\Lambda$.

In other words, the form of the superpotential picks out specific
points in the $\CN=2$ moduli space which correspond to $\CN=1$
vacua. What was shown in \refs{\CachazoIV,\CachazoV} is that this
extremization problem can be rephrased as the problem of factorizing
the Seiberg-Witten curve in the following manner:
\eqn\finalfact{
P_N^2(z) - 4\Lambda^{2N} = {1\over g_{n+1}^2}((W'_{\rm tree}(z))^2+f_{n-1}(z))\, H_{N-n}^2(z) \,.
}
Let us denote the critical points of $W_{\rm tree}$ by $a_i$, i.e., $W'_{\rm tree}(z)=\prod_{i=1}^n\, g_{n+1}(z-a_i)$. We set $g_{n+1}=1$ in what follows.

It is interesting to ask what the ``moduli space" of $\CN=1$ vacua
is, as the parameters $g_k$'s are varied.  The motivation for such a
question was explained in \CachazoYC: semiclassically, as
$\Lambda\rightarrow 0$, the gauge group $U(N)$ can be broken to
$\prod_{i=1}^n U(N_i)$ with $\sum_{i=1}^n N_i=N$ by choosing $\Phi$
to be a diagonal matrix with $N_i$ entries equal to $a_i$.\foot{We
assume that all $N_i\ne 0$. See Appendix C for some comments about
the case when some of the $N_i$ are zero.} It is therefore natural
to ask whether it is possible quantum mechanically to interpolate
between vacua that have the same $n$ but different $N_i$'s.

This was answered in \CachazoYC\ where it was shown that, for
example, vacua with one $U(1)$ factor can be smoothly connected to
vacua with any allowed values of $N_i$'s. All these classical limits
are different corners of a single connected subspace of the $\N=2$
moduli space, which was referred to as an $\N=1$ branch. However, in
\CachazoYC\ it was also discovered that there were other branches
distinguished by order parameters such as the expectation values of
Wilson loops. Branches meet at points where extra massless mutually
local monopoles appear. At these points other branches also emanate
which have the same dimension but where some of the $N_i$'s are
zero.

\medskip
\centerline{\epsfxsize=0.50\hsize\epsfbox{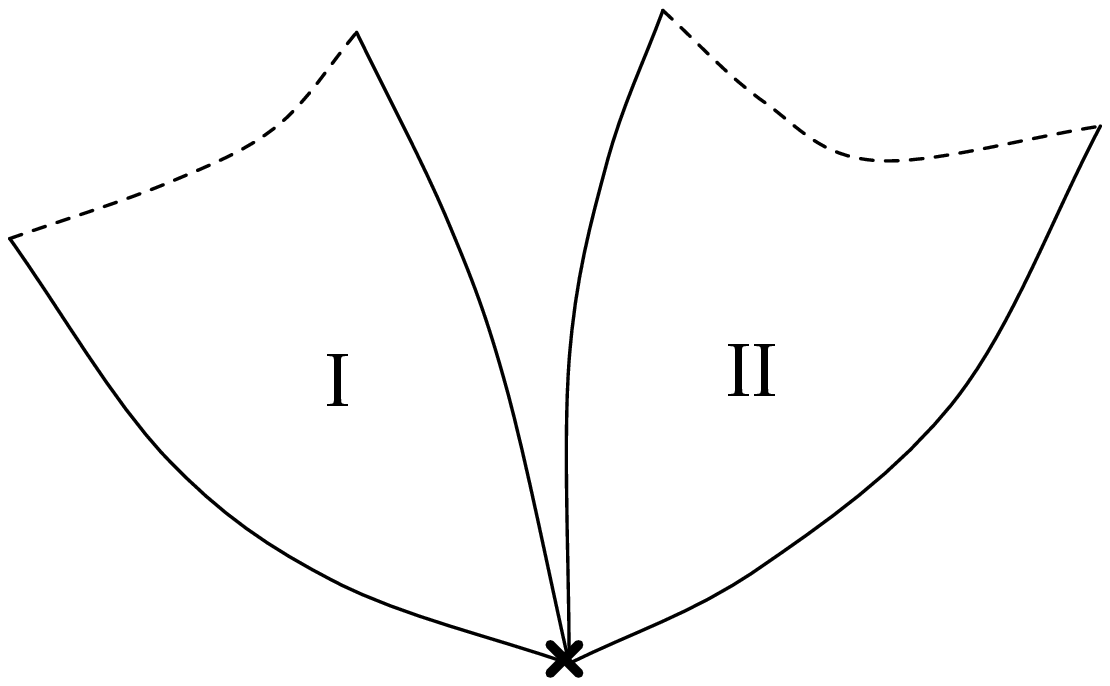}}
\noindent{\ninepoint\sl \baselineskip=3pt {\bf Figure 1}:{\sl $\;$
Two $\N=1$ branches of the $U(4)$ gauge theory obtained by deforming the $\N=2$ theory by a cubic superpotential. Only those branches with two $U(1)$'s in the infrared are shown. The two
branches meet at a point where one more monopole becomes massless.}}
\medskip

The structure of these branches and how they meet can be quite
intricate \CachazoYC. We use the example of a cubic superpotential
in a $U(4)$ gauge theory to represent a region close to one of the
points with three massless monopoles in Figure $1$. At a generic
point the low energy group is $U(1)^2$. At the special point where
the two branches meet, a third branch (not drawn) emanates which consists of vacua with a single $U(1)$ as the low energy group. Along
each of the depicted branches one can take a semiclassical limit and
find vacua corresponding to (classically) unbroken $U(N_1)\times
U(N_2)$ gauge groups.

\subsec{Dessins From Gauge Theory}

Let us continue our analysis of the $U(4)$ example. In Figure 2, we
show a typical configuration of the zeroes of the polynomials that appear in \nonrigid,
which correspond to points in the branches shown in Figure 1 near the semi-classical limits. The line segments that are drawn represent
the branch cuts. Of course, it is not essential to draw
the cuts through the zeroes of $P(z)$ (denoted by $\circ$) but this will have a
mathematical significance when we formally define a dessin. On the
other hand, the cuts naturally pass through the zeroes of $H_2^2(z)$ (denoted by a bivalent $\bullet$ node in the drawing)
since a small deformation away from the $\N=1$ branch will split the
double zeroes. As is clear from the figure, these look like
disconnected branchless trees.

\medskip
\centerline{\epsfxsize=0.60\hsize\epsfbox{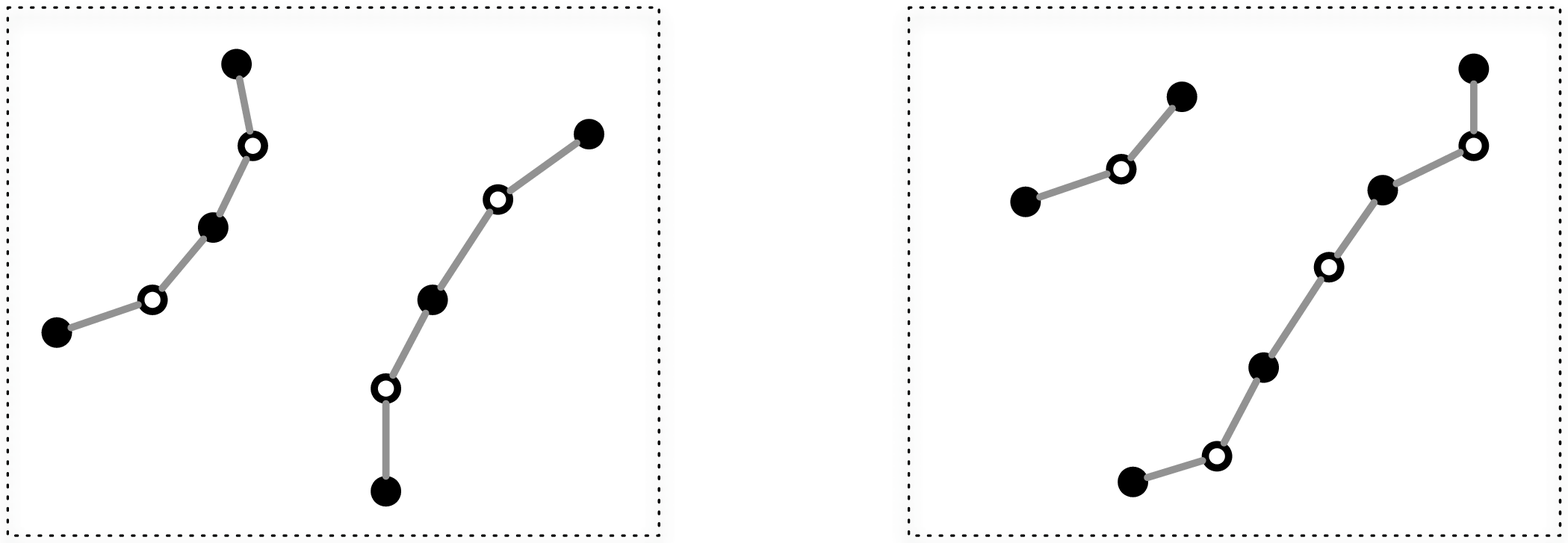}}
\noindent{\ninepoint\sl \baselineskip=3pt {\bf Figure 2}:{\sl $\;$
Zeroes of $P_4(z)$ (denoted by $\circ$), zeroes of $F_4(z)$ (denoted by univalent $\bullet$) and zeroes of $H_2^2(z)$ (bivalent $\bullet$) near the semi-classical limits with $N_1=N_2=2$ ({\it left}) and $N_1=1$,  $N_2=3$ ({\it right}). Edges represent branch cuts.}}
\medskip

It turns out that one of the two branches in Figure $1$ has $U(2)\times U(2)$ as the only semi-classical limit \CachazoYC. We now focus our attention on this branch. We further tune the parameters of the superpotential so that the corresponding $\CN=1$ vacuum is an isolated singular point where $H_2(z)$ develops a double root. There is only one such point in the branch we have chosen and it naturally leads to a connected graph. This is shown in Figure 3. We have omitted the zeroes of $P_4(z)$ so as to not clutter the figure.

It turns out that such connected trees show up in the moduli space
whenever the Seiberg-Witten curve develops isolated singularities.
Examples of such ``rigid curves" include the maximally confining
points \refs{\SWone, \Douglas} and the generalized Argyres-Douglas
points \AD. Such singularities arise when some of the zeroes of
$F_{2n}$ and $H_{N-n}$ coincide, $F_{2n}(z)$ develops double
roots, etc.

\medskip
\centerline{\epsfxsize=0.60\hsize\epsfbox{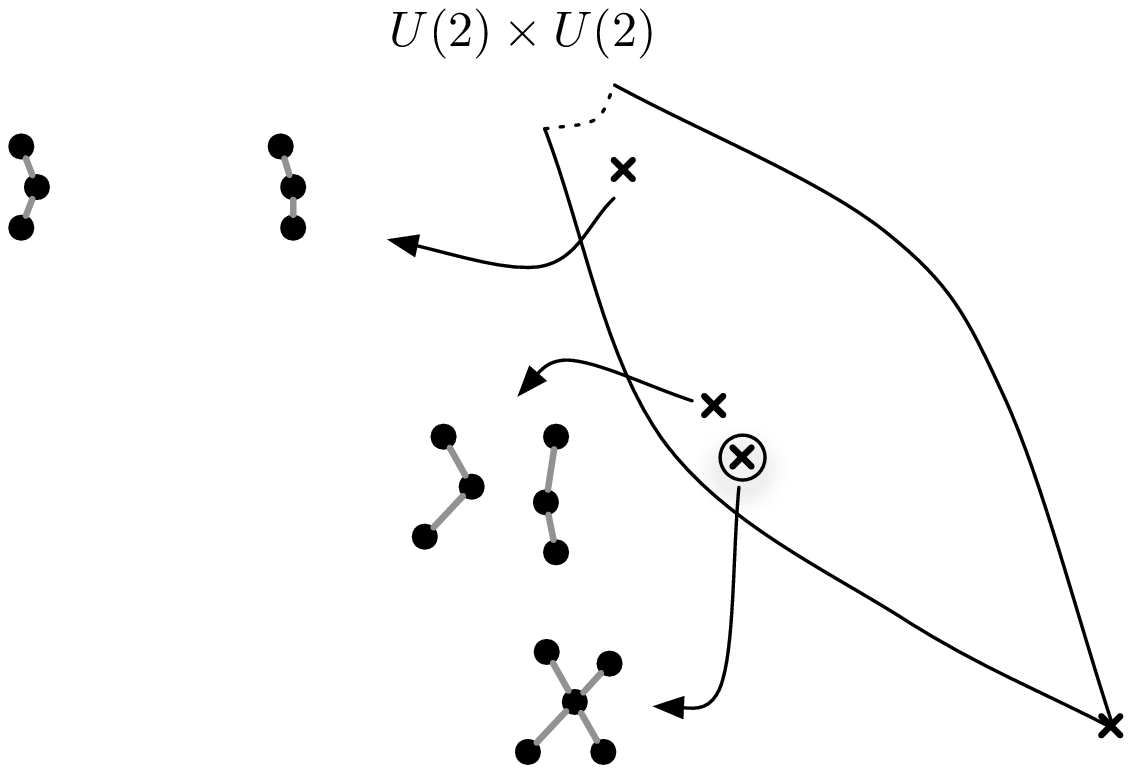}}
\noindent{\ninepoint\sl \baselineskip=3pt {\bf Figure 3}:{\sl $\;$
Evolution of the two branchless trees in one $\CN=1$ branch of the $U(4)$ gauge theory starting near the $U(2)\times U(2)$ semi-classical limit and reaching an isolated singularity where we get a connected tree.}}
\medskip

The connected trees that appear at such special points in the moduli
space are precisely the ``dessins d'enfants" that Grothendieck
introduced as a tool to study the structure of the absolute Galois
group ${\rm Gal}(\bar \Bbb{Q}/\Bbb{Q})$. This is the group of
automorphisms of $\bar\Bbb{Q}$ that leave $\Bbb{Q}$ fixed\foot{Here
$\Bbb{Q}$ is the field of rational numbers and $\bar\Bbb{Q}$ is its
algebraic closure. For more details see Appendix A.}. As mentioned
earlier ${\rm Gal}(\bar \Bbb{Q}/\Bbb{Q})$ acts faithfully on the
dessins. This means that the only element of the group that leaves
every dessin invariant is the identity. The set of dessins is then
partitioned into orbits under the action of the group. (We exhibit
how ${\rm Gal}(\bar \Bbb{Q}/\Bbb{Q})$ acts on the dessins in Section
$2.4$.) One way of learning about the Galois group is to
construct a complete set of invariants that distinguish dessins that
belong to distinct Galois orbits. This is one of the goals in the
study of dessins in mathematics and we will discuss some of the
known Galois invariants in detail in section $3$.

In the following sections we show that the known order parameters that
distinguish different branches of $\CN =1$ vacua can be thought of
as Galois invariants. In particular we prove that the ``confinement
index" introduced in \CachazoYC\ is a Galois invariant and can be
given a purely combinatorial interpretation. Interestingly, we will
find that certain operations on the gauge theory side, such as the
``multiplication map" introduced in \CachazoIV$\,$, have a precise
interpretation as operations on the dessins. We believe that
solving the non-rigid problem in \nonrigid\ before specializing to
singular points might lead to a new and useful perspective in the
study of dessins d'enfants. Conversely we will also see that this
correspondence leads to open questions regarding the interpretation
of interesting mathematical invariants in the gauge theory.

More explicitly, we would like to argue that the special points
where the dessins make their appearance can be thought of as special
phases of the $\CN=1$ gauge theory. This would imply the existence of
appropriate order parameters which distinguish these special points
from generic points in the $\CN=1$ branch to which they belong.
We provide some evidence for this in the examples in Section 5.2.

So far we have seen in an example how dessins can arise at isolated
singular points in the moduli space of a Seiberg-Witten curve. We
will now show that given a dessin, one can associate to it a
polynomial equation which corresponds to a singular Seiberg-Witten
curve, of the type discussed in this section. It turns out that this
is precisely equivalent to the content of the Grothendieck
correspondence which we discuss  next.

\newsec{The Grothendieck Correspondence}

\subsec{Mathematical Preliminaries}

The Grothendieck correspondence is a bijection between classes of dessins and special classes of maps on punctured Riemann surfaces called Belyi maps. At first sight this might seem far removed from gauge theory physics but we will describe a precise route to arrive at the correspondence between Seiberg-Witten theory and the dessins d'enfants. 


The first ingredient in the correspondence is the Belyi map. A Belyi map \Belyi\ is a holomorphic map from any punctured Riemann surface to $\Bbb{P}^1$ with exactly three critical values at $\{0,1,\infty\}$. In \G\ Grothendieck showed that any dessin can be constructed from a Belyi map. For the purposes of our discussion, we will restrict to dessins drawn on a Riemann sphere\foot{All the mathematical discussion in this section can be generalized to a general Riemann
surface and we refer the interested reader to \leila.}. We show two
simple examples in Figure 4. The key result that makes explicit the relation to Seiberg-Witten theory is that Belyi maps are obtained as solutions to certain polynomial equations.

\medskip
\centerline{\epsfxsize=0.50\hsize\epsfbox{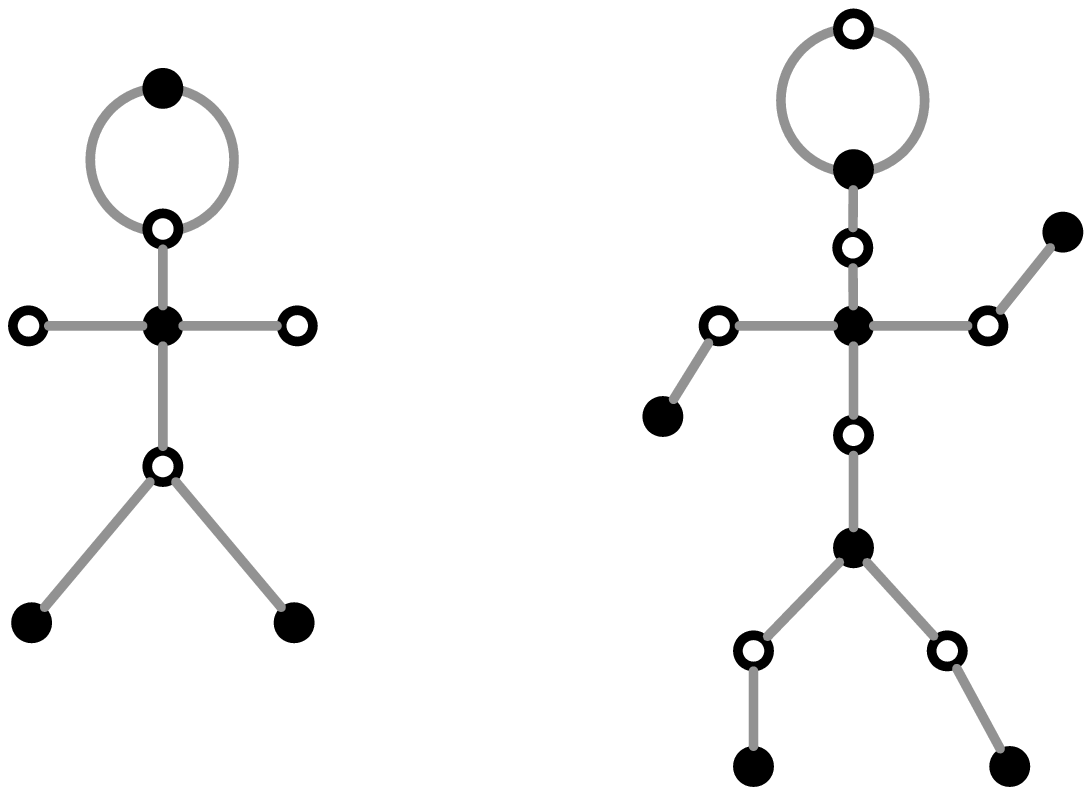}}
\noindent{\ninepoint\sl \baselineskip=3pt {\bf Figure 4}:{\sl $\;$
Examples of dessins; the bipartite structure of the graph is manifestly shown.}}
\medskip

As we will see, these equations have a natural interpretation as Seiberg-Witten curves
that describe particular degenerations of Riemann surfaces, such as
those we have already encountered in the introduction. More
generally, we find that whenever the Seiberg-Witten curve factorizes
so as to give rise to a ``rigid curve", i.e. without free parameters, one can associate a dessin to it\foot{This is up to a shift of $z$ in \nonrigid. In gauge theory, this corresponds to the overall $U(1)$ degree of freedom that decouples from the strong dynamics in the infrared.}. Our goal will be to set up a dictionary that maps the relevant quantities in mathematics, related to the Galois group action on dessins, into
gauge theory language and vice versa.

\subsec{Dessins From Belyi Maps}

We now state without proof some basic mathematical facts which are
crucial to establish the relation between the dessins and
Seiberg-Witten theory. See \leila\ for a thorough discussion and for
a full list of references. The main result we will use is the
Grothendieck correspondence, which connects the theory of dessins
with algebraic curves defined over $\bar{\Bbb{Q}}$, the algebraic
closure of $\Bbb{Q}$. Let us see how this comes about.

Consider an algebraic curve $X$ defined over $\Bbb{C}$. Such a curve
is defined over $\bar{\Bbb{Q}}$ if and only if there exists a
non-constant holomorphic function $f:X\rightarrow \Bbb{P}^1$ such
that all its critical values lie in $\bar\Bbb{Q}$. A theorem by
Belyi \Belyi\ gives a very striking result: $X$ is defined over
$\bar\Bbb{Q}$ if and only if there exists a holomorphic map
$f:X\rightarrow \Bbb{P}^1$ such that its critical values are
$\{0,1,\infty\}$.

A map $\beta: X\rightarrow \Bbb{P}^1$ with all its critical values
in $\{0,1,\infty\}$ is therefore called a {\it Belyi map}. A Belyi
map is called {\it clean} if all ramification degrees over $1$ are
exactly equal to $2$.

Let us give a simple example that will be very relevant in the rest of the paper. Let $X =\Bbb{P}^1$ and $\beta$ a polynomial. To guarantee that all critical points that map to $1$ have ramification degree $2$, we set
\eqn\firstBelyi{ \beta(z) = 1 - P^2(z)\,,}
where $P(z)$ is a polynomial. Let us see under which conditions
$\beta$ is a clean Belyi map. The critical points are computed as
the zeroes of
\eqn\criticalp{ {d\beta(z)\over dz} = -2P(z)P'(z).}
This means that the zeroes of $P(z)$ are critical points. Their
ramification degree is $2$ since $P(z)$ is squared in $\beta$ and
their critical value, i.e., $\beta$ evaluated at a zero of $P(z)$,
is $1$. All we need is that the remaining critical points, which are
precisely the roots of $P'(z)$, have critical value $0$. In other
words, they must also be roots of $1-P^2(z)$. Up to the freedom to shift $z$, these conditions have only a discrete number of solutions. These are Belyi maps.

\medskip
\centerline{\epsfxsize=0.450\hsize\epsfbox{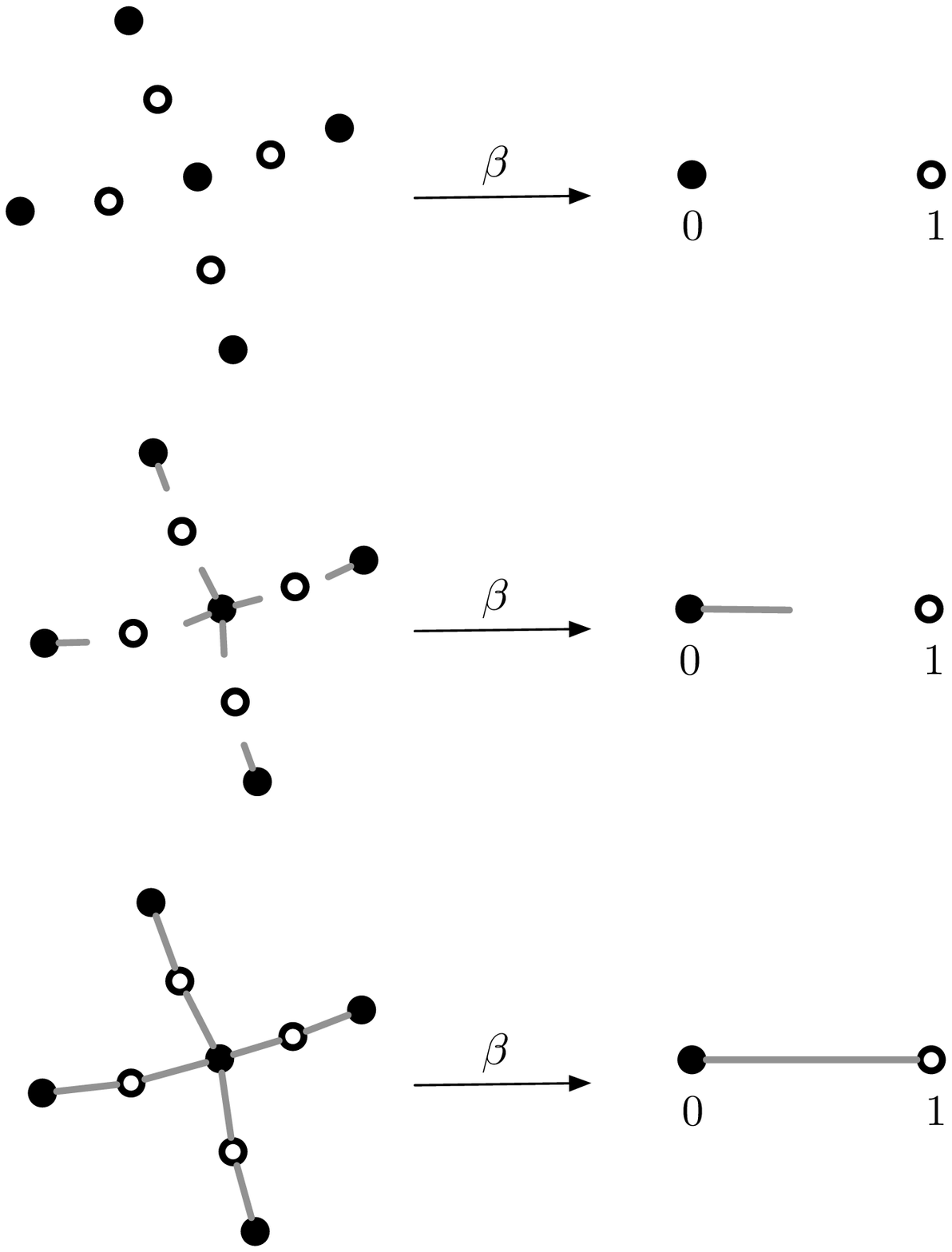}}
\noindent{\ninepoint\sl \baselineskip=3pt {\bf Figure 5}:{\sl $\;$
We show how the dessin is the pre-image of the interval $[0,1]$ under the Belyi map. Note that as we move from $0$ to $1$, the number of lines emanating from a given pre-image of $0$ is given by the ramification degree of the map at those points. Since the map is clean, {\it exactly} two lines meet at each pre-image of $1$.}}
\medskip

We now have the ingredients to formally define a dessin: for the
purposes of this paper we define a {\it dessin d'enfant} on the
sphere as the pre-image under a clean Belyi map of the interval joining $0$ and $1$ in
$\Bbb{P}^1$. In other words, the dessin $D$ associated to a clean Belyi map $\beta$ is
$D=\beta^{-1}([0,1])\subset X$. We show this pictorially in
Figure $5$ for the case $\b(z)=1-P_4^2(z) = F_4(z)\, H_1^4(z)$.

Such a dessin has a natural bipartite structure given by assigning a $\bullet$ to the preimages of $0$ and $\circ$ to the preimages of $1$. We will refer to a pre-image of $0$ as a vertex of the dessin. An edge is a line segment between two vertices that contain exactly one pre-image of $1$. For example the second dessin in Figure $4$ is clean (while the first is not) so that the notion of edges and vertices as we just defined  makes sense: it has $7$ edges and $7$ vertices. In what follows, we will restrict our
attention to clean dessins and refer to them simply as dessins.
Likewise, the corresponding clean Belyi maps will be simply called
Belyi maps.

Also important for characterizing the dessins are the preimages of
$\infty$, denoted by $\times$. There is one pre-image of $\infty$
for each open cell enclosed by a set of edges. For example, a dessin
is a tree if and only if the preimage of $\infty$ is a single point.

The study of dessins on the sphere is important because the absolute
Galois group $\GG$ acts faithfully on them. As mentioned before, the
absolute Galois group is the group of automorphisms of
$\bar{\Bbb{Q}}$ that leaves invariant $\Bbb{Q}$ and it is a
remarkably complex object. We give a basic introduction to the Galois group in
Appendix A. It can also be shown that not only is the action of
$\GG$ on genus-$0$ dessins faithful, but so is the action on the
much smaller set of trees.

The main thing to take away from this section is that one can map
the problem of classifying dessins to the problem of classifying
Belyi maps $\b$. As we have seen, these are a special class of
rational functions on the Riemann sphere that satisfy the conditions
in the definition above. We now turn to the question of how Belyi
maps corresponding to a given dessin can be explicitly constructed. This will naturally lead us to
gauge theory physics.

\subsec{Belyi Maps From Polynomial Equations}

Consider a dessin $D$ on the sphere with $N$ edges. Let
$V=\{u_1,\ldots ,u_k\}$ where $u_i$ is the number of vertices
(pre-images of $0$) of valence $i$. We choose $k$ to be the maximum
vertex valence in $D$. Let $C=\{ v_1,\ldots,v_m\}$ where $v_i$ is
the number of faces with $i$ edges. Again we choose $m$ to be the
maximum face valence in $D$. The lists $V$ and $C$ are called the
{\it valency lists} of $D$.

Let $G_{v_i}(z)$ and $J_{u_i}(z)$ be polynomials of degree $v_i$ and
$u_i$ respectively, with undetermined coefficients. Take one
polynomial for each element in $V$ and in $C$. Pick $i_0$ to be the
valence whose $v_{i_0}$ is the smallest non zero value in $C$. Then
let all polynomials $G_{v_i}(z)$ and $J_{u_i}(z)$ be monic except
for $G_{v_{i_0}}(z)$ which we choose to be of the form
\eqn\kusa{G_{v_{i_0}}(z) = \alpha (z^{v_{i_0}-1} + \ldots ).}
In other words, we have chosen the coefficient of $z^{v_{i_0}}$ to
vanish and we have factored out the coefficient of $z^{v_{i_0}-1}$
which we call $\alpha$.

\noindent
Now construct the two polynomials
\eqn\defPoly{
A(z) =
\prod_{j=1}^k J_{u_j}(z)^j\,, \qquad B(z) = \prod_{i=1}^m G_{v_i}(z)^i \,.
}
Then if $A(z)$ and $B(z)$ are such that there exists a monic
polynomial $P_N(z)$, with $N$ equal to the number of edges of the
dessin, and which satisfies the polynomial constraint
\eqn\sewi{ A(z)-B(z)= P_N^2(z)  \,,}
then
\eqn\bmap{
\beta(z) =1+{P_{N}^2(z)\over B(z)}= {A(z)\over B(z)}
}
is a rational clean Belyi map. Note that the polynomial equation is
rigid, in the sense that there are no coefficients in the
polynomials which are free parameters\foot{Up to a shift and rescaling of $z$. We will return to this point in Section $2.7$.}.
There are thus only a finite number of solutions to
\defPoly. The discussion has been rather abstract so far, so let us illustrate
the various concepts with some simple examples which have already been
discussed in \AshokCD.

Consider the dessin in Figure $6$A which has
$6$ edges. From the figure, we see that each open cell is bounded by
$3$ edges and every vertex is trivalent. The valency lists are
therefore of the form $V=\{0,0,4\}$ and $C=\{0,0,4\}$. From the
discussion above, we need $A(z)=J_4^3(z)$ and $B(z)=G_3^3(z)$ such
that they satisfy the polynomial equation
\eqn\oldex{
P_6^2(z) +G_3^3(z) = J_4^3(z) \,.
}

\medskip
\centerline{\epsfxsize=0.65\hsize\epsfbox{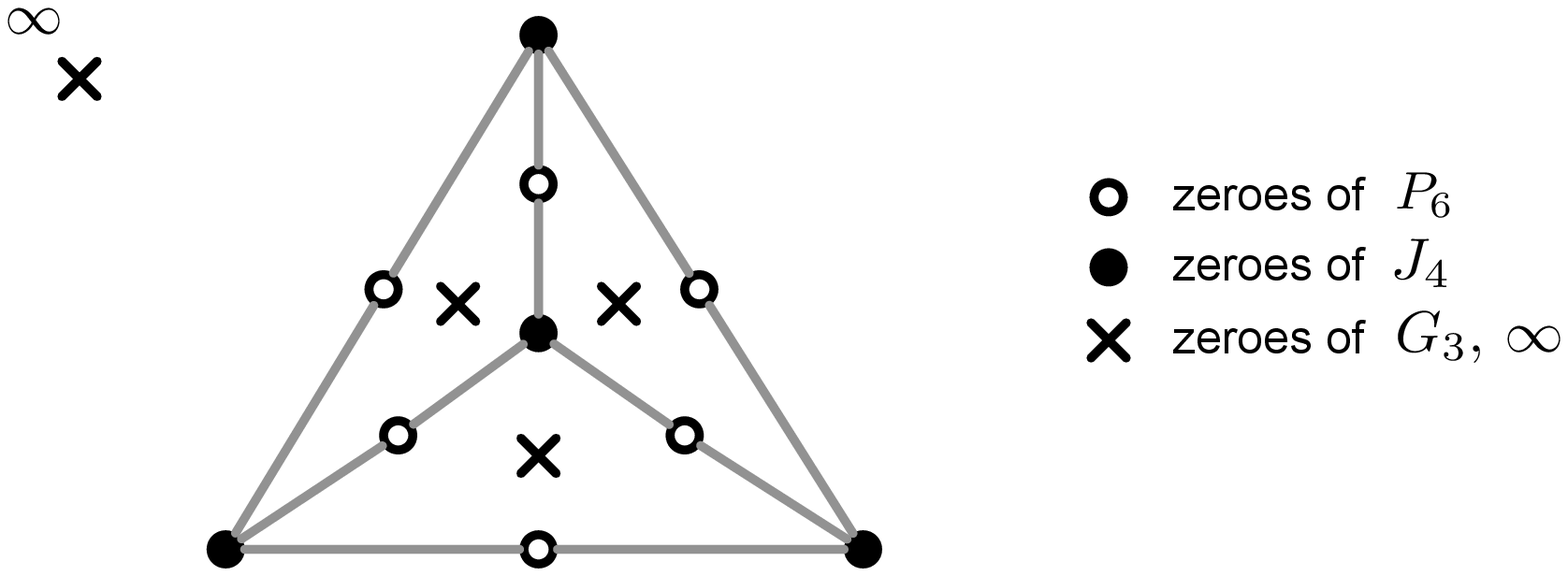}}
\noindent{\ninepoint\sl \baselineskip=3pt {\bf Figure $6$A}:{\sl $\;$
Dessin corresponding to the solution of the factorization problem \oldex.}}
\medskip

It turns out that there is only one solution to the polynomial equation \oldex\ \AshokCD. However, in general, such polynomial equations have more than one solution. For instance, the polynomial equation
\eqn\oldextwo{
P_{10}^2(z) +G_3^5(z) = J_4^3(z)\, \tilde{J}_4^2(z) \,.
}
turns out to have two solutions \AshokCD. This is because for the same valency lists $V=\{0,4,4 \}$ and $C=\{0,0,0,0,4 \}$ (which we can infer from the polynomial equation), there are two dessins one can draw. These are shown below in Figure $6$B. We shall revisit this specific example in Section $6$ in much more detail.

\medskip
\centerline{\epsfxsize=1.0\hsize\epsfbox{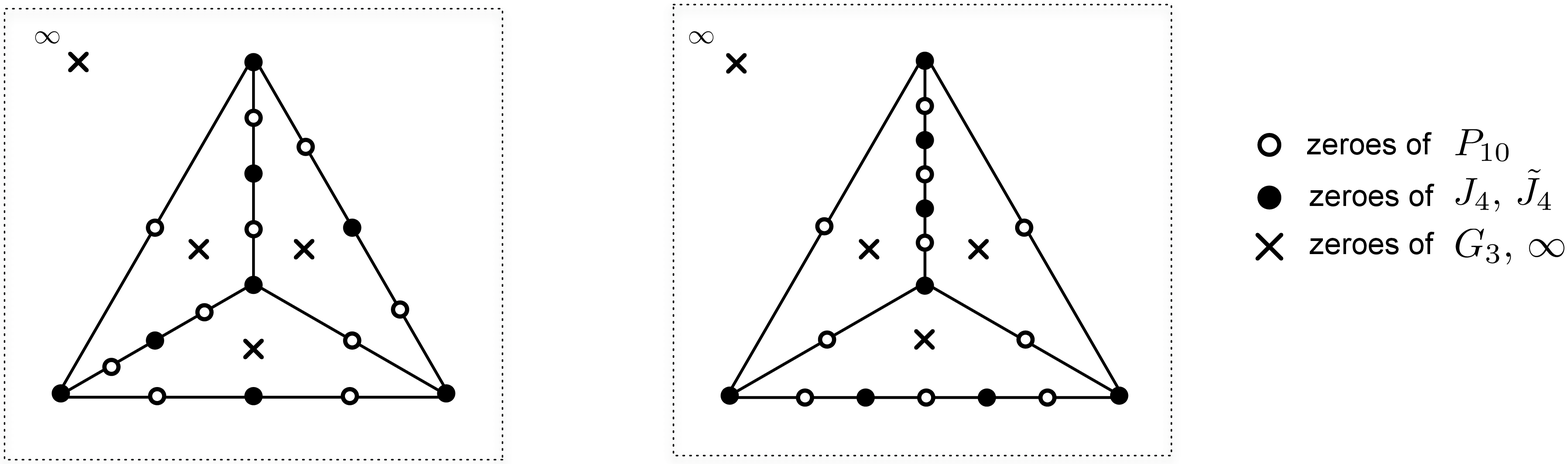}}
\noindent{\ninepoint\sl \baselineskip=3pt {\bf Figure $6$B}:{\sl $\;$
Dessin corresponding to the solution of the factorization problem \oldextwo. The bivalent $\bullet$ vertices are roots of $\tilde{J}_4(z)$ while the trivalent $\bullet$ nodes are the roots of $J_4(z)$.}}
\medskip

The key result which we will use from now on is that for {\it every} dessin $D$ with valency lists $V$ and $C$ there exists a solution to the factorization problem \sewi\ such that the corresponding Belyi function gives $D=\beta^{-1}([0,1])$. This is a simplified version of the Grothendieck correspondence \G.

\subsec{Action Of The Galois Group}

So far we have mentioned repeatedly that the absolute Galois group
$\Gamma = \GG$ acts faithfully on dessins. We now show how $\Gamma$
acts on the dessins via the Belyi map.

Let $D$ be a dessin such that $D = \b^{-1}([0,1])$. Furthermore, let
$\b$ be of the form
$$
\b = {A(z)\over B(z)} = {z^{2N}+a_1 z^{2N-1} + \ldots + a_{2N} \over
z^L + b_1 z^{L-1} + \ldots + b_{L}} \,,
$$
where $A(z)$ and $B(z)$ solve the polynomial equation \sewi, $N$ is,
as before, the degree of $P(z)$ and $L = \sum_{i=1}^m i v_i$. Then
$\Gamma$ acts on $D$ by twisting the coefficients\foot{The $a_i$ (and
$b_j$) are algebraic numbers; i.e. they are solutions to some
polynomial equations with coefficients in $\Bbb{Q}$. The solutions
to such equations include $a_i$ along with other algebraic numbers
which are, by definition, in the Galois orbit of $a_i$. Twisting by
the relevant group element of $\Gamma$ here refers to choosing
another element in the orbit of $a_i$. For a more formal discussion, refer to Appendix A.} in $\b\,$. For $g \in \Gamma$, $D_{g}$ is defined to be the dessin obtained by the action of $g$ on $D$, i.e.
$D_{g}=\beta_{g}^{-1}([0,1])$, where
\eqn\galb{ \beta_{g} = {A_{g}(z)\over B_{g}(z)} =
{z^{2N}+g(a_1) z^{2N-1} + \ldots + g(a_{2N}) \over
z^L+g(b_1) z^{L-1} + \ldots + g(b_{L})} \,. }

Thus, given a solution to the polynomial problem \sewi\ it is easy
to understand how the dessin changes under the action of the Galois
group. However, given two dessins, it is in general very difficult
to tell whether they belong to the same Galois orbit or not. This is
the central problem associated to the dessins d'enfants. Later we
will discuss several Galois invariants that mathematicians have
introduced in order to distinguish dessins that belong to distinct
Galois orbits by studying the combinatorial data associated to each
dessin.

\subsec{The Identification}

We have already discussed the Seiberg-Witten curves for pure gauge
theories in \SW. Recall that for a $U(N)$ gauge theory with $L<2N$
massive flavors with masses given by $m_i$, the Seiberg-Witten curve
that captures the infrared dynamics of the gauge theory is the
following hyperelliptic Riemann surface \refs{\HananyO, \ArgyresPS}:
\eqn\aso{y^2 = \langle{\rm det}(z\Bbb{I}-\Phi)\rangle^2 - 4\Lambda^{2N-L}\prod_{i=1}^L(z+m_i)\,.}

We would like to propose the following identification and argue that
it is a useful one. Let us identify objects in \sewi\ and in \aso\
as follows: $P_N(z) =\langle{\rm det}(z\Bbb{I}-\Phi)\rangle$, $B(z)
=-4\Lambda^{2N-L}\prod_{i=1}^L(z-m_i)$. In particular, $\alpha
=-4\Lambda^{2N-L}$. In each case, the precise form of $A(z)$ in
\defPoly\ defines the special point in the Coulomb moduli space we are
looking at. Since dessins are associated only to rigid
factorizations of the Seiberg-Witten curves, they appear at
isolated singular points in the moduli space of the $\CN =2$ gauge theory.

At this point, it appears as if the relation to Seiberg-Witten
curves is purely at a formal level. We will show in what follows
that this is more than a superficial similarity and we exhibit
features of the gauge theory that have a natural interpretation as
operations on the dessin. For this we have to abandon our $\N=2$
point of view and deform the theory to $\N=1$ by a tree level
superpotential as reviewed in Section 1.1.

We mentioned earlier that the absolute Galois group acts faithfully
on the set of all trees. For most part of the paper we will restrict our discussion to the set of trees and
only in Section $6$ will we discuss dessins with loops.

\subsec{Trees On The Riemann Sphere: Refined Valency Lists}

Since trees have only one open cell (with the associated vertex at
infinity) the Seiberg-Witten curve associated to the dessin is that
of the pure $U(N)$ gauge theory: \eqn\SW{ y^2 = P_N^2(z) -
4\Lambda^{2N}\,. }
By itself, the curve in \SW\ does not correspond to any
dessin, but if we tune the parameters in $P_N(z)$ so that we are at
an isolated singularity in the moduli space, the curve factorizes,
and the zeroes of the polynomials involved will describe vertices of
a dessin. This also means that the Belyi map \bmap\ is a polynomial
\eqn\purebelyi{
\b(z) = {A(z)\over B(z)} = 1- {P_N^2(z) \over 4\Lambda^{2N}} \,,
}
where $P_N(z)$ solves the factorization
\eqn\factprob{
(P_N(z) - 2\Lambda^{N})(P_N(z) + 2\Lambda^N) = \prod_{j=1}^{k} (J_{u_{j}}(z))^{j} \,.
}
From the expression it follows that the two factors on the left cannot have any factors in common. Thus, for trees, the problem always reduces to solving two lower order equations of the form
\eqn\split{\eqalign{ P_N(z) - 2\Lambda^{N} &=  \prod_{j=1}^{k}
(Q_{u^-_{j}}(z))^{j}\cr P_N(z) + 2\Lambda^{N} &=  \prod_{j=1}^{k}
(R_{u^+_{j}}(z))^{j} }}
such that $u^-_{j}+u^+_{j} = u_j$ for every $j$. One can now define
a new bipartite structure on the dessin by assigning a $+(-)$ to
every zero of $R_{u^+_{j}}$ $(Q_{u^-_{j}})$ such that if a given
vertex (pre-image of $0$) is of one sign, every one of its neighbors
is of the opposite sign. The bipartite structure is unique up to an
overall sign flip. This leads to a more refined valency list
$\{V^+,V^-\}$ than the $\{V,C\}$ introduced earlier\foot{$C$
contains just one element and is trivial for the case of trees.}
where $V_{\pm}$ denotes the positive/negative valency list. For
instance, from \split\ $u_j^+$ is the number of $j$-valent vertices
of type ``+". With the refined valency list, the trees can be
redrawn as shown in the figure below.

\medskip
\centerline{\epsfxsize=0.80\hsize\epsfbox{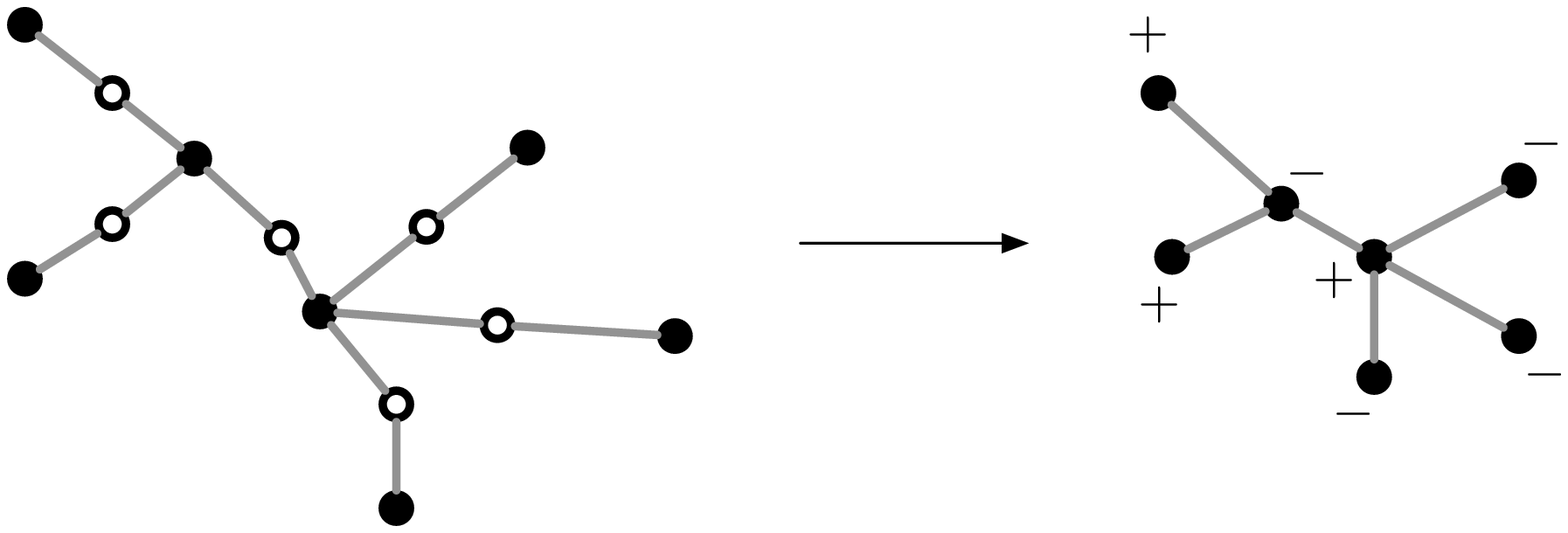}}
\noindent{\ninepoint\sl \baselineskip=3pt {\bf Figure 7}:{\sl $\;$
Refined valency list for the trees. $V^+=\{2,0,0,1\}$ and $V^-=\{3,0,1\}$ for this case.}}
\medskip

Note that in the figure on the right we have removed the preimages
of $1$ depicted as $\circ$ on the left. This is common practice in
the literature when dealing with trees. There is one more common
convention which is to depict elements in $V^+$ by $\bullet$ and
elements in $V^-$ by $\circ$; we have not adopted this convention in
order to avoid confusion and we simply add $\pm$ to the $\bullet$'s
as in Figure $7$.

Clearly, the dessins that arise from different valency lists belong
to distinct Galois orbits; we will comment more about this in the
next section.

Interestingly enough, there is a related splitting of the polynomial
equation in the gauge theory. In \CachazoYC\ while solving the
non-rigid problem \nonrigid\ it was found that the $\CN=1$ branches are classified by the integers $(s_+,s_-)$, where $s_{\pm}$ refers to the number of double roots in each of the factors   on the left
hand side of \split. This is already a hint that the mathematical
goal of classifying dessins according to Galois orbits might be
closely related to the more physical problem of studying the
branches of $\CN=1$ vacua in gauge theory. We will see this in more
detail in Section $4$.

\subsec{Equivalence Classes Of Trees}

Let us consider the equations \split\ in more detail. These
equations have two free parameters corresponding to the scale and
shift of $z$. In other words, if $P^{(1)}_N(z)$ is a solution then
$P^{(2)}_N(z) = P^{(1)}_N(az+b)$ is also a solution. In physics as
well as in mathematics, it is natural to consider monic polynomials.
This means that $a$ is restricted to be an $N^{th}$ root of unity.

The trees constructed using the Belyi map \purebelyi\ with
$P^{(1)}_N(z)$ and $P^{(2)}_N(z)$ are identical except for a
displacement or rotation in the $z$ plane. In the mathematical
literature, such trees are considered equivalent and one considers
equivalence classes of such Belyi maps. The Grothendieck
correspondence is in fact an isomorphism between equivalence classes
of Belyi maps\foot{The equivalence class of Belyi maps is, in fact,
up to any $SL(2,\IC)$ transformation. However, we have used one of
these to put the pole at $\infty$. Thus only shift and scale
transformations remain.} and children's drawings.

Every tree has associated to it a number field (see Appendix A for a
primer on field extensions) \leila, determined by the field of
definition of the polynomial $P_N(z)$ that gives the corresponding
Belyi map \purebelyi. This might be a little puzzling at first,
since the transformation $z\to az+b$ can in general involve
arbitrary algebraic numbers. This means that the number field associated
to trees that differ by translations and rotations can be
different. On the other hand we have just said that such trees are taken to define an equivalence class on which $\GG$ acts.

The resolution to this puzzle is that, although the Galois group
acts nontrivially on all these trees, there is always a way
of choosing the tree with the simplest number field \Shabat\ as a
representative of the equivalence class. It turns out that the
action of $\GG$ on just the representatives of each class is
faithful. Therefore, for the purposes of studying the absolute
Galois group one uses the shift and scale of $z$ to pick the simplest representative.

All these statements have a counterpart in physics. The freedom to
shift by $b$ corresponds to the fact that the underlying theory is
$U(N)$, as opposed to $SU(N)$. The overall $U(1)$ decouples in the
IR and gives rise to this shift degree of freedom. Just like in
mathematics, one can use this shift to bring any tree level
superpotential to a form that displays the $\CN=1$ branches, introduced in
Section 1, most clearly. We will use this in section 5.

More intriguing is the meaning of the rescaling by an $N$-th root of
unity. In physics, this corresponds to different kinds of
confinement distinguished by the behavior of combinations of Wilson
and 't Hooft loop operators. Roughly speaking, the trivial root of
unity corresponds to usual confinement while the other roots
correspond to oblique confinement \CachazoYC. Quite nicely, the mathematical
criterion of choosing the simplest number field corresponds in
physics to choosing the phase with usual confinement.

It would be very interesting to explore the connection between the
``not-so-simple number fields" and the oblique confining phases.
However, since our goal is to establish a connection between dessins
(in terms of equivalence classes) and gauge theory, we will restrict
our study to physics phases with only usual confinement.

\subsec{Example: The Maximally Confining $\CN=1$ Vacua}

For now, let us discuss as an example the simplest tree one can
draw: a branchless linear tree with $N$ edges. Such a tree has $2$
vertices of valence $1$ and $N-1$ vertices of valence $2$. From the
general discussion above, it is easy to write down the corresponding
Seiberg-Witten curve for this case:
\eqn\fofu{ P^2_N(z) -4 = (z^2-4)H^2_{N-1}(z) \,.}
Here we have set $\Lambda^N=1$.

This Seiberg-Witten curve corresponds to points where $N-1$ mutually
local monopoles go massless. The $\CN=1$ vacua are obtained by
perturbing the $\CN=2$ theory by a mass deformation, with $W_{\rm tree}
= \half \Tr\Phi^2$. The condition $\Lambda^N=1$ has $N$ different
solutions that correspond to the $N$ different maximally confining
$\N=1$ vacua\foot{The reason for the name is that the low energy
gauge group is just $U(1)\subset U(N)$.}. However, as discussed
above, we take the simplest solution, i.e., $\Lambda =1$. The
solution to this equation is well known and given in terms of
Chebyshev polynomials\foot{Here the shift symmetry is used to set
$<\Tr \Phi>=0$.}. In Figure $8$ the zeroes of $P_N(z)$ and
$H_{N-1}(z)$ have been depicted showing how the branchless tree
arises.

The importance of the branchless tree lies in the fact that these
appear at the intersection of the $\CN=1$ branches, the point marked
by a cross in Figure $1$. They also appear near the semi-classical limits
($\Lambda\rightarrow 0$) as shown in Figure $2$. Although these
truncated branchless trees cannot be thought of as dessins, they
seem to be ``building blocks" that come together to create a dessin
at an isolated singularity. We discuss this point in more detail in
the conclusions.

So far we have been rather loose in the language employed to discuss
aspects of the factorization problems. Techniques from both the
physics and mathematics literature have been used interchangeably.
We now turn to a more systematic discussion of how the dessins are
classified from a mathematical point of view. We will follow this up
with a review of how the $\CN=1$ vacua are classified from a physics
point of view.

\medskip
\centerline{\epsfxsize=0.65\hsize\epsfbox{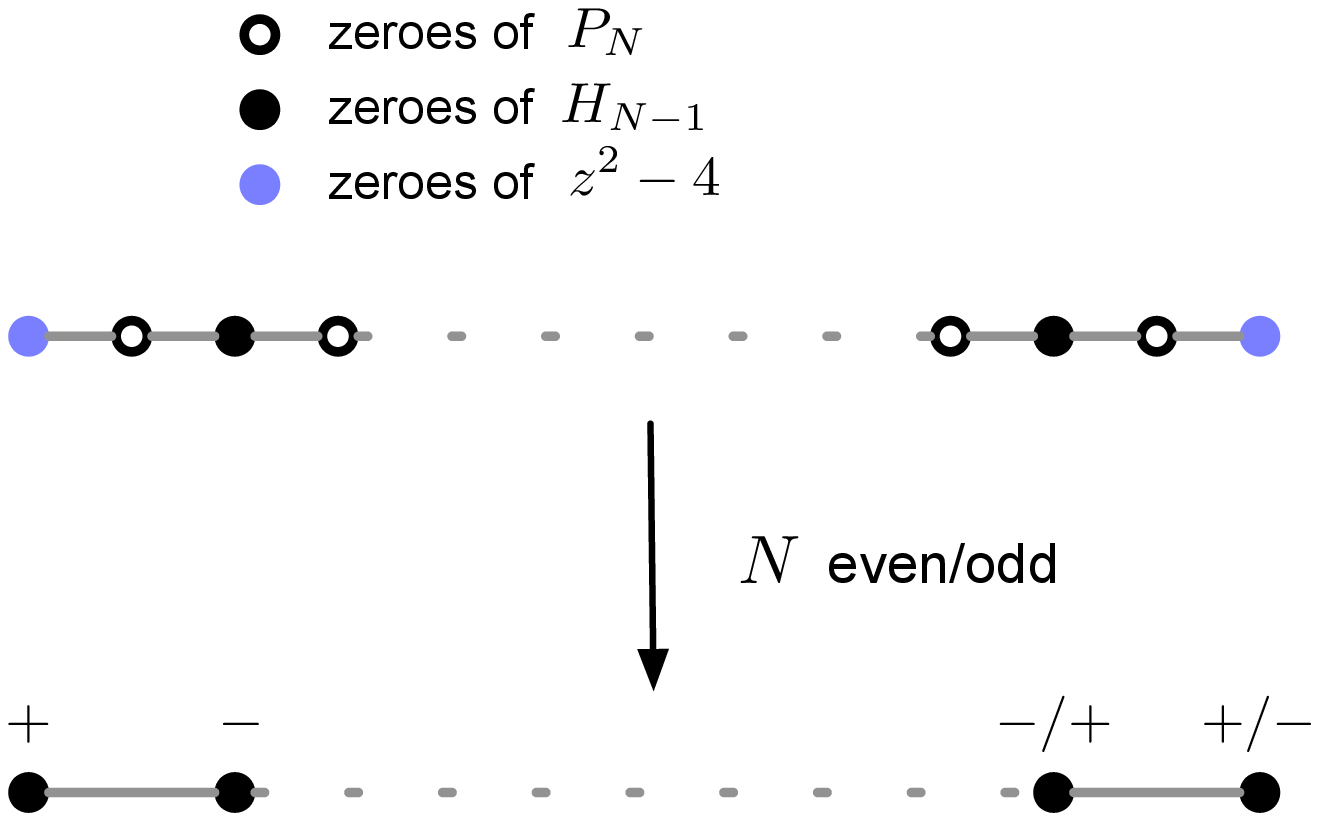}}
\noindent{\ninepoint\sl \baselineskip=3pt {\bf Figure 8}:{\sl $\;$
The dessin that corresponds to the maximally confining vacuum. Case
by case, it is obtained by plotting the zeroes of the polynomials
that solve the factorization problem \fofu. See for example Figure
$17$ for the $N=6$ plot. }}
\medskip

\newsec{Invariants}

\subsec{Invariants From Mathematics}

One way of learning about the structure of the absolute Galois group $\GG$ is by constructing a complete set of invariants under the action of $\GG$ such that any two dessins that do not belong to the same Galois orbit will disagree in at least one invariant. Such a complete list of invariants is currently not known although many invariants have been constructed.
In this section we review the most basic invariants\foot{More invariants than those discussed here are known, but for the purpose of our analysis we will concentrate on those that can be most easily computed explicitly.} associated to dessins that are related by the action of $\GG\,$ \refs{\leila,\ \melanie}.

$\bullet${\it Valency Lists}. The most intuitive invariants are the valency lists $V$ and $C$ introduced in Section $2$. These are clearly invariants, since as we saw they are determined by the form of the polynomial equation which is invariant under the action of $\GG\,$. It is sometimes possible to define more refined valency lists, corresponding to different ways of solving the polynomial equation. We have already seen this for the case of trees, where we introduced the $\{V^+,V^-\}$ valency lists. As we discussed, this possibility of constructing a new invariant has a nice counterpart in physics; it will be further clarified in the examples that follow. Note that by concentrating on dessins coming from the same factorization problem we can forget about the valency list invariant since all dessins constructed this way have the same valence list. Therefore, the search is for other invariants that will distinguish different Galois orbits.

$\bullet${\it Monodromy Group}.
Every dessin is associated to a cover of $\IP^1$, defined by the Belyi map
$$
\b:\Sigma\to U\equiv\IP^1\setminus \{0,1,\infty\}\,,
$$
that maps the edges of the graph on the Riemann surface $\Sigma$
into the open segment $\bar{01}$ on $\IP^1\,$. If the graph has $N$
edges, where we count the number of edges to be equal to the number
of pre-images of $1$, the map $\b$ gives a $2N$-fold cover of $U\,$
(see Figure 9 for an illustration).

Consider $\pi_1(U,\bar{01})\,$, the homotopy group of paths in $U$
that begin and end on a point of $\bar{01}\,$. Since a closed path
based on $\bar{01}$ in $\IP^1$ can be mapped into a path between any
two of the $2N$ segments in the fiber over $\bar{01}\,$, any given
element of $\pi_1(U,\bar{01})\,$ acts on the dessin as a permutation
of the half-edges (that go between a filled and unfilled vertex in
Figure $9$). Therefore, the covering map $\b$ induces a map from
$\pi_1(U,\bar{01})$ to $S_{2N}\,$. Let us denote by $\sigma_0$ the
permutation corresponding to circling once the point $z=0$ on
$\IP^1$ and by $\sigma_1$ the permutation corresponding to circling
once the point $z=1\,$. Recall that the dessins have a bipartite
structure that keeps track of whether a vertex is mapped to $0$ or
$1$ by the Belyi map. One can convince oneself that $\sigma_0$ is
the element of $S_{2N}$ that permutes cyclically the edges incident
on each vertex (that maps to $0$) and, similarly, $\sigma_1$ is a
cyclic permutation of the edges incident on each pre-image of $1$.
The subgroup of $S_{2N}$ generated by $\sigma_0$ and  $\sigma_1$ is
the monodromy group of the dessin and it is a Galois invariant
\JonesS.

Let us consider the example of the dessin in Figure $9$. There are
$8$ half-edges. The monodromy group is generated by the following permutations:
\eqn\monex{\eqalign{
\sigma_0 &= (1,7,6)(2,3)(4,5) \,, \cr
\sigma_1 &= (1,2)(3,4)(5,6)(7,8) \,.
}}
We will give the explicit monodromy groups for the examples we will encounter later.

\medskip
\centerline{\epsfxsize=0.75\hsize\epsfbox{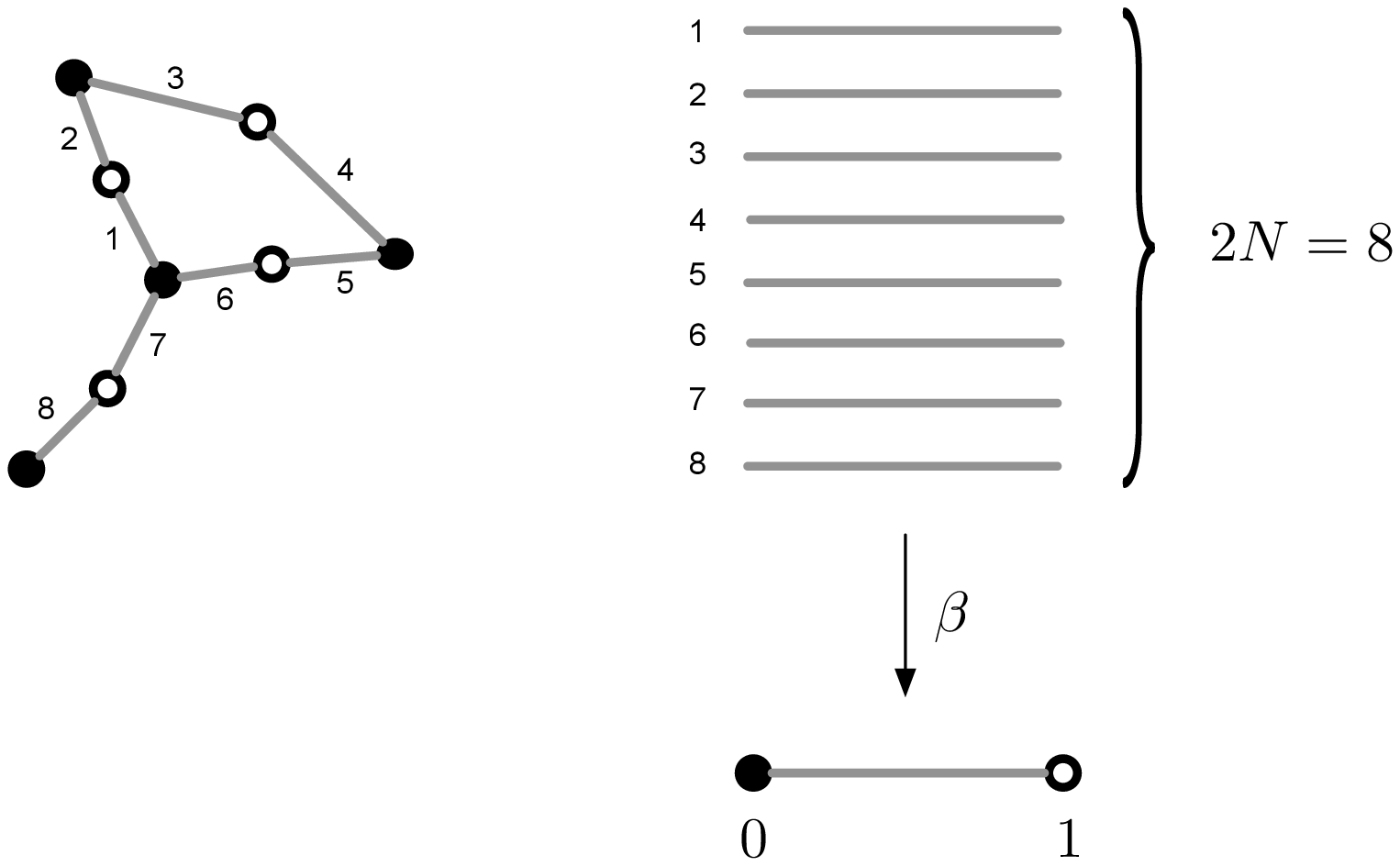}}
\noindent{\ninepoint\sl \baselineskip=3pt {\bf Figure 9}:{\sl $\;$
Left: Example of a dessin with $N=4$ edges. Right: The $8$ half-edges
come from the preimage of the open interval $\bar{01}$, i.e. the
preimage of $0$ and $1$ are not shown.} }
\medskip

$\bullet${\it Belyi Extending Maps}. It is possible to construct new invariants by composing the Belyi map of interest with any Belyi-extending map and then computing the valency lists or the monodromy group of the new dessins obtained this way \melanie. A Belyi extending map is a Belyi map $\alpha: \IP^1\to \IP^1$ defined over $\Bbb{Q}\,$, such that its composition with any other Belyi map does not change the associated number field.  For our purposes, we relax the definition slightly: by a Belyi extending map here we will mean any map $\alpha$ defined over $\Bbb{Q}$ that can be composed with a Belyi map $\beta$ to give another Belyi map $\b_{\a}:=\a\circ \b$.

Let $I$ be any invariant of a dessin $D_{\a}\,$, where $D_{\a} = \b_{\a}^{-1}([0,1])$; the claim \melanie\ is that $I$ is also an invariant of the dessin $D=\b^{-1}([0,1])$. For example, for the Belyi extending map $\alpha_2(z)=4z(1-z)\,$, the monodromy group of $D_{\a}$ is the cartographic group of $D$, which is known to be another Galois invariant. Later we will prove that the multiplication map of \refs{\CachazoIV, \CachazoYC}\ is the physical realization of the Belyi extending map $\a_2$.

\subsec{Invariants From Physics}

As discussed in the introduction, one problem that is very similar to the classification of dessins using Galois invariants is the problem of classifying branches of $\CN=1$ vacua using order parameters, such as Wilson and 't Hooft loops. We will consider those $\CN=1$ vacua that are obtained in the infrared by starting with an $\CN=2$ gauge theory and adding a tree level superpotential $W_{\rm tree}$ for the adjoint scalar $\Phi$. The discussion of the gauge theory order parameters in this section will closely follow that of \CachazoYC. In fact, what follows is just a summary. We refer the reader to \CachazoYC\ for all relevant details and proofs.

$\bullet$ {\it Confinement Index}. In a $U(N)$ gauge theory a natural order parameter is the
expectation value of a Wilson loop $W$ in, say, the fundamental
representation. The Wilson loop in the tensor product of $r$
fundamental representations is $W^r$. Clearly, for $r=N$ there is no
area law, for it is equivalent to a singlet representation (due to
electric screening). A measure of confinement is the smallest value
of $r$, which can only be between $1$ and $N$, for which $W^r$ does
not exhibit an area law. Such a value is denoted by $t$ and it is
called the confinement index. When the gauge group is broken
classically to a product of factors $U(N_1)\times U(N_2)\times
...\times U(N_n)$ one has to also use the 't Hooft loop $H$ to determine the confinement index. By embedding a 't Hooft-Polyakov
magnetic monopole of the full $U(N)$ theory in any two of the $U(N_i)$ factors,
we take into account magnetic screening. If for each $U(N_i)$ we get
that $W^{r_i}_iH_i$ has no area law, this implies that in the full $U(N)$
theory $W^{r_i-r_j}$ has no area law. The relative sign comes from the
fact that the magnetic monopole sits in both groups with opposite
charges.

Therefore, after taking into account electric and magnetic
screening, the confinement index is given by the greatest common
divisor of the $N_i$ and $b_i=r_i-r_{i+1}$. These two sets of
quantities, $N_i$'s and $b_i$'s, will have a very clear
combinatorial meaning, which will allow us to compute the
confinement index just by inspection of any dessin.

As a preparation for that let us mention that both set of quantities
are encoded in the expectation values of the generating function for
chiral operators $\Tr \Phi^s$, given by \CachazoDSW\
\eqn\gene{T(z) = \left\langle\Tr \left( {1\over z\Bbb{I}-\Phi }\right)\right\rangle \,.}
It turns out that the periods of $T(z)dz$, thought of as a meromorphic differential on $y^2=W'_{\rm tree}(z)^2+f_{n-1}(z)$, are related to the $N_i$'s and $b_i$'s as follows: the $N_i$'s are the periods of $T(z)dz$ on the $A$-cycles and the $b_i$'s are the periods of $T(z)dz$ on the $B$-cycles (for an appropriate choice of basis). The $b_i$'s measure the relative theta angle of $U(N_i)$ and $U(N_{i+1})$. Moreover, one can show that in the $\CN=1$ branch with confinement index $t$, $T(z)\,dz =t\, \tilde{T}(z)\, dz$, where $\tilde{T}(z)\, dz$ is the generating function for chiral operators in a Coulomb vacuum (which have $t=1$) of a $U({N\over t})$ theory \CachazoYC. The fact that the two generating functions are related by a multiplication by $t$ has an important consequence: all confining vacua with confinement index $t$ are obtained from Coulomb vacua by using the multiplication map by $t$ \refs{\CachazoIV, \CachazoYC}. The definition and discussion of the multiplication map is given in Appendix B. We also discuss this further in Section $4.1$ where, for the specific case of $t=2$, the multiplication map will be shown to coincide with the Belyi extending map $\a_2$ of [15].

$\bullet$ {\it Holomorphic Invariants}. In cases when the rank of
the low energy gauge group is too high, i.e. when the degree of the
tree level superpotential is large, there is always at least one
$N_i$ which is equal to $1$. The precise condition is ${\rm deg}\,W'(z) > N/2$: when this condition is satisfied, the confinement index is always
one. One might naively think that there is only one branch since we
have exhausted the standard order parameters. However, it is
possible to show that there are many branches, all of them having Coulomb
vacua. This is the problem that motivated the search for
non-conventional order parameters in \CachazoYC. It turns out that the
discussion that follows also applies for superpotentials of any
degree.

The new non-conventional order parameters proposed in \CachazoYC\
are obtained by studying relations between the vacuum expectation
values of different chiral operators that can be defined in the
theory. The expectation values of chiral operators become holomophic
functions on the moduli space of vacua due to supersymmetry. It
turns out that, at least in the examples considered in \CachazoYC,
these functions satisfy different polynomial constraints in
different branches. The problem of whether the existence of these
relations was the reason for the existence of the different branches
or viceversa was left as an open question. For the purposes of
this paper, we take the former as the correct point of view. In
fact, we will see that in the mathematical literature the refined
valency list for trees gives very similar information as the
relations between holomorphic functions found in \CachazoYC.

The chiral operators of relevance are $\Tr \Phi^r W_\alpha W^\alpha$,
whose appropriately normalized vacuum expectation value is denoted by $t_r
=-(1/32\pi^2)\langle\Tr \Phi^r W_\alpha W^\alpha\rangle$. In terms
of the reduced Seiberg-Witten curve $y^2=F_{2n}(z)$, they can be
computed as\foot{Strictly speaking, the curve needed for the
computation is given by the matrix model curve of the Dijkgraaf-Vafa
correspondence. However, in cases when none of the $N_i$ are zero,
the matrix model curve is the same as the reduced
Seiberg-Witten curve.}
\eqn\compu{t_r = {1\over 2\pi i}\, \oint_{\infty} z^r\, y(z)\, dz\,.}
The relations introduced in \CachazoYC\ to distinguish between
different branches are polynomial equations in the $t_r$'s.

The different branches that these relations distinguish are
determined by the distribution of double zeroes of the curve (2.6)
in the two factors of $P_N^2(z)-4\Lambda^{2N}$, i.e. by the pair
$(s_{+},s_{-})$. In other words, if we start with the factorization
problem $P_N^2(z)-4\Lambda^{2N}=F(z) H^2(z)$, we get
\eqn\asli{\eqalign{ P_N(z)-2\Lambda^{N} & = \tilde{R}_{N-2s_-}(z)\tilde{H}_{s_-}^2(z), \cr
P_N(z)+2\Lambda^{N} & = R_{N-2s+}(z)H_{s+}^2(z).}}

In order to derive the relations it is convenient to write
\eqn\expas{ y(z) = \sqrt{\tilde{R}_{N-2s_-}(z)R_{N-2s+}(z)}=
{H_{s_+}(z)R_{N-2s_+}(z)\over {\tilde
H}_{s_-}(z)}\sqrt{1-{4\Lambda^{N}\over H^2_{s_+}(z)R_{N-2s_+}(z)}} \,.}
Since the integral defining the $t_r$'s is around infinity, the computation can be carried out by expanding the square root. It is easy to see that if $0\le r\le s_+ +s_--2$ then only the leading term in the expansion
contributes. It turns out that in order to distinguish different
values of $(s_+,s_-)$ all that is needed are relations between those (restricted)
$t_r$'s. Using the fact that $\Lambda$ does not appear, by matching
dimensions (which for $t_r$ is $3+r$) and by matching the charge
under the $U(1)_\Phi$ symmetry (which for $t_r$ is $r$), one
concludes that the polynomials must be homogeneous in the number of
$\Phi$'s and the number of $W_{\alpha}W^{\alpha}$. Consider for
example the case when $s_-=1$ and $s_+=3$. Then one can show
that $t_0t_2-t_1^2=0$.

We will see in the next section that $(s_+,s_-)$ gives some
information about the refined valency list of a dessin. However, the
refined valency list contains more information. In section 5.2 we
will show by means of examples that the extra information of the
refined valency list can be obtained if one keeps the next to leading order
term in the expansion of the square root. In other words, in the expansion
\eqn\juicy{ \sqrt{1-{4\Lambda^{N}\over H^2_{s_+}(z)R_{N-2s_+}(z)}} =
1 - {2\Lambda^{N}\over H^2_{s_+}(z)R_{N-2s_+}(z)} + {\cal
O}(\Lambda^{2N}/z^{2N}) }
the first term gives information about $(s_+, s_-)$ while the second encodes the whole refined valency list. It would be interesting to find a combinatorial meaning of the higher order terms. It is important to mention that the extra relations we find distinguish the isolated point where the dessin appears from its neighbouring points in the $\N=1$ branch.

\newsec{Cross Fertilization}

Before we discuss some examples to illustrate our ideas, we would like to exhibit part of the dictionary between the mathematical and physical descriptions. First we show how the multiplication map \CachazoIV\ can be interpreted as an example of a Belyi-extending map \melanie; we also show that the information about the refined valency list of trees, described in Section $2.7$, can be recovered by studying the holomorphic invariants. Most importantly, we give a combinatorial interpretation of the confinement index introduced in \CachazoYC. We then speculate on the relation between the classification of dessins and the study of phases of gauge theories in four dimensions and formulate a few precise conjectures.

\subsec{Multiplication Map As A Belyi Extending Map}

Consider a tree with $N$ edges. The Belyi map has the form \purebelyi\
$$
\beta_N(z)=1- {P_N^2(z)\over 4\Lambda^{2N}} \,.
$$
The dessin corresponds to a rigid polynomial equation, which is a
special point in the parameter space of the non-rigid factorization
problem
\eqn\nonrigidN{
P_N^2(z) -4\Lambda^{2N} = F_{2n}(z)\, H^2_{N-n}(z) \,.
}
The multiplication map \CachazoIV\ (with multiplication factor $2$)
guarantees that, if $P_N(z)$ satisfies the non-rigid factorization
equation \nonrigidN, one solution to the factorization equation
$$
P_{2N}^2(z) - 4\Lambda^{4N} = F_{2n}(z)\, \tilde{H}_{2N-n}^2(z)
$$
is given by (see Appendix B for details)
\eqn\nonrigidtwoN{
P_{2N}(z)=2\Lambda^{2N}\,T_2\left({P_N(z)\over 2\Lambda^N}\right)\,,
}
where $T_2(x)=2x^2-1$ is a Chebyshev polynomial of the first kind. Since $P_N(z)$ gives rise to a Belyi map, $P_{2N}(z)$ as defined in \nonrigidtwoN\ also leads to a Belyi map $\beta_{2N}$ whose inverse image of the interval $[0,1]$ leads to a dessin with $2N$ edges. Therefore, by
applying the multiplication map, we get a new Belyi map of the form
\eqn\belyitwo{\eqalign{ \beta_{2N} &= 1-{P_{2N}^2(z)\over
4\Lambda^{4N}} = 1-\left(2\, {P_N^2(z)\over 4\Lambda^{2N}} -
1\right)^2 \cr & = {P_N^2(z)\over \Lambda^{2N}}
\left(1-{P_N^2(z)\over 4\Lambda^{2N}} \right) = 4\,
\beta_{N}(1-\beta_N) \cr &\equiv  \alpha_2\circ\beta_N \,, }} with
$\alpha_2(y)=4y(1-y)$. We thus find that this map coincides with the
Belyi extending map mentioned in Section $3.1$, which relates the monodromy
group to the cartographical group.

In \melanie\ the author gives a prescription to draw the dessin
associated with any Belyi extending map starting from the original
dessin. Roughly, the procedure consists in drawing on $\IP^1$ the
preimage through the Belyi extending map of the $[0,1]$ segment;
then, one substitutes this new drawing in place of each segment of
the original dessin. Let us apply this to our example. The map
$\alpha_2$ is of degree two, so it covers $[0,1]$ twice and we
expect it to double the number of edges of the dessin. More
precisely, the critical point $y=1/2$ is mapped by $\alpha_2$ to the
vertex $z=1$ of the $[0,1]$ segment. Therefore, we infer the
following rule to draw the dessin obtained through the Belyi
extending map $\alpha_2\,$:

\medskip\medskip

\centerline{\epsfxsize=0.50\hsize\epsfbox{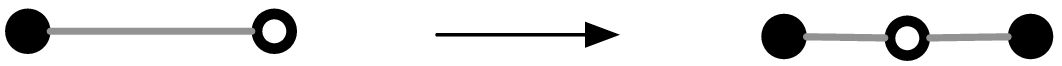}}

\medskip

One could check that for generic $t$ the multiplication map by $t$ is a
Belyi extending map that substitutes to each edge in the original
graph a branchless tree of length $t\,$. Naively, the dessins with
confinement index $2$ or higher would appear to be ``scaled up"
versions of smaller dessins. Indeed this is what one gets if one
applies the multiplication map to the rigid factorizations that lead
to the dessins.

However, from the gauge theory point of view, one can also apply the
multiplication map to the non-rigid problem \nonrigidN\ and {\it
then} impose the constraints that leads to a rigid factorization
problem. In other words, if $\CF_{R/NR}$ is the set of
rigid/non-rigid factorizations and if $M$ is the multiplication map
acting on the factorizations $\CF$,
\eqn\newindex{
M(\CF_{R}) \subset\ \bigg(M(\CF_{NR})\bigg)_{R} \,,
}
where the last subscript $R$ indicates that the factorizations are restricted to be rigid. This shows that the multiplication map is more than an operation to get new Belyi maps from old ones. Starting from a point in the moduli space of a $U(N)$ gauge theory which is {\it not} an isolated singularity (so that there is no associated dessin), one can apply the multiplication map by $t$ and sometimes obtain a singular point in the moduli space of the $U(tN)$ theory where one {\it can} obtain a dessin. We will discuss such examples in Section $5$.
Moreoever, we will also see in Section $4.3$, that applying the multiplication map to non-rigid factorizations is what allows us to prove that the confinement index is a Galois invariant.

\subsec{Refined Valency Lists From (Refined) Holomorphic Invariants}

We have already mentioned in Section $2.7$ that while solving the
non-rigid problem \nonrigid\ it is convenient to classify the
solutions in terms of integers $(s_+,s_-)$, where $s_{\pm}$ refers
to the number of double roots in either of the two factors $(P(z)\pm
2\Lambda^{N})$. We would now like to relate these numbers to the
refined valency list introduced in Section $2.7$. From the
definitions one can check that a given dessin with refined valency
lists $V^+=\{u^{+}_{k}\}$ and $V^-=\{u^-_{k} \}$ will appear in an
$\CN=1$ branch whose $(s_+,s_-)$ values are given by\foot{In assigning a refined valency list to a dessin, there is an overall choice of sign in assigning $+/-$ to the vertices. It follows that this amounts to exchanging $s_+$ and $s_-$. The same choice is also present in gauge theory. }
\eqn\holoval{
s_{\pm}=\sum_{k=1}^{\infty}\, k\, u^{\pm}_{2k} \,.
}
Recall that $u_{k}$ is the number of $k$-valent vertices in the
dessin. Clearly, the refined valency list contains more information
than just the values of $s_{\pm}$.

From the discussion in Section $3.2$, we have seen that the set of
relations between the expectation values of chiral operators $t_r$'s
depends upon the distribution of double roots $s_{\pm}$, i.e. different
branches are defined by the different polynomial relations between the $t_r$'s.
These chiral ring relations are satisfied at any generic point on
that branch. However the dessins appear only at special points in
that moduli space. A simple counting of parameters shows that at
such points there will be more relations that are not generically
satisfied. We will, in the examples to be discussed in Section $5$, write down these extra relations satisfied by the $t_r$'s explicitly.

Since the generic relations seem to distinguish branches of $\CN=1$ vacua, it is tempting to conjecture that these special points are isolated phases. In other words, they are distinct phases
from their neighbours in the $\N=1$ branch. That this is the case is
easy to see in some cases where the corresponding $\N=1$ theory
becomes superconformal in the IR. Moreover it is believed that there is always a choice of superpotential for which the resulting $\CN=1$ vacuum flows to an interacting superconformal theory \refs{\CachazoYC, \EguchiS, \Shih}.

\subsec{Confinement Index As A Galois Invariant}

The physical interpretation of the confinement index $t$ was given
earlier in this section. We also explained how $t$ can be computed
from the periods of a particular meromorphic differential $T(z)dz$
on the Seiberg-Witten curve. Here we first give a purely
combinatorial description of the confinement index $t$. Then, we
show that this is indeed a Galois invariant.

Consider a given tree T, constructed as T$=\beta^{-1}([0,1])$ under
a clean Belyi map $\beta(z) =1 - P(z)^2$, where $P(z)$ is a
polynomial. Let us concentrate on the preimages of $0$ under $\beta$. There can be
vertices with any valency. In particular, there must be vertices
with valence one; this is a simple consequence of the fact that the
Belyi map is clean.

The procedure for computing $t$ is the following: circle all
vertices with odd valence. Choose any of the circled univalent
vertices as the starting point. Move from the chosen vertex to the
next circled vertex, say going clockwise around the tree, and count
the number of edges between the two circled vertices; call it $h_1$.
Move from the second circled vertex to the next circled vertex.
Again count the number of edges between the two vertices; call it
$h_2$. The most important rule to apply when going around the tree
is that each circled vertex can be used only once as a starting
point and only once as an end point. Therefore, if the next vertex
was used previously both as an end point and as a starting point,
one should skip it and go to the next one. Continue around the tree
until there are no more unused circled vertices. The prescription
makes sense because there is an even number of odd-valent
vertices\foot{That there is an even number of odd vertices is clear
from the fact that $\beta$ is a polynomial of even degree.}.

After completing this procedure one is left with a list of integers
${\cal L} = \{ h_1,h_2,\ldots, h_f\}$, where $f$ is the number of
odd vertices in the dessin. Then $t$ is simply given by the greatest
common divisor of the elements of ${\cal L}$. Two simple examples
are shown below.

\medskip
\centerline{\epsfxsize=0.9\hsize\epsfbox{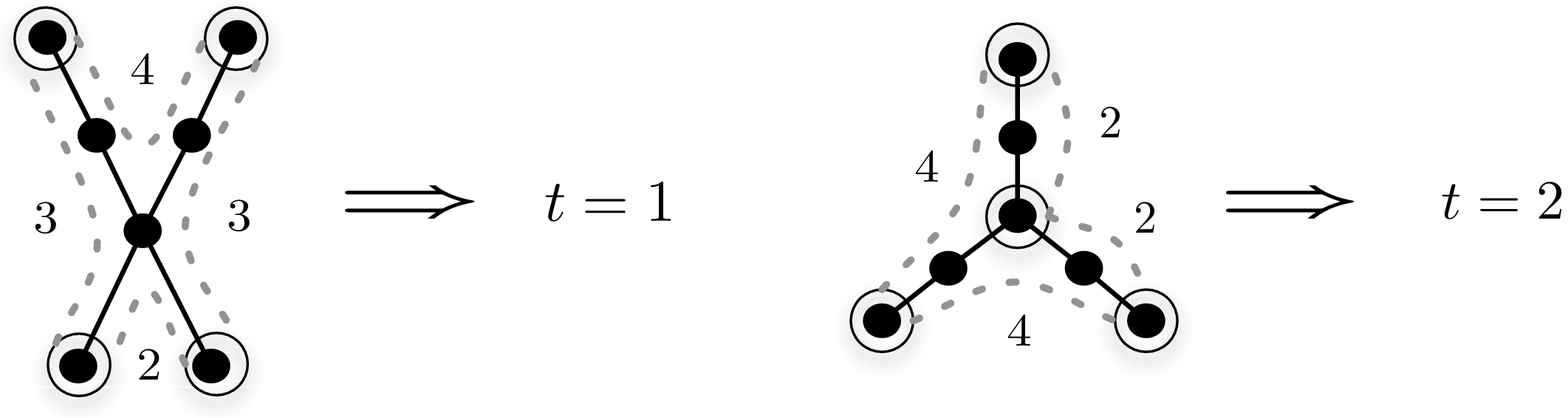}}
\noindent{\ninepoint\sl \baselineskip=3pt
}
\medskip

This rule works because the vertices with odd valence are precisely
the points between which one would draw the cuts in the gauge theory
approach. Then, this definition can be seen to coincide with the
definition of the confinement index introduced in \CachazoYC. This
follows from the fact that the integral of $T(z)\,dz$ between
successive vertices is $1$.

Having given a purely combinatorial definition of $t$ we proceed to show that this is indeed a Galois invariant. The proof involves concepts and terminology reviewed in Appendix A.

Let us consider all possible trees with a fixed number of edges $N$. If $N$ is prime then the only possible values of $t$ are $t=1$ and $t=N$. The only tree with $t=N$ is the branchless tree which is a
tree defined over $\Bbb{Q}$ and hence it is its own Galois
orbit\foot{Here what we have in mind is the branchless tree with
the simplest number field, which in this case it is $\Bbb{Q}$.}. All
other trees have $t=1\,$, so there is nothing further to prove.

Consider an $N$ which is not prime. Let $N=p_1^{r_1}\ldots
p_s^{r_s}$ be the prime decomposition of $N$. Take any $p_i$ and
consider the auxiliary polynomial $P_{p_i}(z)$. Use the
multiplication map by $m=N/p_i$ to produce what we called a
non-rigid curve (for more details on the multiplication map see
appendix B)
\eqn\nonbely{ 1 - T_m^2(P_{p_i}(z)) =
(1-P_{p_i}^2(z))U_{m-1}^2(P^2_{p_i}(z)).}
This depends on the $p_i+1$ coefficients of $P_{p_i}(z)$. As
discussed in section 2.2, the new ${\tilde P}_N(z) =
T_m(P_{p_i}(z))$ gives rise to a Belyi map $\beta(z) = 1 - {\tilde
P}_N^2(z)$ if and only if $\tilde{P}'_N(z)$ divides the right hand side of
\nonbely. This is equivalent to imposing that the right hand side of
\nonbely\ has only $N-1$ distinct roots. These conditions will give
rise to polynomial equations for the coefficients of $P_{p_i}(z) =
a_0 z^{p_i}+\ldots +a_{p_i+1}$. Let the set of polynomials that must
vanish be ${\cal S} = \{ f_1(a),\ldots , f_j(a)\}$. Since all
Chebyshev polynomials $T_m(z), U_{m-1}(z)$ have coefficients in
$\Bbb{Q}$ it follows that $f(a)\in \Bbb{Q}[a_0,\ldots ,a_{p_i+1}]$.
Therefore, there is a splitting number field $K_{\cal S}$ associated to the set
${\cal S}$. It is a finite normal extension of $\Bbb{Q}$ and hence
is left invariant by $\GG$. This means that the solutions form
full Galois orbits. From the relation between the multiplication map
and the confinement index $t$ it follows that all such orbits can
only have values of $t$ which are multiples of $m$ that divide $N$. This means
that either $t=m =N/p_i$ or $t=p_i m = N$. As mentioned above, there
is a single tree with $t=N$ and therefore all other orbits must have
the same value $t=N/{p_i}$.

Consider now $k=p_ip_j$ and the polynomial $P_k(z)$. As before, use
the multiplication map by $m=N/k$. Following the same procedure we
conclude that dessins arising this way can only have values of $t$ which are
multiples of $m$ and that divide $N$. The only possibilities are
$t=N/k, N/p_i, N/p_j, N$. As before, $t=N$ gives a single tree. We
have already proven that dessins with $t=N/p_i$ or $t=N/p_j$ can
only come in full Galois orbits. Therefore the remaining dessins
with $t=N/k$ also arise in full Galois orbits.

One can continue this argument by induction and conclude that any
two dessins in the same Galois orbit must have the same confinement
index $t$. Thus, the confinement index is a Galois invariant.

\subsec{Speculations About Dessins And Gauge Theory: Weak And Strong
Conjectures}

We now have all the ingredients we need to formulate our conjectures
precisely. We have already seen that there exist  $\CN=1$ branches
in pure gauge theory that are classified by order parameters such as
the confinement index. Also, as mentioned in Section $4.2$ (and as
we will show in some simple examples), at the special points where
the dessins appear one has extra chiral ring relations. For an
appropriately chosen superpotential, these points are believed to
give rise to superconformal $\N=1$ theories in the IR and thus
define new phases. However, it is known that for some
superpotentials, these theories might not be singular. This does not
exclude the possibility that these points might be new phases,
perhaps distinguished by less exotic behavior, such as extra
massless states or smaller rank of the gauge group.

Given the earlier discussion regarding the theory of dessins and the
phases of supersymmetric gauge theory, our first conjecture should
be fairly well motivated: all points where dessins appear correspond
to special phases embedded in the $\N=1$ branches which we call
``isolated phases".

Our second conjecture, relating the phases to Galois orbits of
dessins has a weak and a strong form. The strong form is easily
stated: all Galois invariants are physical order parameters that can
be used to distinguish the isolated phases of supersymmetric gauge
theory with a given superpotential.

On the other hand, based on the examples we work out in Section $5$,
we find that all the order parameters used to distinguish
branches of gauge theories that meet a particular $U(1)$ branch,
defined below, are Galois invariants. This is what we refer to as
the weak form of the conjecture.

Some comments are in order. By a $U(1)$ branch we mean a branch
where the low energy group is a single $U(1)\subset U(N)$. This can
be called the maximally confining branch. This branch has the same dimension as the other
branches \CachazoYC. Any generic branch meets these $U(1)$ branches
at points where the corresponding dessin is a branchless tree with
$N$ edges.

We believe that the weak form of the second conjecture is very
likely to be correct and we provide evidence for it in the examples.
The strong form is on much less firm ground. In particular, it
relies on the correctness of the first conjecture and on assumptions that require much more study.

It would be very important to gather more evidence for the strong
form since, if true, it provides a striking connection between
Grothendieck's program of unveiling the structure of $\GG$ via its
action on dessins and the physics problem of classifying phases of
supersymmetric gauge theories. Some of the most striking
consequences would be for gauge theories with matter where physics
order parameters are scarce. Almost all known Galois invariants
would become new gauge theory order parameters. In the next
sections, we will provide some evidence in support of these
conjectures.

\subsubsec{A More General $\N=1$ Viewpoint And A Global $\N = 2$
Viewpoint}

\medskip

The possibility that there is always an ``extremal" superpotential
for which the $\N=1$ $U(N)$ gauge theory at one of the isolated
singular points becomes superconformal in the IR motivates the
following point of view. Up to now we have studied theories with a
superpotential of a given degree. However, if we fix $U(N)$ and vary
the degree of the superpotential, the theories arising at the points
where a dessin $D$ appears can go from being non singular to
singular in the IR. Let us denote by $d(D)$ the smallest degree of an
extremal superpotential for $D$. It is tempting to conjecture that
two theories that arise at points corresponding to two dessins $D$
and $D'$ in the same Galois orbit will necessarily have
$d(D)=d(D')$. In other words, $d(D)$ might be a Galois invariant. We
leave this problem as an interesting direction for future work.

Finally, let us comment on yet another point of view. Suppose that
we set the tree level superpotential to zero. Then we recover an
$\N=2$ gauge theory. The Seiberg-Witten curve of section 2.5 becomes
the curve describing the physics in the moduli space of vacua of a
single theory. The valency lists $C$ and $V$ of section 2.3 simply
encode information about the masses of particles in the theory. More
explicitly, $C$ determines the distribution of masses of fundamental
hypermultiplets. $V$ determines, up to modular transformations, the
charges of the various monopoles and dyons that are massless in
addition to the $U(1)^N$ vector multiplets present at generic
points. According to the cases studied in the literature, there is
reasonable evidence to suspect that all such points are $\N=2$
superconformal field theories\foot{Of course, the point with $N-1$
mutually local massless monopoles is not a superconformal field
theory. In this case one can write down, using $S-$duality, a local
lagrangian describing the full behavior of the theory in the IR.}.

It would be very interesting to explore the relation between the
classification of dessins into Galois orbits and the classification
of such $\N=2$ superconformal field theories. A
natural possibility, worth exploring, is that field theories giving
rise to dessins in the same Galois orbit might be dual theories in
some sense.

\newsec{Examples: $U(6)$ Pure Gauge Theory}

In this section we would like to illustrate by means of examples the
concepts we have covered up to now. We consider pure $\CN=2$ $U(6)$
gauge theory broken to $\CN=1$ by a tree level superpotential. From
the general discussion about dessins and polynomial equations, it
follows that the $\CN=2$ moduli space contains an isolated
singularity for every connected tree with $6$ edges. It turns
out that {\it all} such dessins can be obtained by just using a
quartic superpotential. We discuss why higher degree superpotentials
are not needed and list all dessins with their corresponding
factorization problems in Appendix C.

Here, we restrict our study to a cubic superpotential. The dessins
we obtain are, of course, a subset of the general quartic
superpotential but they turn out to exhibit all the relevant points
of the physics-mathematics dictionary we have established. All
$U(N)$ gauge theories with $N=2,\ldots, 6$ were studied in detail in
\CachazoYC\ where one parameter solutions to the
factorization problem
\eqn\extwo{ P_N(z)^2 -  4\Lambda^{2N}  = F_4(z)\,H^2_{N-2}(z) }
are listed.  In fact the analysis in this section can be easily
repeated for all these cases. We choose $U(6)$ because
it is the simplest case that exhibits four different values of the
confinement index, i.e. $t=1,2,3,6$.

The rigid factorizations corresponding to dessins are obtained by imposing suitable conditions on the solutions found in \CachazoYC\ along the lines we described in the introduction. In Sections $5.1$ and $5.2$ we classify the dessins according to the $\CN=1$ branches to which they belong and specify the order parameters that distinguish the special points where the dessins appear as isolated phases.

From a mathematical point of view, in order to find the explicit Belyi maps, it is not necessary to start from the non-rigid problem \extwo. Instead, one solves the rigid problems directly. In Section $5.3$ we will present the solution to all possible rigid factorization problems that can be derived from \extwo\ along the lines of \AshokCD\ by using differentiation tricks. This analysis shows explicitly the classification of trees into distinct Galois orbits. In Section $5.4$ we will reproduce the same classification of dessins using some of the known Galois invariants. We will find that this parallels the classification of phases in gauge theory.

We mentioned in Section $2.7$ that, both in the physics and mathematical analysis, there is the freedom to shift and scale the $z$ variable. From a physics perspective, it is natural \CachazoYC\ to shift the $z$ variable in order to bring the superpotential to the canonical form $W'_{\rm tree}(z)=z^2-\Delta$ and then analyze the $\CN=1$ branches obtained by varying $\Delta$. However, since our primary goal is to exhibit the dessins and where they appear in the gauge theory moduli space, in the examples that follow, we have followed the mathematical strategy to shift and scale $z$ to put the solution in the simplest form possible, so that the number field associated to the tree is the simplest.

\subsec{$U(6)$ Gauge Theory: A Physicist's Point Of View}

Let us now review the solution of \CachazoYC\ in detail. We then
specialize to rigid factorizations by tuning the one free parameter
available in the solutions to \extwo.

As described in \CachazoYC\ one can, first of all, classify $\CN =1$ branches by the number of double roots in either factor $(P_6(z) \pm 2\Lambda^6)$. If we denote the number of double roots in either factor as $(s_{+},s_{-})$, in our case, these can take the values $(3,1)$, $(1,3)$  and $(2,2)$. All these branches meet at vacua that have $(s_+,s_-)=(3,2)$ or $(2,3)$ at which the branchless tree discussed in Section $2.7$ appears. The $\CN=1$ branches are further classified by the confinement index  and the non-conventional order parameters which we defined earlier in Section $3.2$. Let us consider each value of $(s_+,s_-)$ in turn.

\medskip

$\bullet$ {\it The $(3,1)$ And $(1,3)$ Confining Vacua}

\medskip

The general factorization problem \extwo\ is solved by the polynomials\foot{Refer to equation $(3.47)$ in \CachazoYC.}

\eqn\gaugeone{\eqalign{
P_6(z) + 2\eta\Lambda^6 &= \big((z-a)^2 (z-b)-2\epsilon\Lambda^3\big)^2 \equiv R^2(z) \cr
P_6(z)- 2\eta\Lambda^6 &= (z-a)^2(z-b)\big((z-a)^2 (z-b)-4\epsilon\Lambda^3\big) \equiv S(z)}}
with $\eta^2=1$ and $\epsilon^2=\eta$. These polynomials can by
obtained by the ``multiplication by $2$" map acting on either of the
polynomials $P_3(z) = (z-a)^2(z-b) \mp 2\Lambda_0^3$:
\eqn\Psix{
P_6(z)  = 2 \Lambda^6 \kappa^2 \, T_2\left({P_3(z)\over 2\kappa\Lambda^3} \right) \,,
}
with $\Lambda_0^6 = \kappa^2\Lambda^6$, $\kappa^4=1$ and
$\epsilon=\pm\kappa$. The various signs and phases in these
expressions are crucial so that all the $\CN=1$ vacua are taken into
account.  However, as discussed in Section $2.7$, since these lead
to trees in the same equivalence class, we will  drop such phase factors in what
follows.

\medskip

{\it The Rigid Quartic Factorization}

\medskip

If we require that the polynomial $R(z)$ has a double root (which is to
say that its discriminant vanishes) we get the rigid factorization
\eqn\exone{ P_6(z)^2 - 4\Lambda^{12} = F_4(z)\,H^2_2(z)\,
Q^4_1(z) \,. } This fixes $\Lambda$ to be
\eqn\gaugetwo{
\Lambda^3 = {2\over 27}(a-b)^3 \,.
}
Substituting this into the polynomials in \gaugeone\ and using the shift and scaling symmetry to set $a=-5$ and $b=10$, we get
\eqn\gaugethree{\eqalign{
R^2(z) &= (z-5)^4(z+10)^2 \,, \cr
S(z) &= (z+5)^2(z-10)(z^3-75z+750) \,.
}}
From this, we see that the polynomials that solve the equation
\eqn\quarticagain{
P_6^2(z) - 4(250)^4= F_4(z)\, H_2^2(z)\, Q_1^4(z)
}
are given by
\eqn\gaugefour{\eqalign{
Q_1(z) = z-5\,, \quad& H_2(z) =  (z+5)(z+10)\,, \quad F_4 = (z-10)(z^3-75z+750)\,, \cr
P_6(z)&=z^6-150z^4+500 z^3+5625z^2-37500z-62500\,.
}}
Plotting the zeroes of the polynomials leads to the tree in Figure $10$. Let us make a few comments about the solution. All polynomials are defined over $\Bbb{Q}$. From the discussion in Section $2.4$ about the action of the Galois group, we see that the tree is left invariant; in other words, it is the only element in its Galois orbit. From \Psix\ we see that the tree has confinement index $2$. However, observe that the tree is {\it not} a scaled up version of a smaller tree. This illustrates the point made in Section $4.1$ and especially equation \newindex. One can moreover check that the combinatorial method for computing the confinement index, as explained in Section $4.2$, also gives the correct answer $t=2$.

\medskip

\centerline{\epsfxsize=0.84\hsize\epsfbox{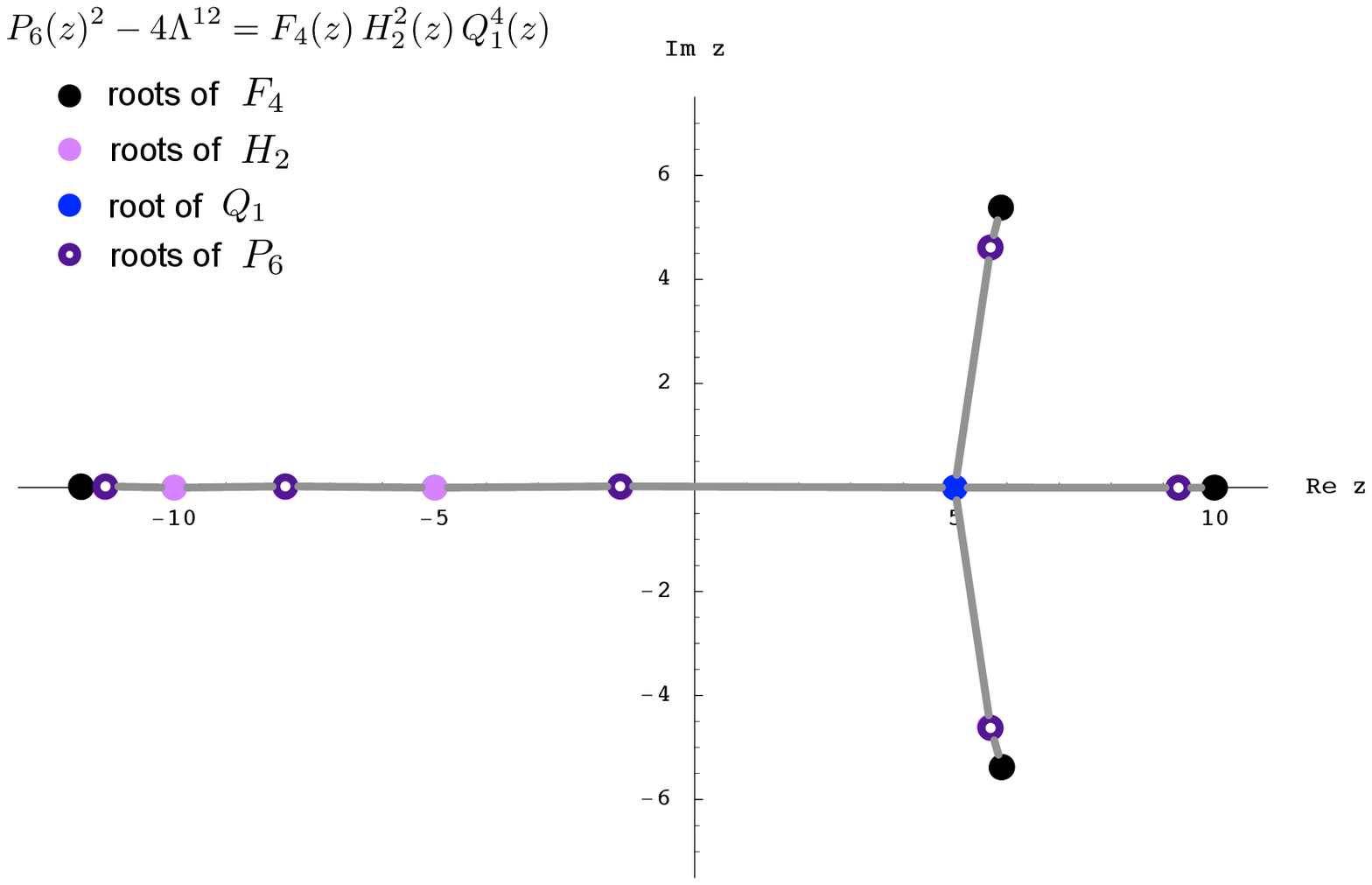}}
\noindent{\ninepoint\sl \baselineskip=3pt {\bf Figure 10}:{\sl $\;$
The tree obtained in the $(3,1)$ confining branch with a four-valent vertex. It has confinement index $t=2$.}}

\medskip

{\it The Rigid Cubic Factorization}

\medskip

Starting from \extwo\ one can also get another rigid factorization by tuning one of the zeroes of $F_4(z)$ to coincide with a zero of $H_4(z)\,$:
\eqn\excubic{
P_6^2(z) - 4\Lambda^{12} = F_3(z)\, H^2_3(z)\,Q_1^3(z) \,.
}
This leads to the condition $a=b$ in \gaugeone. This implies that $S(z)$ has a cubic root at $z=a$. Using the shift and scale symmetry, we can set $a=0$ and $\Lambda=1$. This leads to a very simple solution of \excubic\ :
$$\eqalign{
P_6(z) = z^6-4z^3+2\,, \qquad H_3(z) = z^3-2\,, \qquad F_3(z) = z^3-4 \quad\hbox{and}\quad Q_1(z) = z \,.
}$$
The tree associated to the factorization is drawn below in Figure $11$.

\medskip
\centerline{\epsfxsize=0.85\hsize\epsfbox{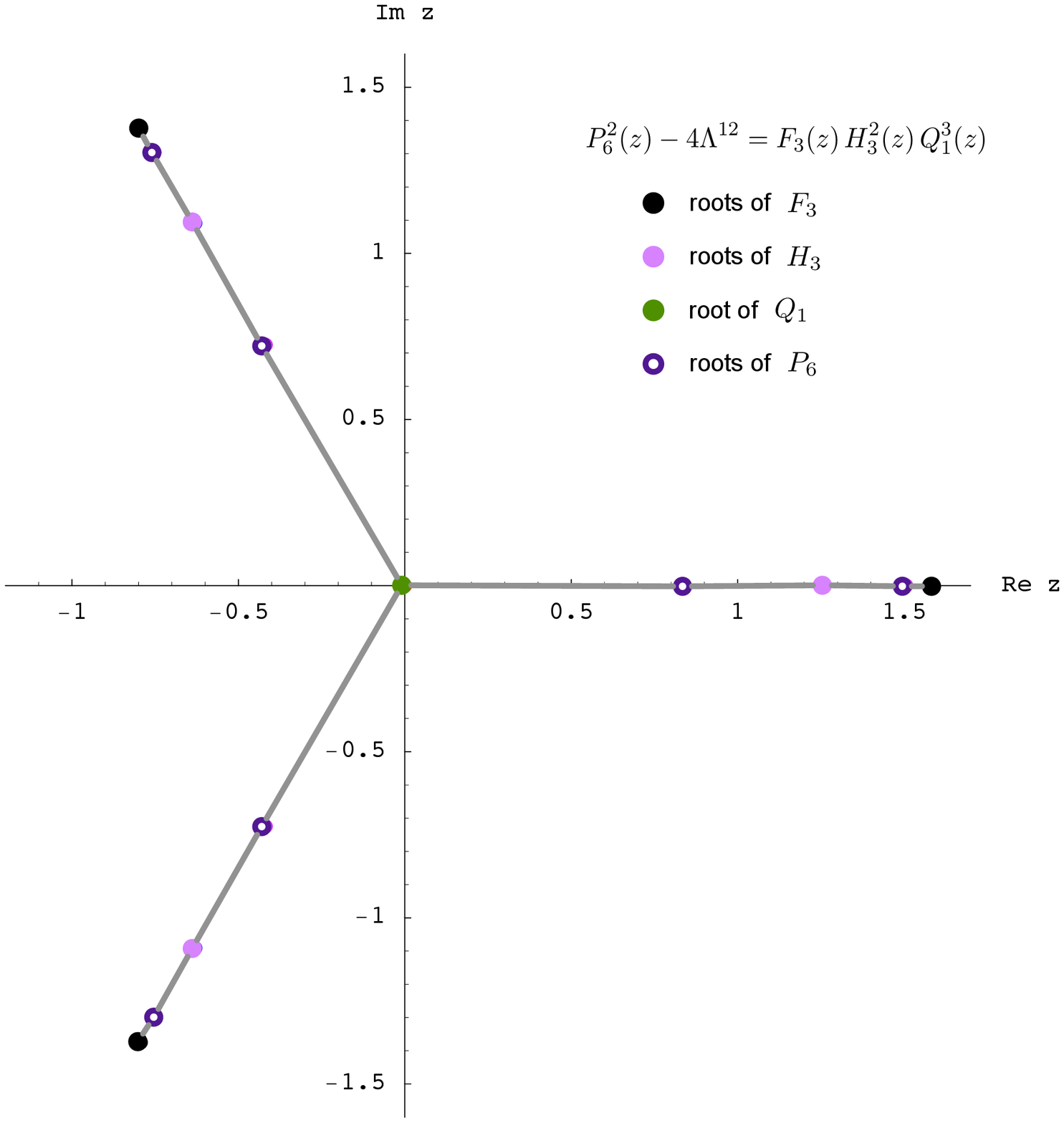}}
\noindent{\ninepoint\sl \baselineskip=3pt {\bf Figure 11}:{\sl $\;$
Tree obtained in the $(3,1)$ confining branch with a trivalent vertex. It is a scaled up dessin, obtained by applying the multiplication map on a smaller dessin with $3$ edges.}}
\medskip

Note that, unlike the quartic case, the multiplication by $2$ is easily understood: this particular solution can also be obtained by first solving the rigid factorization problem
\eqn\olddessin{
P_3^2(z) - 4\Lambda^6 = Q_1^3(z)\, F_3(z) \,,
}
and then applying to the resulting solution the multiplication map. The solution is once again defined over $\Bbb{Q}$ and the tree in Figure $11$ is the lone element in its Galois orbit.

\medskip

$\bullet${\it The $(2,2)$ Confining Vacua}

\medskip

In this sector, the factorization problem \exone\ is solved by the polynomials\foot{Refer equation $(3.50)$ in \CachazoYC.}
\eqn\gaugefive{\eqalign{
P_6(z)+2\Lambda^6 &= (z^2+g-\Lambda^2)^2(z^2+g+2\Lambda^2) \,, \cr
P_6(z)-2\Lambda^6 &= (z^2+g+\Lambda^2)^2(z^2+g-2\Lambda^2) \,.
}}
These polynomials are obtained by the ``multiplication by $3$" map; modulo phase factors, $P_6(z)$ in \gaugefive\ is given in terms of $P_2(z) = z^2+g$ as
\eqn\multthree{
P_6(z) = 2\Lambda^6\,T_3\left({P_2(z)\over 2\Lambda^2} \right) \,.
}
The trees in this branch will therefore have confinement index $3$.

\medskip

{\it The Rigid Quartic Factorization}

\medskip

One can set $g = \Lambda^2$ to get the quartic factorization \exone, while one can rescale $z$ to set  $\Lambda = 1$. The resulting polynomials that solve \exone\ are
\eqn\gaugesix{\eqalign{
P_6(z) &= z^4(z^2+3)-2\,, \quad H_2(z)=(z^2+2)\,, \cr
Q_1(z) &= z\quad \hbox{and}\quad F_4(z)= (z^2+3)(z^2-1)  \,.
}}
Plotting the roots of the polynomials leads, this time, to the tree in Figure $12$. The polynomials are defined over $\Bbb{Q}$ and so the tree is the only element in its Galois orbit.

\medskip

{\it The Rigid Cubic Factorization}

\medskip

Note that it is not possible to get the cubic factorization equation \excubic\ by tuning the available free parameter. Thus, we do not find any trivalent tree in this branch of the moduli space. One can also show this using the combinatorial definition of the confinement index by trying to construct a trivalent tree with six edges and $t=3$. 

\medskip
\centerline{\epsfxsize=0.9\hsize\epsfbox{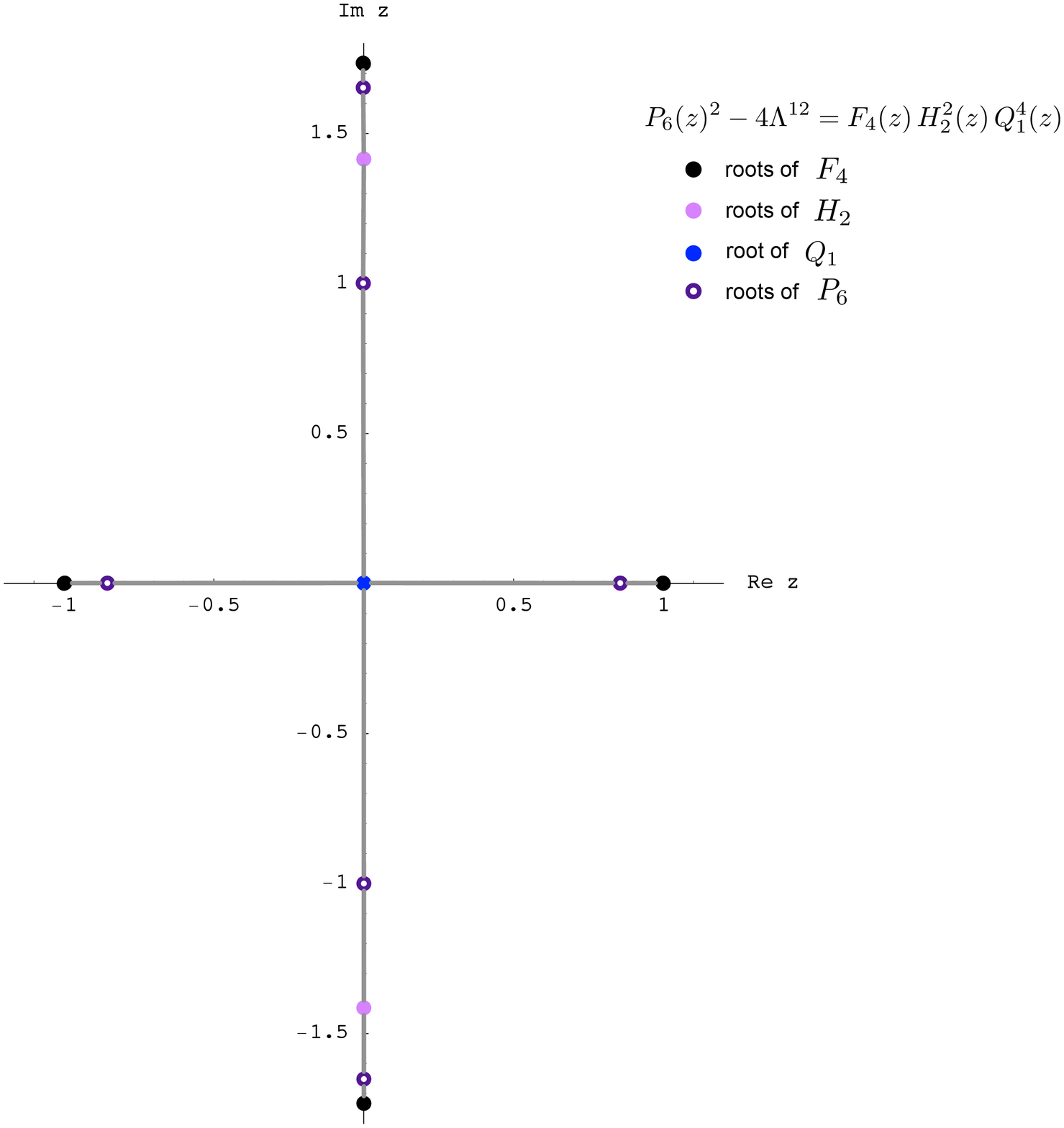}}
\noindent{\ninepoint\sl \baselineskip=3pt {\bf Figure 12}:{\sl $\;$
The tree obtained in the $(2,2)$ confining branch with a four-valent vertex. It has confinement index $t=3$. }}

\medskip

$\bullet${\it The $(2,2)$ Coulomb Vacua}

\medskip

The solution of the factorization problem \exone\ in this branch is parametrized as\foot{Refer to equation $(3.53)$ in \CachazoYC. We have set $g\rightarrow \sig\, h$, $z-\rightarrow z\, h$ and $\Lambda \rightarrow \Lambda\, h$ in that equation.}

 \eqn\gaugeseven{\eqalign{
 &P_6(z)+2\Lambda^6= \left[z^2 + (1+\sig)z+{(3+\sig)(9+15\sig -\sig^2+\sig^3)
 \over 108 }\right]^2\cr
 &\qquad\qquad
 \qquad\qquad \left[z^2 -{(1-\sig)(3-\sig)^2(3+\sig)\over 108}\right]\cr
 &P_6(z)-2\Lambda^6= \left[(z+{2\sig\over 3})^2 + (1-\sig)(z+{2\sig\over 3})
 +{(3-\sig)(9-15\sig -\sig^2-\sig^3) \over 108}\right]^2\cr
 &\qquad\qquad \qquad\qquad \left[(z+{2\sig\over 3})^2
 -{(1+\sig)(3+\sig)^2(3-\sig)\over 108 }\right]}}
with $\sig$ and $\Lambda$ satisfying the constraint
 \eqn\constraint{\sig^5(\sig^2-9)^2 = 27^3\Lambda^6.}

\medskip

{\it The Rigid Quartic Factorization}

\medskip

Requiring that the first factor in the either of the two equations in \gaugeseven\ has a double root leads to the quartic factorization \exone. We get the condition
\eqn\gaugeeight{
\sig^2 - 25 = 0\,.
}
For $\sig=5$, the polynomials that solve the equation \quarticagain\ are given by
\eqn\gaugefour{\eqalign{
Q_1(z) &= z-2\,, \quad H_2(z) =  z^2-{2\over 3} z+{128\over 27}\,, \quad F_4 = \left(z^2+{64\over9}\right)\left(z^2-{20\over 3}z+{332\over 27}\right)\,, \cr
P_6(z)&=z^6-8z^5+{280\over 9}z^4-{800\over 9} z^3+{560\over 3}z^2-{2048\over 9}z+{3839488\over 19683}\,.
}}
Plotting the zeroes of the polynomials lead to the tree in Figure $13$. Since the tree is found in the Coulomb branch, it has confinement index $1$. This can also be checked directly using the combinatorial definition: we get $t=GCD(3,4)=1$. For $\sig=-5$, we get an equivalent tree but reflected about the $\Re(z)=0$ axis.  Note that the solution in \gaugefour, like the ones we have obtained earlier, are polynomials defined over $\Bbb{Q}$. This is why the Galois orbits in each case consist of only a single tree. We now discuss a set of trees whose associated number field is non-trivial and therefore constitute a larger Galois orbit.

\medskip
\centerline{\epsfxsize=0.85\hsize\epsfbox{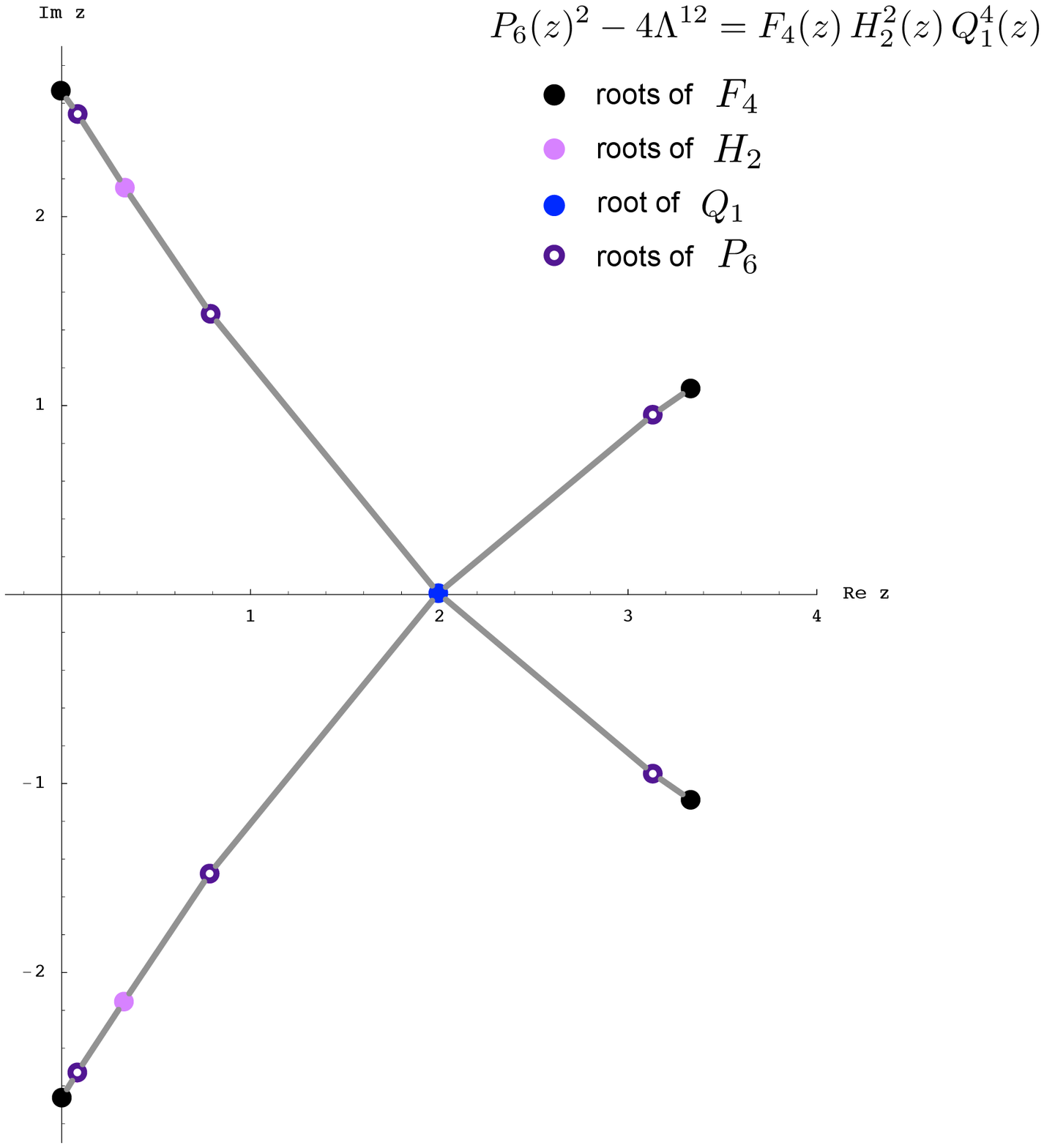}}
\noindent{\ninepoint\sl \baselineskip=3pt {\bf Figure 13}:{\sl $\;$
The tree obtained in the $(2,2)$ coulomb branch with a four-valent vertex.}}
\medskip

\medskip

{\it The Rigid Cubic Factorization}

\medskip

Requiring that the two factors in the first of the two equations in \gaugeseven\ have a root in common leads to the non-trivial condition\foot{The similar constraint for the second equation does not change the number field and we get the same set of trees.}
\eqn\gaugenine{
\sig^3-3\sig^2+3\sig+15 = 0 \,.
}
Solving for $\sig$ leads to three solutions
\eqn\gaugeten{
\sig = \left\{\matrix{\sig^{(0)} = (1-2\, (2)^{1\over 3}) \cr \cr
\sig^{(+)} = (1+2^{1\over 3}(1-i\,\sqrt{3})) \cr\cr
\sig^{(-)} = (1+2^{1\over 3}(1+i\,\sqrt{3})) }\right. }
Substituting these results into the polynomials and plotting their roots lead to the trees in Figure $14$, $15$ and $16$ respectively. All of these have confinement index $t=1$.

From the fact that all three trees are obtained from a single polynomial \gaugenine\ irreducible over $\Bbb{Q}$, it follows that these three trees belong to the same Galois orbit.  From \gaugeten, we observe that unlike the earlier solutions which were all defined over $\Bbb{Q}$, the number field associated to these  trees is non-trivial.

\medskip
\centerline{\epsfxsize=0.80\hsize\epsfbox{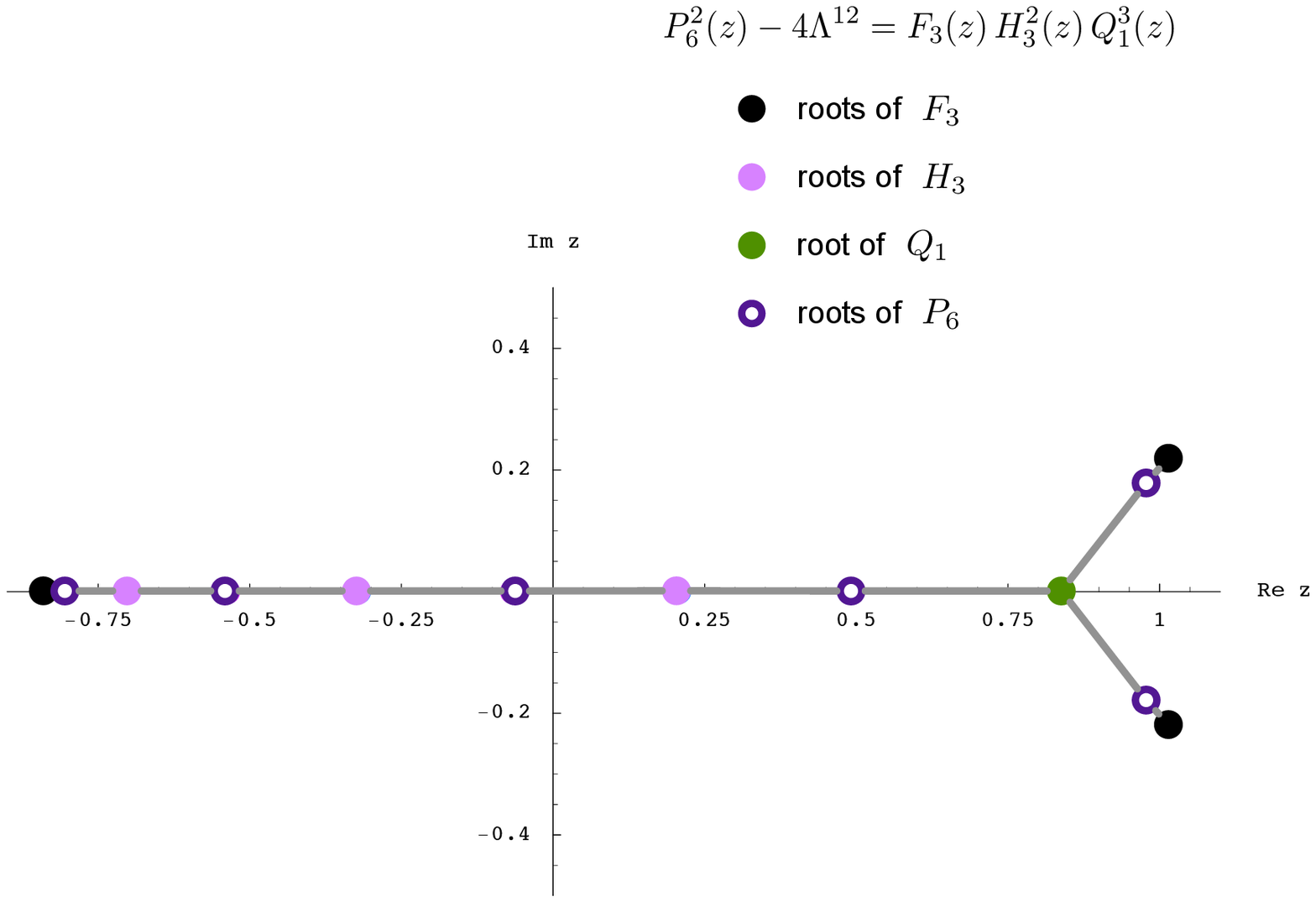}}
\noindent{\ninepoint\sl \baselineskip=3pt {\bf Figure 14}:{\sl $\;$
Tree for the case $\sig^{(0)}$ in \gaugeten. It has $t=1$.}}
\medskip

Let us study this example in more detail. The terminology is
explained in Appendix A. By inspection of \gaugeten, we find that $\sigma^{(0)}$ and 
$\sigma^{(\pm)}$ belong to the field $K=\Bbb{Q}(2^{1\over
3},\omega)$ where $\omega = \half (1+i\sqrt{3})$ (a cube
root of unity). It is easy to see that $K$ is also the splitting
field of the polynomial $x^3-2$, whose associated Galois group ${\rm
Gal}(K/\Bbb{Q})$ is the group of permutations of three elements,
$S_3$, which is non-abelian.

\medskip
\centerline{\epsfxsize=0.85\hsize\epsfbox{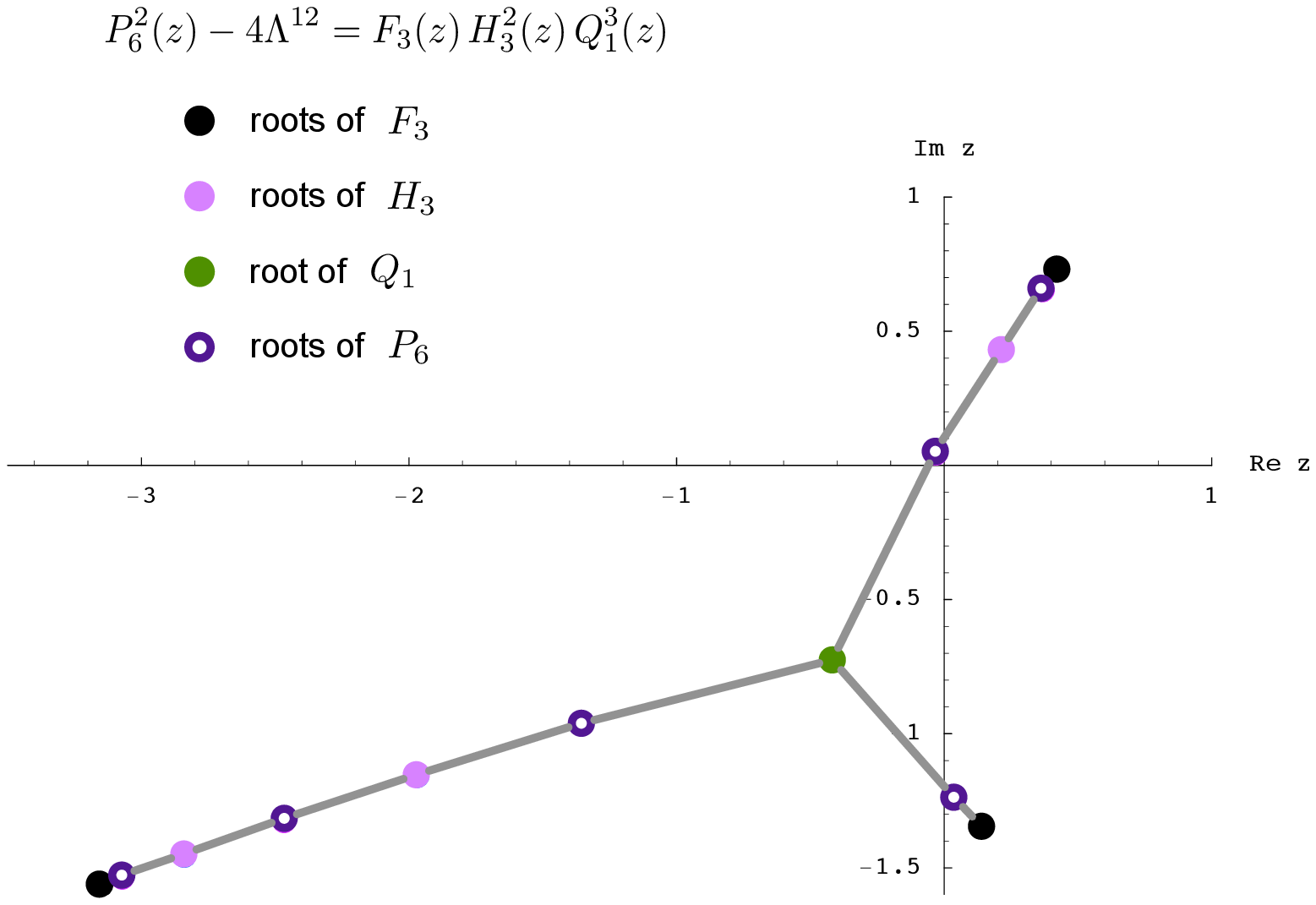}}
\noindent{\ninepoint\sl \baselineskip=3pt {\bf Figure 15}:{\sl $\;$
Tree for the case $\sig^{(-)}$ in \gaugeten. It has $t=1$.}}

\medskip

\centerline{\epsfxsize=0.85\hsize\epsfbox{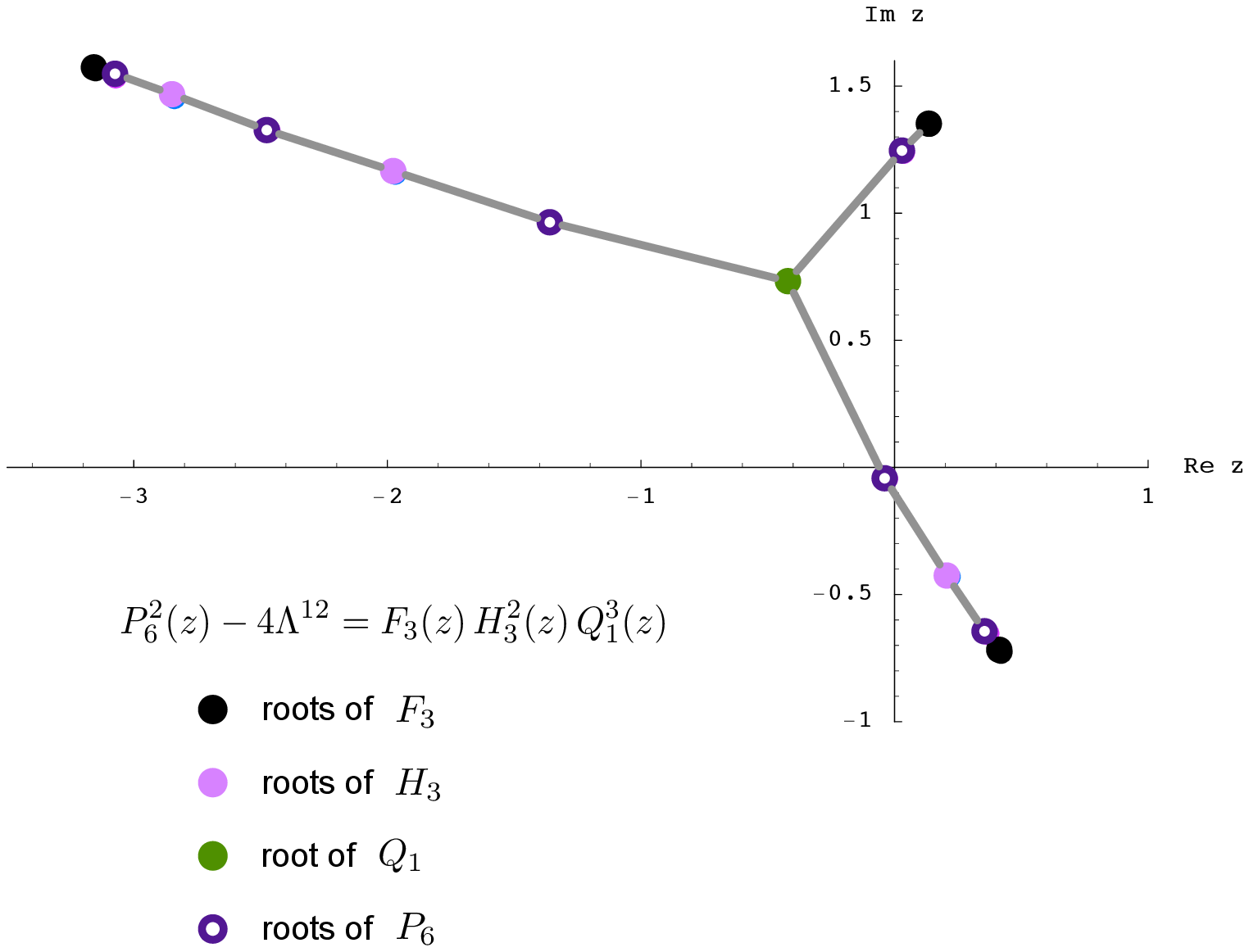}}
\noindent{\ninepoint\sl \baselineskip=3pt {\bf Figure 16}:{\sl $\;$
Tree for the case $\sig^{(+)}$ in \gaugeten. It has $t=1$.}}

\eject

$\bullet${\it The $(3,2)$ And $(2,3)$ Vacuum}

\medskip

If two roots of $F_4(z)$ coincide in \extwo\ we get the factorization problem
\eqn\exDS{
P_6^2(z) - 4\Lambda^{12} = F_2(z)H_{5}^2(z) = (z^2-4\Lambda^2)H_{5}^2(z) \,,
}
where, in the second equality, we have suitably shifted and scaled
$z$. Setting $\Lambda$ to one, \exDS\ is solved by the polynomials \Douglas\
$$
P_6(z)= 2\, T_6\left({z\over 2}\right) \qquad \hbox{and}\qquad H_{5}(z) = U_{5}\left({z\over 2} \right) \,.
$$
Plotting the zeroes of these polynomials leads to the branchless
tree discussed in Section $2.8$. In Figure $17$, we show the tree
that arises for the particular case of $U(6)$. As explained in the
earlier more general discussion, this is the singularity at which
the $\CN=1$ branches meet. This is the only case we consider that
has one of the $N_i=0$. Therefore this can be explained as the
intersection of $U(1)$ and $U(1)^2$ branches. See Appendix C for a
more complete discussion. From the combinatorial definition of the
confinement index, the dessin has $t=6$, the largest value for the
case with $N=6$ edges.

\medskip

\centerline{\epsfxsize=1.0\hsize\epsfbox{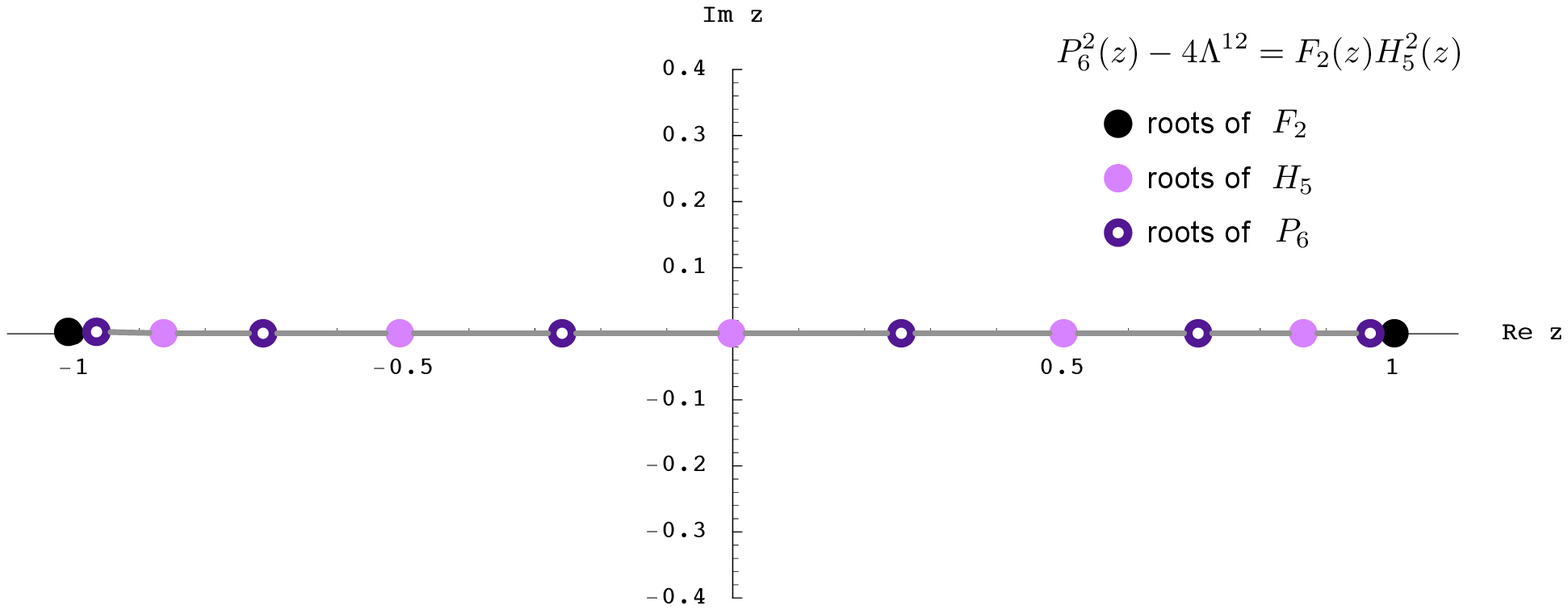}}
\noindent{\ninepoint\sl \baselineskip=3pt {\bf Figure 17}:{\sl $\;$
The tree obtained at the maximally confining point. It has $t=6$}}

\medskip

\subsec{Classifying Dessins From Gauge Theory}

So far we have started with the non-rigid factorization problem and tuned the parameters to get isolated singularities where dessins appear. We have seen how the dessins fall into different Galois orbits. We now classify them according to the $\CN=1$ branches to which they belong using gauge theory order parameters. We summarize all our findings from the gauge theory point of view in Figure $18$. Given that the confinement index is a Galois invariant, we find each of the three trees with a $4$-valent
vertex to belong to distinct Galois orbits as shown in Figure $18$. Similarly, using the confinement index, we find that the trivalent trees fall into at least two distinct Galois orbits: the trivalent dessin with $t=2$ is left invariant under the action of $\GG$.

\medskip

\centerline{\epsfxsize=0.90\hsize\epsfbox{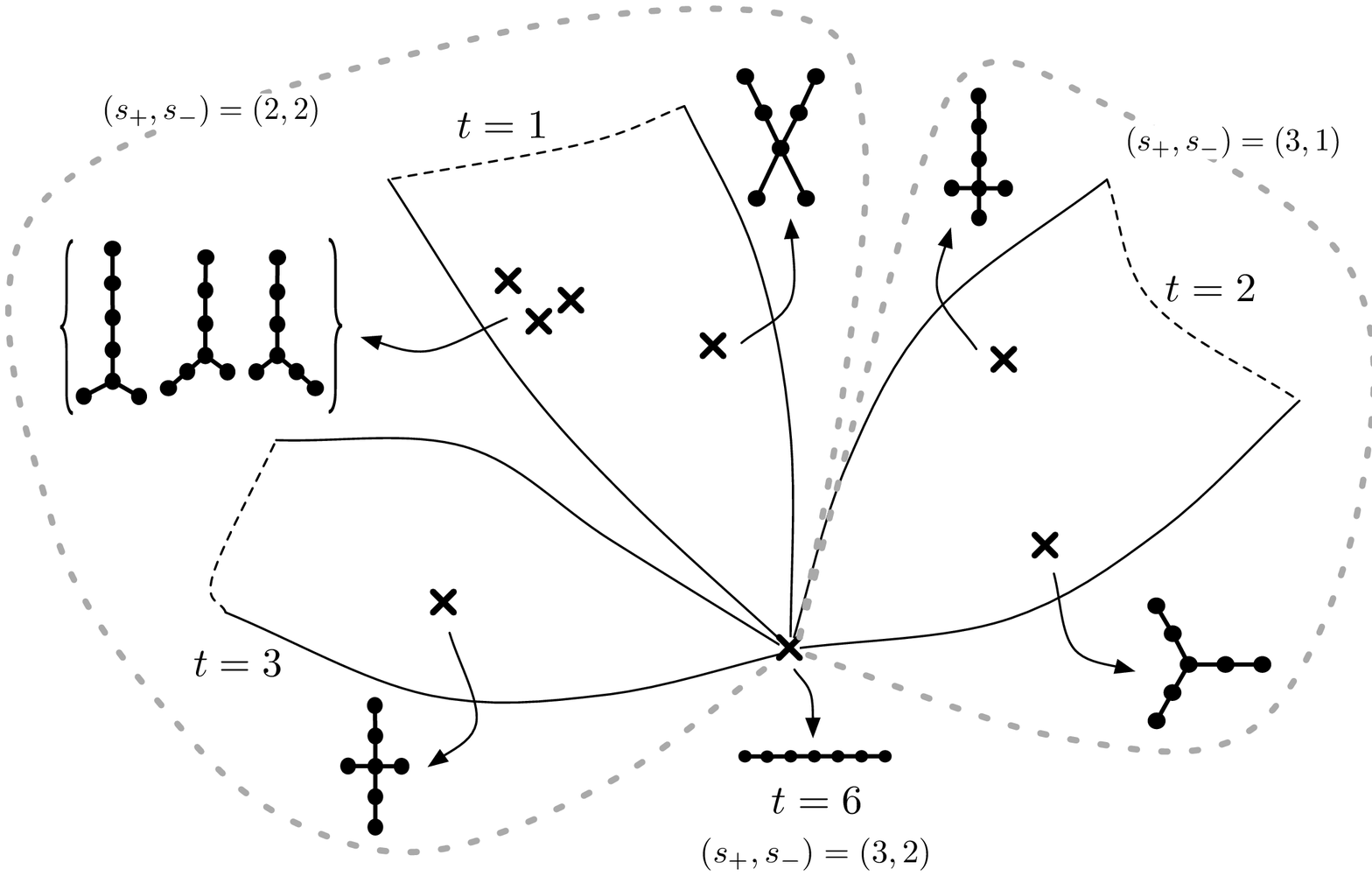}}
\noindent{\ninepoint\sl \baselineskip=3pt {\bf Figure 18}:{\sl $\;$
Summary of  the analysis of the $U(6)$ gauge theory and the location
of the trees in the various branches of the gauge theory moduli
space. The dotted lines indicated a coarse-grained classification of
$\CN=1$ branches based on the $(s_+,s_-)$ values. In this case the
branches are more finely distinguished by the confinement index $t$.
Branches meet at the point where another monopole becomes massless. }}

\medskip

Note that in the $t=1$ and $t=2$ branches we have both trivalent and
quartic dessins. From our discussion about valency lists, it follows
that they are in distinct Galois orbits. However, in each case, they
have the same value of $t$ and of $(s_+,s_-)$. This is where extra
gauge theory criteria are needed in order to distinguish these
isolated phases.

Let us concentrate on the pair of dessins with $t=2$. A simple way
to see that these correspond to two different phases is by tracing
them back to the problem in $U(3)$ broken to $U(1)\times U(2)$ and
then ``multiplying by 2". The trivalent dessin comes from the unique
dessin with three edges. Such a special point is where $P_3^2(z)-1 =
z^3(z^3-1)$. This is actually known to be a superconformal field
theory in the IR (see \refs{\AD, \Shih}\ and references therein). On the other
hand, the quartic dessin comes from a generic point in the $U(3)$
theory and therefore it is a different phase.

This discussion proves that the two dessins are in different phases.
However, we want to go further and show that even the theory
corresponding to the quartic dessin is a distinct phase from its
neighbors in the $t=2$ branch. Recall that in Section $4.2$ we
argued that there might be extra chiral ring relations that
characterize the corresponding special points within each branch.
Let us see this in detail in this case.

For the quartic factorization \exone\ in the branch defined by $s_+=s_-=2$, we take
\eqn\deffunc{ \tilde{H}_{2}(z)= (z-c)^2 \,, \qquad  R_2(z)=z^2+\a
z+\b\,, \qquad H_2(z)=z^2+\gamma z+ \delta \,. }
where we have adopted the notations in \asli. Using these in the definition of the $t_r$ in Section 3.2 we find the following chiral ring relation that is satisfied {\it only} at the special (quartic) point in the $\CN=1$ branch:
\eqn\newholo{ 4\, t_0\,  t_2^3-3\, t_1^2\, t_2^2+4\, t_1^3\, t_3-6\,
t_0\, t_1\, t_2\,t_3+t_0^2\,t_3^2-4t_1^3\, \Lambda^{6}+6\,
t_0\,t_1\,t_2\,\Lambda^{6}-2t_0^2\,t_3\,
\Lambda^{6}+t_0^2\,\Lambda^{12} =0 \,. }
Note that this relation uses the $\Lambda^6$ term in \juicy. In this
polynomial, each term has the same R-charge, or equivalently the
same dimension, and also the same $Q_{\Phi}$. To see this, note that
$\Lambda$ naturally has $Q_{\Phi}=1$. However, in \newholo\ we have
set the coupling of the cubic term in the superpotential to $1$.
Such a coupling $g$ is dimensionless, has $Q_{\Phi}=-3$ and shows up in \newholo\ in
the combination $g\Lambda^6$ which then has $Q_{\Phi}=3$.

Our new relation \newholo\ is not satisfied at any other point in the $t=2$ branch apart from the special point under consideration. At any other point in the same branch one can show that all $t_r$'s with $r=0,\ldots , 5$ are independent. In other words, one can at most find a relation similar to \newholo, which gives $t_6$ in terms of the other six. Assuming that our conjectures about the physical order parameters being Galois invariants are correct, this concludes our discussion of the trees from the physics point of view as we have, using purely gauge theory criteria, managed to classify the dessins into distinct Galois orbits.

\subsec{$U(6)$ Gauge Theory: A Mathematician's Point Of View }

We now exhibit how a mathematician would tackle the same problem of classifying dessins into Galois orbits. We start with a particular valency list and find all solutions to the associated polynomial equations using differentiation methods. In the end, one generically finds a polynomial that factors over $\Bbb{Q}$. Each factor corresponds to a different Galois orbit.  

For the $U(6)$ gauge theory perturbed by a cubic superpotential, there are three distinct valency lists possible for the rigid factorization problem:

$\bullet$ Consider the branchless tree shown in Figure $17$. From the valency list, one gets the polynomial equation
\eqn\exDS{
P_6^2(z) - 4 = F_2(z)H_{5}^2(z) = (z^2-4)H_{5}^2(z) \,,
}
where, in the second equality, we have suitably shifted and scaled $z$. In Appendix A of \AshokCD\ we have already shown how to obtain the solution to this equation using the differentiating trick. The solutions are Chebyshev polynomials. Plotting the roots of the polynomials, we get back the tree in Figure $17$.

$\bullet$ The trees shown in Figures $10$, $12$ and $13$ have the same valency list and arise from the  polynomial equation
\eqn\exone{
P_6(z)^2 - 4 = F_4(z)\,H^2_2(z)\, Q^4_1(z) \,.
}
Differentiating \exone\ we get
\eqn\exthree{
2\,P_6(z)P'_6(z) = H_2(z)\,Q_1^3(z)\big(F'_4(z)H_2(z)Q_1(z)+2\,F_4(z)H'_2(z)Q_1(z)+4\,F_4(z)H_2(z)Q'_1(z)\big)\,.
}
Since all polynomials involved are monic, it is easy to see that this leads to two equations
\eqn\exfour{\eqalign{
P'_6(z) &= 6\,H_2(z)\, Q_1^3(z)\,, \cr
12\, P_6(z) &= F'_4(z)H_2(z)Q_1(z)+2\,F_4(z)H'_2(z)Q_1(z)+4\,F_4(z)H_2(z)Q'_1(z) \,.
}}
After scaling and shifting the $z$ variable, one can write
\eqn\exfive{\eqalign{
H_2(z) = z^2-1\,, &\qquad Q_1(z) =z+q_1\,, \qquad \cr
P_6(z) = z^6+\sum_{i=1}^6\, p_i\, z^{6-i}\,, &\qquad F_4(z) = z^4+ \sum_{i=1}^{4} f_i\, z^{4-i} \,.
}}
The first equation in \exfour\ leads to linear equations for the $p_i$, which we can easily solve to obtain
\eqn\exsix{\eqalign{
p_1 &= {18 \over 5}q_1\,, \qquad p_2 = {3\over 2}(3q_1^2-1)\,,  \cr
p_3 &= 2q_1(q_1^2-3)\,, \qquad p_4 = -9 q_1^2\,, \qquad p_5 = -6q_1^2 \,.
}}
Substituting this in the second of the two equations in \exfour\ leads to
\eqn\exseven{\eqalign{
f_1 &= {16\over 5}q_1\,, \qquad f_2 = {1\over 25}(79q_1^2-25)\,, \qquad f_3 = {2\over 25}q_1(7q_1^2-55) \,,\quad  \cr
f_4 &= {1\over 100}(-75-718 q_1^2-35 q_1^4)\,, \qquad p_6 = {1\over 100}(25+276 q_1^2+7 q_1^4)\,,
}}
such that $q_1$ satisfies the equation
\eqn\exeight{
q_1(q_1^2-25)(5 q_1^2+3) = 0 \,.
}
Each inequivalent solution of \exeight\ leads to a dessin. Thus, each dessin has associated to it a specific number field \leila. In \exeight, there are solutions obtained by an overall sign flip: these do not lead to inequivalent trees.

Plotting the roots of the polynomials for each of the cases $q_1=\{5,0,i\sqrt{3\over 5}\}$ respectively leads to the trees in Figures $10$, $12$ and $13$. Since each of the values of $q_1$ is a Galois orbit in itself (up to an overall sign), the three solutions lead to dessins that belong to different Galois orbits. Thus, they should have a different set of Galois invariants. That this is so can be checked by computing the monodromy groups. We will postpone further analysis to the discussion in Section $5.3$.

$\bullet$ The trivalent trees in Figures $11$, $14$, $15$ and $16$ all have the same valency list and arise from the polynomial equation
\eqn\excubic{
P_6^2(z) - 4 = F_3(z)\, H^2_3(z)\,Q_1^3(z) \,.
}
Differentiating the equation as before
%
leads to two equations
\eqn\cubictwo{\eqalign{
P'_6(z) &= 6 H_3(z)Q_1^2(z) \cr
12 P_6(z) &= F'_3(z)H_3(z)Q_1(z)+2F_3(z)H'_3(z)Q_1(z)+3F_3(z)H_3(z)Q'_1(z) \,.
}}
We can choose to parametrize the polynomials as
\eqn\cubicthree{\eqalign{
P_6(z) = z^6+ \sum_{i=1}^6\, p_i z^{6-i}\,, &\qquad H_3(z) = z^3+\sum_{i=1}^{3}\, h_i z^{3-i}\,, \cr
Q_1(z) = z \,, &\qquad F_3(z) = z^3+\sum_{i=1}^{3}\, f_i z^{3-i} \,,
}}
where we have used the shift symmetry to set the constant coefficient of $Q_1$ to be zero. We will not discuss the solution in detail here, as the analysis is similar to the one we did for the quartic factorization.  The solution set is parametrized by $(h_1,h_2)$ that satisfy the relation
\eqn\cubicfour{
24\,h_1^6-156\, h_1^4\, h_2+450\,h_1^2\,h_2^2-625\,h_2^3 = 0\,.
}
We find two branches of solutions :

$a)$ $h_1=h_2=0$ : This leads to the simple solutions
\eqn\cubicfive{
P_6(z)=z^6+2z^3+\half \,, \qquad H_3(z) = z^3+1 \,,\qquad f_3(z)=z^3+2\,.
}
The tree that corresponds to this solution is shown in the Figure $11$.

$b)$ $h_1, h_2 \ne 0$ : One can use the scaling symmetry to set $h_1=1$ and there are three solutions which are solutions to the cubic equation for $h_2$ in \cubicfour. These are given by
\eqn\htwo{
h_2 =  \left\{\matrix{ h_{2}^{(0)}=-{2\over 25}(-3-2\, (2)^{1\over 3}+2^{2\over 3}) \cr\cr
h_{2}^{(+)}={1\over 25}(6+2^{2\over 3}(1-i\,\sqrt{3})-2\, (2)^{1\over 3}(1+i\,\sqrt{3})) \cr\cr
h_{2}^{(-)}={1\over 25}(6+2^{2\over 3}(1+i\,\sqrt{3})-2\, (2)^{1\over 3}(1-i\,\sqrt{3})) }\right. \,.\qquad}
The three trees associated to $h_2^{(0)}$, $h_2^{(-)}$ and
$h_2^{(+)}$ are shown in the Figures $14$, $15$ and $16$
respectively. Since $h_2^{(0)}$, $h_2^{(\pm)}$ are solutions to the
polynomial equation \cubicfour\ which is irreducible over $\Bbb{Q}$,
the corresponding dessins are part of the same Galois orbit.
Moreover, the number field is $\Bbb{Q}(2^{1/3},w)$ with $w^3=1$ as
it should be from our discussion in Section 5.1.

For $h_2=h_2^{(0)}$ (Figure $14$), we present the polynomials that solve \cubictwo\ :
\eqn\cubicsix{\eqalign{
P_6(z)&=z^6+{6\over 5}z^5+{11250+7500\, 2^{1\over 3}-3750\, 2^{2\over 3}\over 31250}\,z^4 +{1000+4500\, 2^{1\over 3}-3000\, 2^{2\over 3}\over 31250}\,z^3 \cr
&\qquad\qquad+ {44+30\, 2^{1\over 3}-51 \, 2^{2\over 3} \over 31250} \cr
H_3(z) &= z^3+z^2 + {30+20\, 2^{1\over 3} -10\, 2^{2\over 3} \over 125}\, z+{2+9\, 2^{1\over 3}-6\, 2^{2\over 3}\over 125} \cr
F_3(z) &= z^3+{2\over 5}z^2 + {-15+20\, 2^{1\over 3} -10\, 2^{2\over 3} \over 125}\, z+{8-6\, 2^{1\over 3}\over 125}\,.
}}

By direct computation, we have therefore classified into Galois orbits the class of trees with $6$ edges that we considered in Section $5.1$. The results of the mathematical analysis of the rigid factorizations  are summarized in Figure 19. These coincide with the classification we obtained from the gauge theory analysis.

\medskip
\centerline{\epsfxsize=0.75\hsize\epsfbox{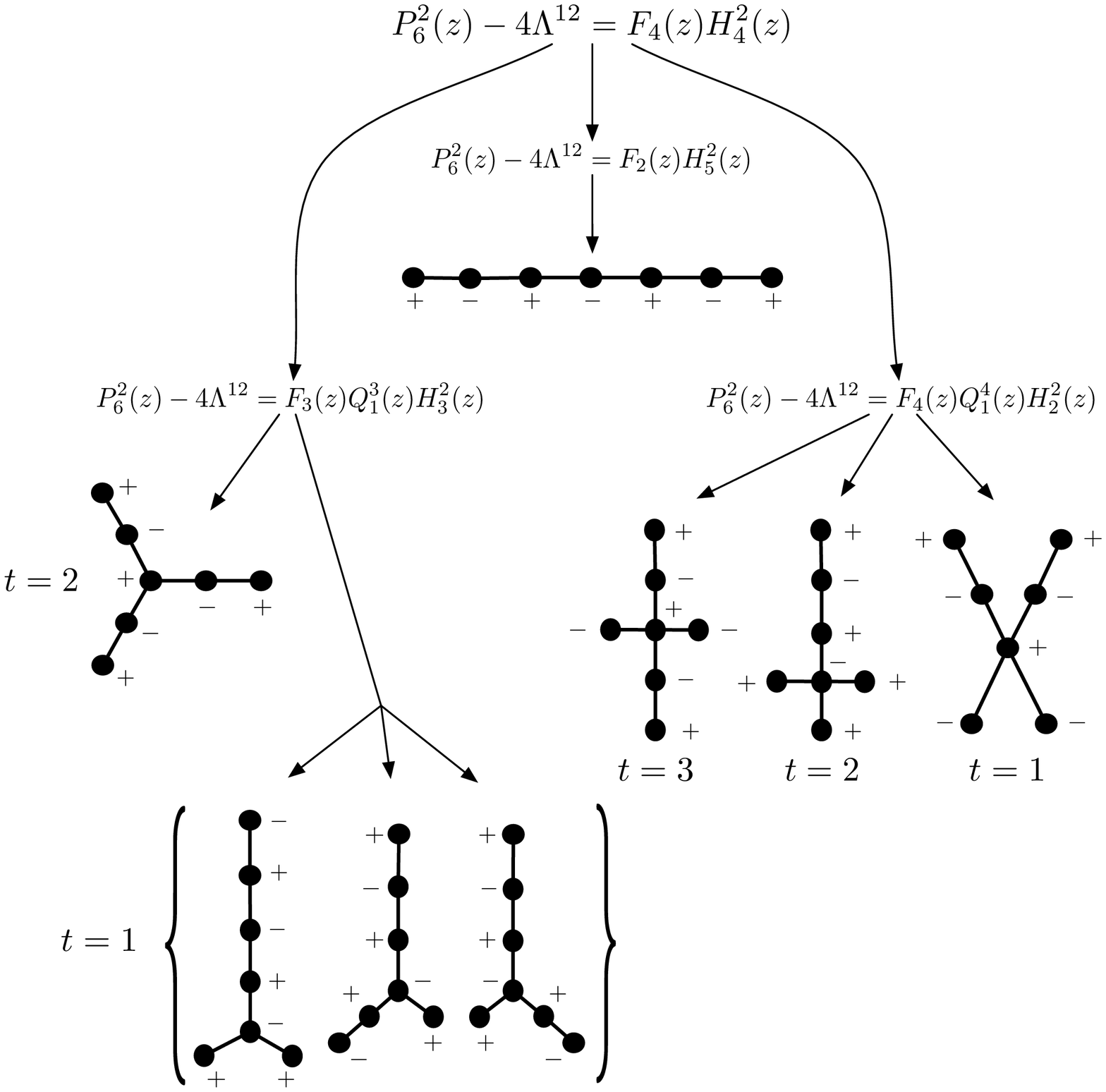}}
\noindent{\ninepoint\sl \baselineskip=3pt {\bf Figure 19}:{\sl $\;$
Summary of  the analysis of the $U(6)$ gauge theory and the
associated dessins. We have also included the results of gauge
theory analysis and indicated the confinement index and refined valency list of each figure.}}
\medskip

\subsec{Using Galois Invariants To Classify Dessins}

In the previous section, we have shown explicitly how dessins are organized into Galois orbits. We now attempt to rediscover the classification using the Galois invariants discussed in Section $3.1$, focusing on the factorizations in \exone\ and \excubic.

\medskip
$\bullet${\it Quartic Case}
\medskip

From the direct solution of the factorization problem, we see that there are three distinct Galois orbits that correspond, respectively, to the three distinct trees in Figures $10$, $12$ and $13$. These three trees can been partially distinguished by the more refined valency list introduced in Section $2.6$ for trees. Let us see this in detail.

Depending on how one distributes the roots between the two factors $(P(z) \pm 2)$, there are two distinct possibilities:
\eqn\quarticsplit{
P_6(z) - 2 =  \left\{\matrix{ \tilde{H}_1^2(z)Q_1^4(z) \cr \cr
F_2(z) Q_1^4(z) }\right.\qquad
}
If we assign negative valences to each of the zeroes of the
polynomials appearing on the right, then, the first choice singles
out the tree in Figure $12$ as the only possibility. On the other
hand, the second possibility is satisfied by the trees in both Figures
$10$ and $13$. We assign signs $+/-$ to the vertices in Figure $19$
to indicate these two possibilities.

In order to distinguish the remaining two dessins, we can compute the monodromy group of the trees\foot{All the monodromy groups have been obtained using the GAP software downloaded from http://www.gap-system.org.}. Here, we compute the monodromy groups of the trees with the refined bi-partite structure, as in Figure $7$ of Section $2.6$. Taking the difference of the two equations in \split\ we find an auxiliary polynomial equation that leads to a non-clean Belyi map, whose pre-images of $1$ are the vertices with $-$ valency and whose pre-images of $0$ are vertices with $+$ valency. The monodromy group for trees is therefore generated by $\sigma_{+/-}$, which correspond to the permutation of edges around the $+/-$ vertices respectively. 

\bigskip

\centerline{
\vbox{\hsize=2truein\offinterlineskip\halign{\tabskip=2em plus2em
minus2em\hfil#\hfil&\vrule#&\hfil#\hfil\tabskip=0pt\cr
Figure &depth6pt& Monodromy group\cr \noalign{\hrule}
$10$&height14pt&$(S_3\times S_3)\ltimes C_2$\cr
$12$&height14pt&$C_2\times S_4$\cr $13$&height14pt&$S_5$\cr
$11$&height14pt&$C_3\times S_3$\cr $14,15,16$&height14pt&$S_6$
\cr\noalign{\medskip} }}} \nobreak \centerline{ \vbox{\hsize=3truein
\noindent Table 1. Monodromy groups for dessins that occur in the
pure $U(6)$ gauge theory perturbed by a cubic superpotential. }}

\bigskip

The groups of all the trees we have encountered in the $U(6)$ example have been collected in Table $1$. $S_n$ is the permutation group of $n$ elements while $C_n$ is the cyclic group of $n$ elements. The monodromy group turns out to be different for the trees in Figures $10$ and $13$: for Figure $10$, we get the monodromy group $(S_3\times S_3)\ltimes C_2$, of order $72$, while for Figure $13$, we get the monodromy group $S_5$, of order $120$.

$\bullet${\it Cubic Case}

\medskip
The discussion parallels the one for the quartic factorization. The
two possibilities of distributing the roots
\eqn\quarticsplit{
P_6(z) - 2 =  \left\{\matrix{ F_3(z) Q_1^3(z)   \cr \cr
F_2(z) \tilde{H}_2^2(z)}\right.\qquad }
correspond to two distinct refined valency lists that distinguishes the tree in Figure $11$ from any one of the trees in Figures $14$, $15$ or $16$. In this case, no further invariant is required to distinguish them.

Thus, we find that the classification of dessins into Galois orbits agrees with what we obtained in Section $5.2$ regarding the classification of isolated phases in gauge theory. So far we have only considered dessins that are trees. We now generalize our discussion and consider more general dessins. This will highlight some open questions related to the phases of gauge theories with flavour.

\newsec{Gauge Theories With Flavour}

In this section we turn to a discussion of gauge theories with matter. The dessins that appear at isolated singularities in the moduli space will no longer be trees. We will mostly focus on the curves that were discussed in \AshokCD, with isolated Argyres-Douglas singularities in the moduli space. 

We start with a general discussion of the non-rigid factorization
\eqn\nrcurve{ P_N^2(z) + B_L(z) = Q_{2n}(z)H_{N-n}^2(z) \,,}
where we have exhibited the degrees of the polynomials explicitly. This curve arises from a $\N=2$ $U(N)$ gauge theory with $N_f$ massive flavors broken to $\CN=1$ by a tree level superpotential
\eqn\masski{ W_{\rm tree} = \Tr W(\Phi) + \tilde Q_{\tilde f}m_f^{\tilde
f}(\Phi)Q^f \,,}
where $f$ and $\tilde f$ run over the number of flavors $N_f$ and
\eqn\defina{W(z) =\sum_{k=1}^{n+1}{g_k\over k}z^{k}, \qquad
m_f^{\tilde f}(z) = \sum_{k=1}^{l+1}m_{f, k}^{\tilde f} z^{k-1}.}
The $\CN=1$ vacua, as before, are those for which $Q_{2n}(z) = W'(z)^2 + f(z)\,,$ where $f(z)$ is a polynomial such that ${\rm deg}\,(f) ={\rm deg}\,(W'(z))/2 -1$. $m(z)$ is a matrix of polynomials of size $N_f\times N_f$.

It turns out that the only information about the superpotential $\tilde Q_{\tilde f}m_f^{\tilde f}(\Phi)Q^f$ which is relevant for the curve \nrcurve\ is the polynomial \refs{\Kapustin, \SeibergF, \CachazoSWtwo}
\eqn\detti{ B_L(z)  = {\rm det}\, m(z).}
Clearly, plenty of choices of $m(z)$ can lead to the same $B_L(z)$.

The particular class of dessins we are interested in arise when
$B_L(z)$ has only $n+1$ distinct roots. We use our shift and scale
symmetry to set $B_L(z)$ to be of the form
\eqn\agapo{ B_L(z) = \alpha\,
z^{m_0}(z-1)^{m_1}\prod_{j=2}^{n}(z-p_j)^{m_j}\,.}
Two natural ways of obtaining such $B(z)$'s are the following:

\item {$\bullet$} $N_f= L={\rm deg}\, B_L(z)$ and
$m_f^{\tilde f}(z)$ a constant diagonal mass matrix with $m_0$
masses equal to $0$, $m_1$ masses equal to $1$, and $m_j$ masses
equal to $p_j$.

\item {$\bullet$} $N_f= n+1$ and $m_f^{\tilde f}(z)$ a diagonal matrix with
polynomial entries $z^{m_0}$, $(z-1)^{m_1}$, and $(z-p_j)^{m_j}$.

The former leads to a theory with unbroken $\N=2$ supersymmetry if
there is no $W(z)$. Moreover, it has a large flavor symmetry
classically. The latter, on the other hand, has a very small number
of flavors and generically no special flavor symmetry.

What we now do to obtain the Argyres-Douglas (AD) dessins studied in
\AshokCD\ is to further tune the masses of the flavors and the
parameters of the superpotential to set 
\eqn\finetune{
H_{N-n}(z)=Q_{2n}(z)\, R_{N-3n}(z) \,.
}
This leads to the rigid factorization problem \AshokCD\
\eqn\adcurve{ P_N^2(z)+B_L(z) = Q_{2n}^3(z)\, R_{N-3n}^2(z)\,. }
We will focus on such factorizations in the rest of the section since many explicit solutions to this class of factorization problems have already been obtained in \AshokCD. We will exploit these solutions to discuss some interesting issues in gauge theory. 

\subsec{$U(10)$ Gauge Theory With Flavour}

Consider the specific case of dessins arising from the factorization
problem of the second example in Section $2.3$:
\eqn\juha{ P^2_{10}(z) + \alpha z^5(z-1)^5(z-t)^5 =
Q_4^3(z)R^2_4(z).}
This problem was completely solved in \AshokCD. There
are two inequivalent solutions to \juha. Borrowing the explicit
solutions from \AshokCD\ one can plot the zeroes of the polynomials
as before; we have drawn the corresponding dessins in Figure $20$
and $21$.

It turns out that the monodromy group distinguishes between the two
dessins and therefore they belong to different Galois orbits. For
the two dessins considered here, the relations $\sigma_1^2 =1$ and
$\sigma_0^6=1$ are satisfied. Let us denote the two monodromy groups
corresponding to the two dessins by $M_{1}$ and $M_{2}$
respectively. In order to explicitly write down the generators, it
is useful to number the half-edges of the dessin, as in the figures $20$
and $21$.

From the definitions, one can check that

$-$ $M_{1}$ is generated by
\eqn\figten{\eqalign{
\sigma_1 &= (1,2)(3,4)(5,6)(7,8)(9,10)(11,12)(13,14)(15,16)(17,18)(19,20) \cr
\sigma_0 &= (1,10,11)(2,15,3)(4,5)(6,19,7)(8,9)(12,13)(14,20,18)(16,17) \,.
}}
It has order $30720$ and has $84$ conjugacy classes.

$-$ $M_{2}$ is generated by
\eqn\figeleven{\eqalign{
\sigma_1 &=(1,2)(3,4)(5,6)(7,8)(9,10)(11,12)(13,14)(15,16)(17,18)(19,20) \cr
\sigma_0 &= (1,10,11)(2,13,3)(4,19,5)(6,7)(8,9)(12,20,18)(14,15)(16,17) \,.
}}
It has order $30720$ and has $63$ conjugacy classes.

Since $M_1\neq M_2$, the two dessins belong to distinct Galois orbits. We can now ask if it is possible, from the gauge theory analysis, to distinguish between them. Note that the holomorphic invariants introduced in section 3.2 do not give any information in this case, as it is not possible to factorize \juha\ as $P(z)\pm f(z)$, for some polynomial $f(z)$, unlike the pure gauge theory case. This is reflected in the fact that from the mathematical point of view, it is not possible in general to define a refined valency list when the dessin is not a tree.

\medskip
\centerline{\epsfxsize=1.0\hsize\epsfbox{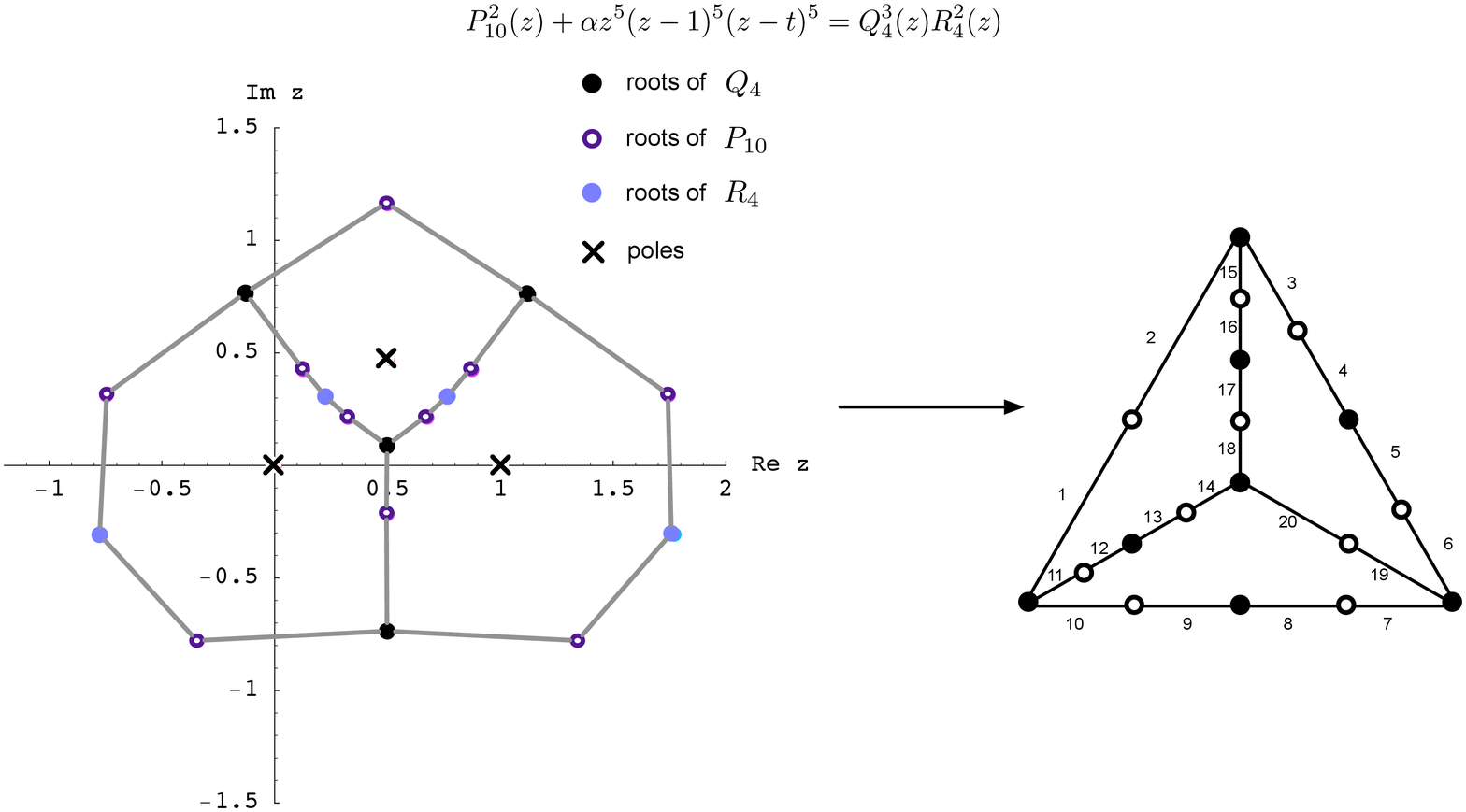}}
\noindent{\ninepoint\sl \baselineskip=3pt {\bf Figure 20}:{\sl $\;$
One of the two dessins arising from the factorization \juha. Each face is bounded by three line segments containing $1,2$ and $2$ edges respectively. The figure to the right is a schematic version of the dessin,  with the edges numbered to aid the computation of the monodromy group.}}

\medskip
\centerline{\epsfxsize=.95\hsize\epsfbox{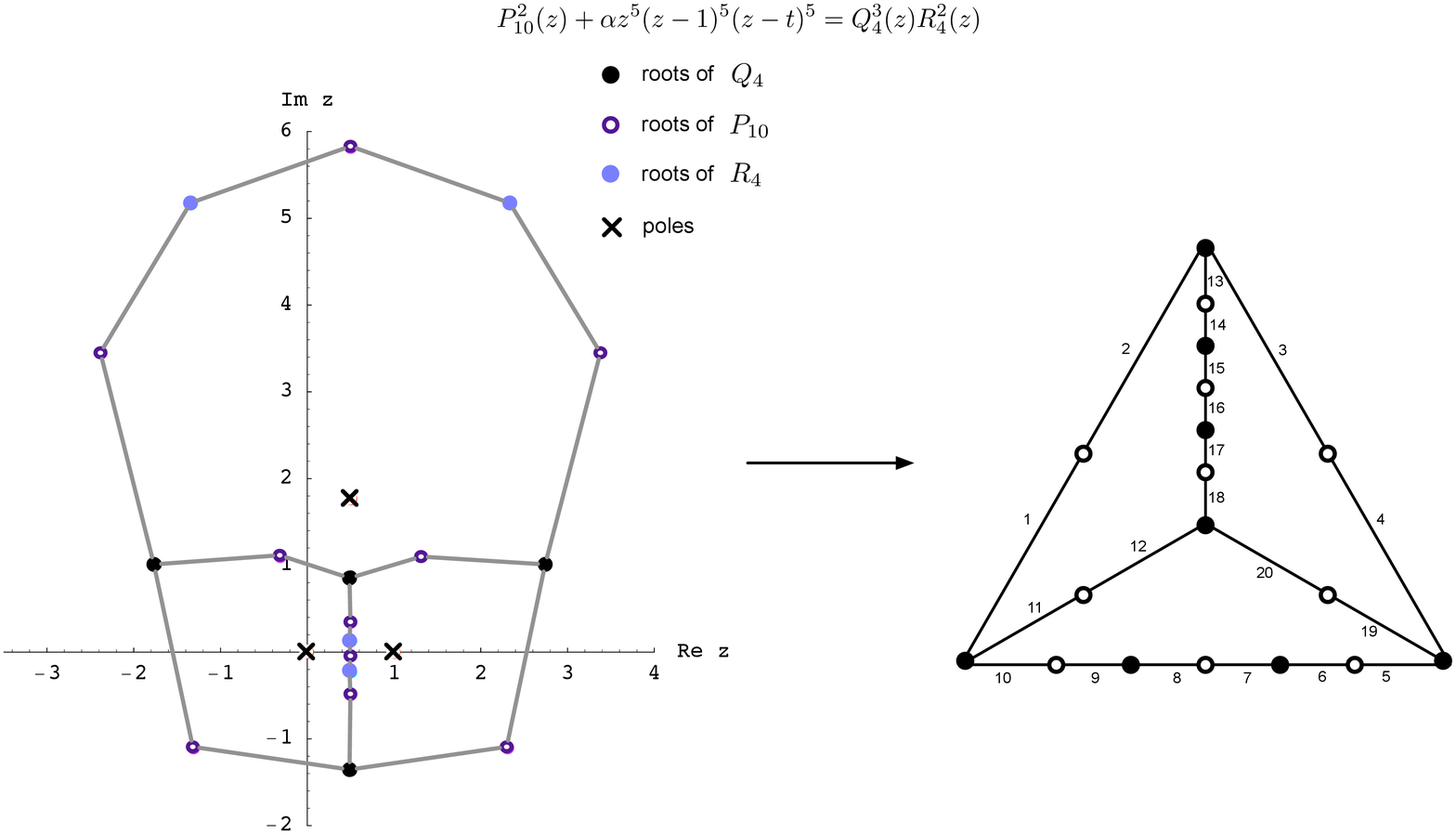}}
\noindent{\ninepoint\sl \baselineskip=3pt {\bf Figure 21}:{\sl $\;$
The other dessin arising from \juha. Each face is bounded by $1,1$ and $3$ edges between trivalent vertices.}}
\medskip

Similarly, although the combinatorial definition of the confinement index still makes sense, there is no sense in which confinement is a good order parameter for gauge theories with matter. So although $t$ might still be a good Galois invariant, its physical interpretation is unclear.
Some preliminary analysis of the phases have already been attempted in \Bala. We leave a more detailed study of theories with flavor  for future work.

\newsec{Conclusions And Open Questions}

The central theme of this work has been the relation between dessins d'enfants and supersymmetric gauge theory. In this section, we discuss some of the results we have obtained and list some of the immediate questions and future directions of research that have emerged from our analysis.

We have seen that any clean dessin with $N_c$ edges and $N_f+1$ faces can be found at an isolated singularity in the moduli space of an $\CN=2$ $U(N_c)$ gauge theory with $N_f$ flavours. The particular rigid factorization of the Seiberg-Witten curve that corresponds to the isolated singularity is determined by the valency lists of the dessin. Typically there are many solutions to this factorization problem. For each such solution, the rigid curve determines a rational Belyi map $\b(z)$ whose inverse image of the $[0,1]$ interval gives a dessin $D$ on the sphere. Such an $\CN=2$ gauge theoretic perspective matches very closely the mathematical point of view of obtaining and classifying dessins into Galois orbits by direct solution of the polynomial equation.

From a physics point of view, it is also natural to study what we called the non-rigid factorizations \nonrigid. The solutions to these equations are interpreted as the space of vacua that preserve $\CN=1$ supersymmetry when the $\CN=2$ theory is perturbed by a tree level superpotential. One can associate to these vacua disconnected graphs (which are not dessins), that come together and join to form a dessin at any of the isolated singularities mentioned in the previous paragraph. The difference lies in the fact that these are now looked upon as $\CN=1$ vacua.

Interestingly, it is this $\CN=1$ point of view that has a nice counterpart in the more refined mathematical approach to the study of dessins, which is to define a complete list of Galois invariants that distinguish the dessins that belong to different Galois orbits. Based on the examples we have worked out, we have been led to conjecture a relation between the mathematical programme of classifying dessins and the physics programme of classifying phases of $\CN=1$ gauge theory.
The strongest form of the conjecture states that every Galois invariant is a physical order parameter that distinguishes different phases in the gauge theory.

One example of a Galois invariant that might give new physical information is the monodromy group associated to the dessins. In gauge theory, the group usually discussed in the context of Seiberg-Witten theory is the S-duality group: for a genus $g$ Seiberg-Witten curve, it is an $Sp(2g,\IZ)$ group that acts on the $A_i$ and $B_i$ cycles of the Riemann surface. These cycles correspond to the edges that go between the filled vertices of the dessins (the pre-images of $0$ under the Belyi map). However, in general the monodromy group of the dessin involves action on half-edges that go between the pre-images of $0$ and $1$ and involves the zeroes of $P(z)$, apart from the zeroes of $y_{SW}$. It is therefore different from the $S$-duality group, but what exactly the group signifies in the gauge theory is an open question.

In the other direction, one can ask whether the $\CN=1$ gauge theoretic way of finding the dessins, first by solving a more general (non-rigid) factorization problem and {\it then} imposing suitable constraints so that one approaches an isolated singularity in the moduli space, is useful from a mathematical point of view. From our proof that the confinement index is a Galois invariant, it seems that the answer is yes. More generally we have seen that all the order parameters have a simple interpretation in the mathematical literature as Galois invariants. We believe that in this direction the correspondence is on much firmer ground. 

In Section $6$, we discussed dessins that appear in the moduli space of a $U(N)$ gauge theory with matter. Much less is known about the possible phases of the corresponding $\CN=1$ vacua. We exhibited two dessins in a $U(10)$ theory with flavour, which, from the mathematical point of view belong to distinct Galois orbits as they have different monodromy groups. However, from the physics point of view, with the available order parameters, it seems that the points where the two dessins appear describe the same phase. By this we mean that none of the known order parameters are of any use to distinguish between them. If our stronger conjecture that Galois invariants map to order parameters is correct, the monodromy group should correspond to an order parameter in physics that can distinguish the two special points where the dessins appear.

Many possible generalizations of our work present themselves. Since the dessins can be drawn on any two dimensional topological surface, it should be possible to extend the correspondence we have found to dessins drawn on genus $g \ge 1$ Riemann surfaces. On the gauge theoretic side, it would be very interesting to classify the dessins that appear in the moduli spaces of more general gauge theories, such as $SO/Sp$ gauge groups, quiver gauge theories with products of $U(N)$ factors, etc.

The study of the $\CN=1$ branches in \CachazoYC\ uses the Dijkgraaf-Vafa relation between gauge theory and matrix models \refs{\DVone, \DVtwo, \DVthree}. In this relation, two distinct hyperelliptic Riemann surfaces emerge \CachazoDSW: the Seiberg-Witten curve and the spectral curve of the matrix model. In the examples considered in the text, the spectral curve was equivalent to the reduced Seiberg-Witten curve in \reduced. However this is not true in general as discussed in the examples in Appendix C. It is conceivable that one can tune the parameters of the tree level superpotential so that the spectral curve develops an isolated singularity, leading to a Belyi map. It might be very interesting to study the dessins that arise this way.

As we have stressed throughout, the main objective of this article was to provide the rudiments of a dictionary between the physics of supersymmetric gauge theory and the mathematics related to the action of the absolute Galois group on the children's drawings of Grothendieck. Much more work needs to be done to fully understand the correspondence between these two fascinating fields of study, which we hope will lead to a deeper and fuller understanding of the relevant physical and mathematical problems.

\medskip

\centerline{\bf Acknowledgments}

\medskip

The authors would like to thank Emanuel Diaconescu, Jaume Gomis, Kentaro Hori, Sheldon Katz, Kristian Kennaway, Gregory Moore, Sameer  Murthy, Rob Myers, Christian R\"omelsberger, Nathan Seiberg and David Shih for discussions. We would especially like to thank Edward Witten for
very useful discussions and suggestions on the manuscript. ED would like to thank Perimeter Institute for hospitality while most of the work was completed. The research of SA and FC at the Perimeter Institute is supported in part by funds from NSERC of Canada and MEDT of Ontario.

\appendix{A}{Field Theory And The Absolute Galois Group}

The main mathematical object that appears in this paper is the
absolute Galois group ${\rm Gal}(\bar{\Bbb{Q}}/\Bbb{Q})$. It is 
of central importance in many areas of mathematics. Since this
group is not very familiar to physicists, in this appendix we give a
short description of the definition stated in the text: ${\rm
Gal}(\bar{\Bbb{Q}}/\Bbb{Q})$ is the group of automorphisms of the
field $\bar{\Bbb{Q}}$ of algebraic numbers that leaves $\Bbb{Q}$
fixed.

Instead of directly studying $\GG$, which is a group that cannot
even be finitely generated, we will study the relevant concepts by
first reviewing Galois groups of finite order. Just to give an idea
of the complexity of $\GG$ it is nice to mention that mathematicians
are considering the possibility that any finite group can arise as a
projection of $\GG$; this is the so-called ``Inverse Problem of
Galois Theory"\foot{The more precise statement is that any simple
group $S$ might be the Galois group of a finite normal extension of
$\Bbb{Q}$ and hence there will be a natural restriction map from
$\GG$ onto $S$.}.

Before going into the details about the different elements that
enter in the definition of Galois groups let us lay down some field
theory basis. In this review we will assume familiarity with
definitions of fields, rings of polynomials and basic group theory
(for a very basic introduction and more details of the main example
in this appendix see \fraleigh).

Let us start by recalling some basic definitions. We say that a
field $E$ is an extension of a field $F$, denoted by $F\leq E$, if
$E$ has a subfield isomorphic to $F$. Examples of fields and
extensions are $\Bbb{Q}\leq\Bbb{Q}(2^{1/4},i)\leq\bar\Bbb{Q}$. Our
first goal is to review the meaning of expressions such as
$\Bbb{Q}(2^{1/4},i)$.

Given a field $F$, a natural object to study is the ring of
polynomials, $F[z]$, with coefficients in $F$. From now on we will
assume that $F$ is a field of characteristic zero or a finite field.
This is to avoid certain pathologies that can happen otherwise. Of
course, our final target, which is $\Bbb{Q}$, has characteristic
zero.

An element $a\in E$ is called algebraic over $F$ if it is a zero of
a polynomial $p(z)$ in $F[z]$. $E$ is called an algebraic extension
of $F$ if all its elements are algebraic over $F$. We will only
consider algebraic extensions from now on. It turns out that there
exists a unique monic irreducible (over $F$) polynomial such that
$p(a)=0$ (since $F$ is of characteristic zero or finite, $a$ can
only be a zero of order one. This is an example of one of the
possible pathologies we have avoided). Since such a $p(z)$ is unique
we call it $p_a(z)$. The degree of $p_a(z)$, ${\rm deg}(p_a(z))=n$
is also called the degree of $a$ in $F$.

A special class of algebraic extensions are those with the structure
of a vector space with basis $\{ 1, a,\ldots , a^{n-1}\}$ and
coefficients in $F$. These are called {\it simple extensions} and
are denoted by $F(a)$. In general, if $F\leq E$ and $E$ is of finite
dimension $n$ as a vector space over $F$, we say that $E$ is a {\it
finite extension}\foot{Quite nicely, in our case (with $F=\Bbb{Q}$),
all finite extensions are also simple! This is called the primitive
element theorem.} of degree $|E:F| = n$ over $F$.

An example is $F=\Bbb{Q}$ and $a=2^{1/4}$. Then $\Bbb{Q}(2^{1/4})$
is generated by $\{ 1, 2^{1/4},2^{1/2},2^{3/4}\}$ since $p_{a}(z) =
z^4- 2$ has degree $n=4$.

Note that $z^4-2$ is reducible over $\Bbb{Q}(2^{1/4})$, in fact,
$z^4-2=(z-2^{1/4})(z+2^{1/4})(z^2+2^{1/2})$. The last factor,
$(z^2+2^{1/2})$, is irreducible of degree 2 in $\Bbb{Q}(2^{1/4})$.
So if we {\it adjoin} a root of $z^2+2^{1/2}$ to $\Bbb{Q}(2^{1/4})$
then $z^4-2$ {\it splits} over this new field.

The new element we need is $2^{1/4}i$. However, multiplying by
$2^{-1/4}\in \Bbb{Q}(2^{1/4})$ we get $i$. Therefore, the new field
is $(\Bbb{Q}(2^{1/4}))(i) = \Bbb{Q}(2^{1/4},i)$. The latter notation
shows the fact that the order in which we adjoint $2^{1/4}$ and $i$
to $\Bbb{Q}$ is irrelevant.

We have achieved our first goal: $\Bbb{Q}(2^{1/4},i)$ is called the
{\it splitting field} of $z^4-2$. More generally, an extension $E$
of $\Bbb{Q}$ is a splitting field if there is an irreducible
polynomial in $\Bbb{Q}[z]$ such that $E$ is the smallest field that
contains all its roots.

Now we need to introduce the concept of the algebraic closure of a
field $F$. A field $K$ is called algebraically closed if every
non-constant polynomial in $K[z]$ has a root in $K$. Such a $K$ is
called an algebraic closure of $F$ if $K$ is an algebraic extension
of $F$, and it is denoted by $\bar F$. $\bar F$ is unique up to
isomorphisms.

The next goal is to study automorphisms of fields. Splitting fields
are important because given any one of them, say $E$ such that
$F\leq E\leq \bar F$, any automorphism of $\bar F$ that fixes $F$
maps $E$ onto itself and induces an automorphism of $E$ leaving
fixed $F$. Moreover, splitting fields are the only ones with this
property. The basic automorphisms are quite simple: if $a$ and $b$
are roots of the same irreducible polynomial, then the map
$\phi(a)=b$ with $\phi(q)=q$ if $q\in F$ is an automorphism. $a$ and
$b$ are called conjugates and $\phi$ is a conjugation. Automorphisms
of $E$ that leave fixed $F$ form a group under composition denoted
by $G(E/F)$. If $E$ is a splitting field, then $G(E/F)$ is called
the Galois group of $E$ over $F$ and it is denoted by ${\rm
Gal}(E/F)$.

Let $E$ be a finite extension of $F$. The number of isomorphisms of
$E$ into $\bar F$ leaving $F$ fixed is the index $\{E:F\}$ of $E$
over $F$. It turns out that for the fields we consider
\eqn\kalos{ \{E:F\} = |E:F| = |G(E/F)|}
where $|G(E/F)|$ is the order of the group.

The next step in any algebra book would be to define separable
extensions. However, over $\Bbb{Q}$, all extensions are separable
and we do not need to worry about that. A very special role is
played by (separable) splitting fields which are then called {\it
finite normal extensions}\foot{The parenthesis around ``separable"
are there to indicate that it can freely be removed.}.

Now we are ready to discuss the {\it Fundamental Theorem of Galois
Theory}. The theorem states that if $E$ is a finite normal extension
of $F$ then there is a one to one correspondence between
intermediate extensions of $F$ and subgroups of ${\rm Gal}(E/F)$.
The correspondence is the following: to each extension $B$ of $F$
such that $B\leq E$, one associates the largest subgroup $G_B$ of
${\rm Gal}(E/F)$ that leaves $B$ fixed. Moreover, $B$ is a finite
normal extension of $F$ if and only if $G_B$ is a normal subgroup.
In fact, ${\rm Gal}(B/F)$ is isomorphic to the factor (or quotient)
group ${\rm Gal}(E/F) / G_B$.

Let us apply this to our example $E = \Bbb{Q}(2^{1/4},i)$. As
discussed above, $E$ is the splitting field of $z^4-2$. It has a
basis $\{ 1,a,a^2,a^3, i, ia, ia^2, ia^3\}$ where $a=2^{1/4}$.

Since $|E:\Bbb{Q}|=8$ we must have $|{\rm Gal}(E/\Bbb{Q})| = 8$. It
is a simple exercise to exhibit the eight automorphisms of $E$
leaving $\Bbb{Q}$ invariant (see section 47.2 of \fraleigh).
Studying the composition table one discovers that the group is
nonabelian. Moreover, ${\rm Gal}(E/\Bbb{Q})= D_4$, the dihedral
group (the symmetry group of a square). If we denote rotations by
$k\pi/2$ (with $k=0,1,2,3$) as $\rho_k$, mirror images (reflections)
as $\mu_i$ and diagonal flips as $\delta_i$, then the identification
with automorphisms is collected in the table below (only the action
on $a$ and $i$ is needed).

\medskip
\smallskip

\centerline{\vbox{\offinterlineskip \hrule \halign{\vrule # &
\strut\ \hfil #\ \hfil &\vrule # & \strut\ \hfil # \ \hfil  &\vrule
# & \strut\ \hfil #\ \hfil &\vrule # & \strut\ \hfil #\ \hfil  &
\vrule # & \ \hfil # \hfil \ & \vrule # & \ \hfil # \hfil \ & \vrule
# & \ \hfil # \hfil \ & \vrule # & \ \hfil # \hfil  \ & \vrule #& \
\hfil # \hfil  \ & \vrule #   \cr
&  && $\rho_0$ && $\rho_1$ && $\rho_2$ && $\rho_3$ && $\mu_1$ && $
\delta_1$ && $\mu_2$ && $\delta_2$  & \cr
\noalign{\hrule}
& $a \rightarrow$  && $a$ && $i\, a$ && $-a$ && $-i\, a$&& $a$&&
$i\,
a$  && $-a$&& $-i\, a$ &\cr 
\noalign{\hrule}
&   $i \rightarrow$ &&$i$ && $i$ && $i$ && $i$ &&$-i$ && $-i$ &&
$-i$ && $-i$ &\cr
}\hrule }}

\medskip

The lattice of all subgroups of $D_4$ is well known (see section
47.2 of \fraleigh). According to Galois theory there must be one and
only one intermediate extension of $\Bbb{Q}(2^{1/4},i)$ for each
subgroup. This gives rise to the lattice of intermediate extensions
of $\Bbb{Q}(2^{1/4},i)$. Let $K_{H}$ denote the subfield of
$\Bbb{Q}(2^{1/4},i)$ left fixed by the subgroup $H$ of $D_4$. For
example, it is easy to check that $ K_{\{ \rho_0,\rho_2\}} =
\Bbb{Q}(\sqrt{2},i)$. Note that $\Bbb{Q}(\sqrt{2},i)$ is also a
splitting field and hence a finite normal extension. One can easily
check that $\{\rho_0,\rho_2\}$ is indeed a normal subgroup of $D_4$!
Likewise, consider $K_{\{ \rho_0,\mu_1 \}} = \Bbb{Q}(2^{1/4})$. This
is not a splitting field and one can check that $\{ \rho_0,\mu_1\}$
is not a normal subgroup of $D_4$.

Now we can go back to our object of interest: $\GG$. Here we will
follow very closely an explanation given in \JonesS\ and we will
illustrate the main ideas using our example of $\Bbb{Q}(2^{1/4},i)$.
We have already explained the meaning of the algebraic closure of a
field. Here $\bar\Bbb{Q}$ is then the algebraic closure of
$\Bbb{Q}$, the field of algebraic numbers. This is clearly a
complicated object that can be constructed as the union of all
splitting fields over $\Bbb{Q}$, which as we know, are finite normal
extensions of $\Bbb{Q}$. Let $E$ denote a generic one, then
\eqn\que{ {\bar\Bbb{Q}} = \bigcup_{E\in {\cal E}} E}
where ${\cal E}$ is the set of all such extensions. For each
extension $E$ we have the corresponding Galois group, ${\rm
Gal}(E/\Bbb{Q})$, of $E$ over $\Bbb{Q}$.

Consider our favorite example, $E = \Bbb{Q}(2^{1/4},i)$. Its Galois
group over $\Bbb{Q}$ is ${\rm Gal}(E/\Bbb{Q}) = D_4$. Consider
$L=\Bbb{Q}(\sqrt{2},i)$. As we saw, $L\leq E$. Now, every
automorphism of $E$ leaves $L$ invariant. This is because $L$ is a
splitting field.
The Galois group of $L$ over $\Bbb{Q}$ is then the factor group
${\rm Gal}(L/\Bbb{Q}) = {\rm Gal}(E/\Bbb{Q}) / \{ \rho_0,\rho_2\}$.

Now there is a natural group epimorphism $\rho_{E,L}: {\rm
Gal}(E/\Bbb{Q})\to {\rm Gal}(L/\Bbb{Q})$ given by the restriction
map. That this is an epimorphism, i.e, an onto map, follows from the
fact that every automorphism of $L$ that fixes $\Bbb{Q}$ can be
extended to an automorphism of $E$ in $|E:L|$ ways. In our example
$|E:L|=2$. Consider the following automorphism of $L$:
$(\sqrt{2},i)\to (\sqrt{2},-i)$. This can then be extended to $E$ in
two ways a follows: $(2^{1/4},i)\to (\pm 2^{1/4},i)$.

Consider now $\Bbb{Q}(i) = K_{\{ \rho_0,\rho_1,\rho_2,\rho_3\}}$.
This is also a splitting field. We then have the following sequence
of finite normal extensions $\Bbb{Q}\leq \Bbb{Q}(i) \leq
\Bbb{Q}(\sqrt{2},i)\leq \Bbb{Q}(2^{1/4},i)$. From this sequence we
can make an observation that will be very important in the
definition of $\GG$: an element $(g_1,g_2)$ of the cartesian product
${\rm Gal}(\Bbb{Q}(i)/\Bbb{Q})\times {\rm
Gal}(\Bbb{Q}(\sqrt{2},i)/\Bbb{Q})$ can be extended to an element of
${\rm Gal}(\Bbb{Q}(2^{1/4},i)/\Bbb{Q})$ if and only if
$\rho_{L,\Bbb{Q}(i)}(g_2) = g_1$. This is because if $g\in {\rm
Gal}(\Bbb{Q}(2^{1/4},i)/\Bbb{Q})$ is one of the possible extensions
then it has to have a consistent action on each of the subfields.

The absolute Galois group $\GG$ can now be constructed in a very
similar way. It is a subgroup of the cartesian product of the Galois
groups of all finite normal extensions of $\Bbb{Q}$
\eqn\cartt{\GG < \prod_{E\in {\cal E}}{\rm Gal}(E/\Bbb{Q}) }
consisting of all elements $(g_{{\cal E}})\in \prod_{E\in {\cal
E}}{\rm Gal}(E/\Bbb{Q})$ (this is an infinite ``array'' with one
entry for each $E\in {\cal E}$) satisfying the constraint that
$\rho_{K_2,K_1}(g_{K_2}) = g_{K_1}$ whenever $K_1\leq K_2$. The
identification of each $g\in \GG$ with the element $(g_{\cal E})$
implies that $g_{E}$ is the restriction of $g$ to $E$. That this set
of restrictions is consistent follows from the condition involving
$\rho$.

Finally, the action of $g$ on $\bar\Bbb{Q}$ is determined by the
action on each finite normal extension via $g_E$. This is the basic
result we used in section 2.4 where we discussed the action of $\GG$
on dessins. We said that if $g\in \GG$ then $\eta$ acts on a
dessin by acting on the coefficients of the Belyi map. In other
words, the coefficients of the Belyi map, being found as solutions
to some set of polynomial equations, belong to a splitting field $E$
and $g$ acts via its restriction $g_E$.

\subsec{Glossary Of Terms In The Text}

\item {$\bullet$} An algebraic number is an element $a\in \Bbb{C}$
that generates a finite extension $\Bbb{Q}(a) \geq \Bbb{Q}$.

\item {$\bullet$} $\bar\Bbb{Q}$ is the field of all algebraic
numbers and it is also the algebraic closure of $\Bbb{Q}$.

\item {$\bullet$} A number field is a finite algebraic extension of $\Bbb{Q}$.

\item {$\bullet$} A monic polynomial is one whose monomial of highest degree has coefficient $1$.

\appendix{B}{The Multiplication Map As A Belyi-Extending Map}

It was shown in \refs{\CachazoIV, \CachazoYC}\ that once a solution
to the factorization problem \nonrigid\ is known for $U(N)$, then it
is possible to construct a solution to a similar factorization
problem for $U(tN)$. Let us first review this construction.

Consider the factorization problem
\eqn\appCone{
P_t^2(z) - 4\Lambda^{2t} = F_2(z)H^2_{t-1}(z) \,.
}
The solution is given by
\eqn\appCtwo{
P_t(z) = 2\Lambda^t\eta^tT_t\left({z\over 2\eta\Lambda}\right)\,, \quad F_2(z) = z^2-4\eta^2\Lambda^2\,, \quad H_{t-1}(z) = \eta^{t-1}\Lambda^{t-1}U_{t-1}\left({z\over 2\eta\Lambda}\right) \,,
}
where $\eta^{2t}=1$. $T_t(z)$ and $U_{t-1}(z)$ are the Chebyshev polynomials of the first and second kind respectively, defined by setting $z=\cos\theta$ and
\eqn\appCthree{
T_t(z) = \cos(t\theta) \qquad U_{t-1}(z) = {1\over t} {dT_{t} \over dz}(z) ={\sin(t\theta)\over \sin\theta}\,.
}
This implies that they satisfy the relation
\eqn\cheb{
1-T_{t}^2(z) = (1-z^2)\, U_{t-1}^2(z) \,.
}
Now suppose we have a solution to the factorization problem
\eqn\appCfour{
P_N^2(z)-4\Lambda_0^{2N} = F_{2n}(z)H^2_{N-n}(z) \,.
}
Then we can use the solution to \appCtwo\ to construct a solution to
\eqn\appCfive{
P_{tN}^2(z)-4\Lambda_0^{2tN} = \tilde{F}_{2n}(z)\tilde{H}^2_{tN-n}(z)
}
as follows:
\eqn\appCsix{\eqalign{
P_{tN}(z) &= 2\Lambda^{tN}\eta^{t}T_t\left({P_N(z)\over 2\eta\Lambda^N}\right)\,, \quad \tilde{F}_{2n}(z) = F_{2n}(z)\cr \quad H_{tN-n}(z) &= \eta^{t-1}\Lambda^{N(t-1)}H_{N-n}(z)\,U_{t-1}\left({P_N(z)\over 2\eta\Lambda^N}\right) \,,\qquad \Lambda_{0}^{2N} = \eta^2\,\Lambda^{2N}\,.
}}
This procedure to get exactly $t$ solutions to the $U(tN)$ theory from a given solution of the $U(N)$ theory was referred to as the multiplication map (by $t$) in \CachazoIV.

Let us now show that the multiplication by $t$ map can be used to construct new Beyi maps from old ones. The simple example of $t=2$ has already been discussed in Section $4$. As discussed in that section, the main point is to use the multiplication map to define the new Belyi map as
\eqn\belyiconjec{
\tilde{\b}_{t}(z) = 1- {P_{tN}^2(z)\over 4\Lambda_0^{2tN}} \,,
}
where $P_{tN}(z)$ is given by the first equation in \appCsix\ and $P_N(z)$ gives rise to a Belyi map
\eqn\belyiold{
\b(z) = 1- {P_{N}^2(z)\over 4\Lambda_0^{2N}} \,.
}
We would now like to exhibit \belyiconjec\ as the composition of a Belyi-extending map with the Belyi map \belyiold. For this purpose, let us define the polynomials
\eqn\evenodd{
T_{2k}(z)= M_{k}(z^2) \qquad\hbox{and}\qquad T_{2k+1}(z)=z\, S_{k}(z^2) \,.
}
This follows from the form of the Chebyshev polynomials \appCthree. If we now define the Belyi-extending maps
\eqn\belyiext{
\a_{2k}(u)=1-M_k^2(1-u) \qquad \hbox{and}\qquad \a_{2s+1}(u)=1-(1-u)S_k^2(1-u) \,,
}
one can check that
\eqn\newbelyi{
\tilde{\b}_t(z)=\a_t(\b(z))
}
is a Belyi map for all integer values of $t$. As for $\a_2$, one check that geometrically, the multiplication by $t$ replaces a given edge in a dessin by a branchless tree of length $t$.

\appendix{C}{Complete List Of Trees For $U(6)$}

In this appendix we study the problem of realizing all possible
connected trees with $6$ edges. Along the way we
will mention some interesting points about the structure of $\N=1$
vacua from the matrix model point of view \CachazoDSW.

In section $5$, we studied examples a pure $U(6)$ gauge theory
deformed by a cubic superpotential. However, a cubic superpotential
does not allow enough flexibility in the non-rigid Seiberg-Witten
curve to realize all possible trees with $6$ edges. We mentioned in the text that if we consider a superpotential $W(z)$ of degree $n+1$, then the curve describing the $\N=1$ vacua of
$U(N)$ is given by
\eqn\kass{ y^2 = P^2_N(z) - 4\Lambda^{2N} = (W'(z)^2 +
f_{n-1}(z))H^2_{N-n}(z)}
where $W'(z)$ has degree $n$.

However, this is true only for vacua with all values of $N_i$
different from zero. In other words, if $U(N)$ is classically
broken to $U(N_1)\times \ldots \times U(N_n)$. It turns out that if $n-s$
of the $N_i$'s are zero then the description of $\N=1$ vacua is more
subtle and it was elucidated in \CachazoDSW\ by using matrix model techniques inspired by the
Dijkgraaf-Vafa relation \refs{\DVone, \DVtwo, \DVthree}. The way to treat all cases at once is by
introducing another curve, called the matrix model curve,
$y_m^2=W'(z)^2 + f_{n-1}(z)$. Then, if $U(N)$ is broken classically
to $U(N_1)\times \ldots\times U(N_s)$ the factorization problems to be
solved are
\eqn\toge{\eqalign{ y^2 & = P^2_N(z) - 4\Lambda^{2N} = F_{2s}(z)
H^2_{N-s}(z), \cr y_m^2 & = W'(z)^2 + f_{n-1}(z)
=F_{2s}(z)R^2_{n-s}(z). }}

This means in particular, that the $(3,2)$ vacua in Section $5.1$ for the $U(6)$ example is the intersection of the $s=2$ branch with the $s=1$ branch, where the low energy gauge groups are $U(1)^2$ and $U(1)$ respectively. This intersection was studied in great generality in \Shih.

Note however, that the trivalent and four-valent factorizations in $U(6)$ do
not correspond to points where the $s=2$ branch intersects the $s=1$
branch. This is because both values of $N_i$ are nonzero. 
More generally, it is clear that isolated singular points of a theory
with $s=p$ are particular cases of those of a theory with $s=n$ for any 
$p<n$, since the corresponding branches can intersect.

The set of interesting values of $n$ in the case of $U(6)$ is
$n=1,\ldots, 6$.\foot{There is a subtlety when $n=6$ which is related
to the fact that $\tr \Phi^7$ in the superpotential is not an
independent quantity but this will not affect our discussion (see
\CachazoYC\ for more details).} We now turn to the construction of all possible dessins associated
to $U(6)$. It turns out that one only needs to consider a quartic
superpotential, i.e., $n=3$.

Let us prove that $n=3$ suffices. Instead of starting with $n=6$ and
descending all the way to $n=3$ it is best to use the fact that
isolated singular points give rise to all possible connected trees
with 6 edges. These trees must have
seven vertices.
All possible valency lists are the
following $(6,0,0,0,0,1)$, $(5,1,0,0,1,0)$, $(5,0,1,1,0,0)$,
$(4,2,0,1,0,0)$, $(4,1,2,0,0,0)$, $(3,3,1,0,0,0)$, and
$(2,5,0,0,0,0)$. Recall that a valence list $(u_1,\ldots, u_6)$
means that there are $u_k$ vertices with valence $k$. These were
obtained by requiring that the sum of all $u_k$'s is always seven
and that the sum of all $u_k$'s times $k$ is always $12$.

The previous valence lists lead to factorization problems of the
Seiberg-Witten curve of the form
$$ \eqalign{
 F_6(z)H_1^6(z),\qquad F_5(z)Q_1^2(z)H_1^5(z),\qquad
F_5(z)Q_1^3(z)H_1^4(z), \cr F_4(z)Q_2^2(z)H_1^4(z), \qquad
F_4(z)Q_1^2(z)H_2^3(z),\qquad  F_3(z)Q_3^2(z)H_1^3(z),}$$ and
$F_2(z)Q_5^2(z)$ respectively.

Now we can prove our claim by simple inspection. All these factorization problems are particular points in the space of curves (with $n=3$) given by
\eqn\kass{ y^2 = P^2_6(z) - 4\Lambda^{12} = F_6(z) H^2_{3}(z) \,.}

It is also easy to check that $F_6(z)H_1^6(z)$,
$F_5(z)Q_1^2(z)H_1^5(z)$, $F_5(z)Q_1^3(z)H_1^4(z)$ and
$F_4(z)Q_1^2(z)H_2^3(z)$ cannot possibly be obtained from the cubic
superpotential (the $n=2$ case considered in the text), while all other factorizations --
$F_4(z)Q_2^2(z)H_1^4(z)$, $F_3(z)Q_3^2(z)H_1^3(z)$ and
$F_2(z)Q_5^2(z)$ -- were considered in Section $5$.

The inherently $n=3$ factorizations give rise to only six different trees:

\medskip
\centerline{\epsfxsize=0.80\hsize\epsfbox{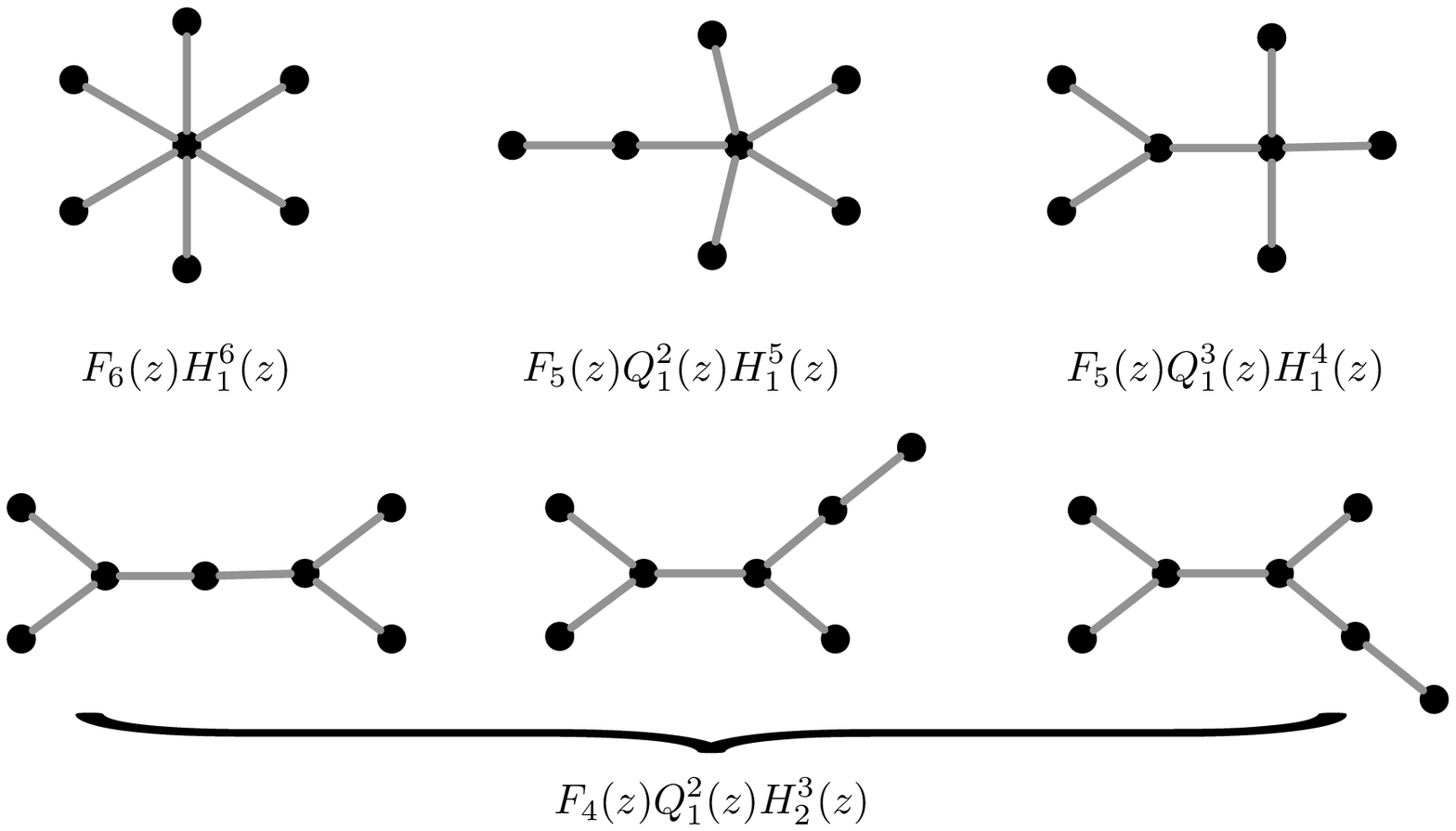}}
\noindent{\ninepoint\sl \baselineskip=3pt {\bf Figure 22}:{\sl $\;$
All possible dessins (and the associated factorizations) that appear
in the $\CN=2$ moduli space of the $U(6)$ gauge theory, apart from
the ones already discussed in the main body of the article. We have
omitted the vertices corresponding to the zeroes of $P_6(z)$.}}
\medskip

\listrefs
\end